\pdfoutput=1
\documentclass[11pt, reqno]{article}
\usepackage{jheppub}
\usepackage{epsfig}
\usepackage{amssymb}
\usepackage{amsmath}
\usepackage{mathrsfs}
\usepackage{hyperref}
\usepackage{multirow}
\usepackage{feynmp-auto}
\usepackage[utf8]{inputenc}

\def\be{\begin{equation}}
\def\ee{\end{equation}}
\def\ba{\begin{eqnarray}}
\def\ea{\end{eqnarray}}
\newcommand{\bea}{\begin{eqnarray}}
\newcommand{\eea}{\end{eqnarray}}

\def\Li{\textrm{Li}}
\def\l{\langle}
\def\r{\rangle}

\newcommand{\del}{\partial}
\def\fig#1{fig.~{\ref{#1}}}

\def\eqn#1{eq.~(\ref{#1})}
\def\Eqn#1{Equation~(\ref{#1})}
\def\eqns#1#2{eqs.~(\ref{#1}) and~(\ref{#2})}

\newcommand{\fwboxL}[2]{\text{\makebox[#1][l]{$#2$}}}

\newcommand{\cP}{{\cal P}}

\def\OL{{\cal O}}
\def\WL{{\cal W}}
\def\VL{\mathcal{V}}
\def\VLt{\tilde{\mathcal{V}}}
\def\cL{{\cal L}}
\def\zb{\bar{z}}

\def\lr{\leftrightarrow}
\def\Omt{\tilde{\Omega}}
\def\Omto{\tilde{\Omega}_{\rm o}}
\def\Omte{\tilde{\Omega}_{\rm e}}
\def\cDisc{{\rm Disc}_c}

\newcommand{\lsim}{\mathrel{\hbox{\rlap{\lower.55ex \hbox{$\sim$}} \kern-.3em \raise.4ex \hbox{$<$}}}}

\title{The Double Pentaladder Integral to All Orders}

\author{Simon~Caron-Huot,$^1$}
\author{Lance~J.~Dixon,$^{2,3,4,5}$}
\author{Matt~von~Hippel,$^{4,6}$}
\author{Andrew~J.~McLeod$^{2,3,6}$}
\author{and Georgios~Papathanasiou$^{3,7}$}

\affiliation{$^1$ Department of Physics, McGill University, 
3600 Rue University, Montr\'eal, QC Canada H3A 2T8}

\affiliation{$^2$ SLAC National Accelerator Laboratory,
Stanford University, Stanford, CA 94309, USA}

\affiliation{$^3$ Kavli Institute for Theoretical Physics, 
UC Santa Barbara, Santa Barbara, CA 93106, USA}

\affiliation{$^4$ Perimeter Institute for Theoretical Physics, 
Waterloo, Ontario N2L 2Y5, Canada}

\affiliation{$^5$ Laboratoire de physique th\'eorique,
\'Ecole normale sup\'erieure, 75005 Paris, France}

\affiliation{$^6$ Niels Bohr International Academy, Blegdamsvej 17,
2100 Copenhagen, Denmark}

\affiliation{$^7$ DESY Theory Group, DESY Hamburg, Notkestra{\ss}e 85,
D-22607 Hamburg, Germany}

\abstract{We compute dual-conformally invariant ladder integrals that are
capped off by pentagons at each end of the ladder.  Such integrals appear
in six-point amplitudes in planar ${\cal N}=4$ super-Yang-Mills theory.
We provide exact, finite-coupling formulas for the basic double pentaladder
integrals as a single Mellin integral over hypergeometric functions.
For particular choices of the dual conformal cross ratios, 
we can evaluate the integral at weak coupling to high loop orders in terms
of multiple polylogarithms.  We argue that the integrals are exponentially
suppressed at strong coupling.  We describe the space of functions
that contains all such double pentaladder integrals and their derivatives,
or coproducts.  This space, a prototype for the space of Steinmann
hexagon functions, has a simple algebraic structure, which we elucidate by considering a particular discontinuity of the functions that localizes the Mellin integral and collapses the relevant symbol alphabet.
This function space is endowed with a coaction, both perturbatively
and at finite coupling, which mixes the independent solutions of the hypergeometric
differential equation and constructively realizes a coaction principle of the
type believed to hold in the full Steinmann hexagon function space.}

\emailAdd{schuot@physics.mcgill.ca}\emailAdd{lance@slac.stanford.edu}\emailAdd{mvonhippel@nbi.ku.dk}\emailAdd{amcleod@nbi.ku.dk}\emailAdd{georgios.papathanasiou@desy.de}

\begin{document}
\hypersetup{pageanchor=false}

\begin{flushright} DESY 18--041 \\ SLAC--PUB--17228
\end{flushright}

\maketitle
\hypersetup{pageanchor=true}
\begin{fmffile}{feyndiags}

\section{Introduction}

Despite substantial progress, our understanding of particle scattering in perturbative quantum field theory remains incomplete. One might think that this is to be expected, and that perturbation theory inherently limits us to order-by-order progress in the number of loops. However, the last decade has seen the development of powerful new methods to address scattering at the multi-loop level in both gauge and gravity theories --- see, for example, refs.~\cite{Bern:2008ap,ArkaniHamed:2009dn,ArkaniHamed:2010kv,Bourjaily:2011hi,ArkaniHamed:2012nw,ArkaniHamed:2009sx,Arkani-Hamed:2013jha,Arkani-Hamed:2013kca,Lipstein:2012vs,Lipstein:2013xra,Carrasco:2011hw,Bern:2012uf,Bern:2014kca,Bern:2015ple} and refs.~\cite{Bern:2017ucb,Bern:2017yxu,He:2016mzd,Bern:2010ue,Bern:2007xj,Bern:2012uf}.  Many of these methods are expected to function up to any desired order in perturbation theory. In the case of planar ${\cal N}=4$ super-Yang Mills (SYM) theory~\cite{Brink:1976bc,Gliozzi:1976qd} there is even an all-orders geometric formulation~\cite{Arkani-Hamed:2013jha}. 

Rather, our understanding is incomplete because most of these methods supply, not scattering amplitudes, but integrands depending on loop momenta. Evaluating the multi-loop Feynman integrals produced by these methods is a substantial endeavor in its own right, with two loops only just beginning to yield to systematic analysis~\cite{Gluza:2010ws,Kosower:2011ty,CaronHuot:2012ab,Johansson:2012zv,Badger:2012dp,Badger:2015lda,Ita:2015tya,Larsen:2015ped,Georgoudis:2016wff,Badger:2017jhb,Bourjaily:2017bsb,Broedel:2017kkb,Chicherin:2017dob,Abreu:2017idw,Abreu:2017xsl,Abreu:2017hqn,Boehm:2018fpv,Bourjaily:2018ycu,Bourjaily:2018aeq}. In general, perturbative scattering amplitudes are complicated transcendental functions of momentum invariants. If we want to understand these amplitudes to all orders, then we need to understand how to compute these functions order by order, and further, how to sum them into all-orders expressions.

It is not obvious that this is possible in general.  However, in the planar limit of ${\cal N} = 4$ SYM we have unique evidence that it should be, due to the presence of integrability~\cite{Beisert:2010jr}. Integrability has been used to compute the theory's cusp anomalous dimension for finite coupling~\cite{Beisert:2006ez} and has been instrumental in the Pentagon Operator Product Expansion, which calculates finite-coupling amplitudes in an expansion around a kinematic limit \cite{Basso:2013vsa,Basso:2013aha,Basso:2014koa,Basso:2014nra,Basso:2014hfa,Basso:2015rta,Basso:2015uxa}. Notably, the perturbative expansion of these formulas has a finite radius of convergence in the coupling. The kinematic dependence of four- and five-particle amplitudes in planar ${\cal N}=4$ SYM is also captured to all loop orders by the BDS ansatz~\cite{Bern:2005iz}, which is uniquely dictated by the theory's dual conformal symmetry \cite{Drummond:2006rz, Bern:2006ew, Bern:2007ct, Alday:2007hr, Bern:2008ap, Drummond:2008vq}. While this symmetry does not uniquely fix the form of amplitudes involving more than five particles, it does restrict the problem to a special class of dual conformally invariant (DCI) integrals~\cite{Drummond:2007aua,Drummond:2006rz,Nguyen:2007ya,Paulos:2012nu}, and by extension restricts the form and kinematic dependence of these amplitudes at finite coupling. 

As a consequence, a great deal is known about the space of functions that can contribute to six- and seven-particle perturbative amplitudes in planar ${\cal N} = 4$ SYM.  When these amplitudes are normalized by the BDS ansatz, they can be written in terms of dual superconformal invariants (that encode the helicity structure) multiplied by multiple polylogarithms that have known kinematic dependence and branch cuts only in physical channels~\cite{Gaiotto:2011dt, Dixon:2011pw, Dixon:2013eka, Golden:2013xva, Drummond:2014ffa}.  Additional physical constraints come from the Steinmann relations, which imply that double discontinuities of amplitudes must be zero when taken in overlapping channels~\cite{Steinmann,Steinmann2,Cahill:1973qp}.  These relations are obeyed by the polylogarithmic part of the amplitude when it is normalized by the BDS-like ansatz (which contains only two-particle kinematic invariants)~\cite{Alday:2009dv, Dixon:2015iva, Caron-Huot:2016owq}. We refer to this space of multiple polylogarithms as the space of Steinmann hexagon functions (${\cal H}$) and Steinmann heptagon functions for six- and seven-point kinematics, respectively.  These function spaces have proven sufficient to describe maximally helicity violating (MHV) and next-to-MHV (NMHV) amplitudes at six points through six loops~\cite{Dixon:2011pw, Dixon:2011nj, Dixon:2013eka, Dixon:2014voa, Dixon:2014xca, Dixon:2014iba, Dixon:2015iva, Caron-Huot:2016owq, Caron-Huot:six_loops}, and at seven points through four loops~\cite{Drummond:2014ffa, Dixon:2016nkn}. While the fact that only multiple polylogarithms show up in these amplitudes remains conjectural, there exists evidence that it holds to all loop orders~\cite{ArkaniHamed:2012nw}.  Also, the specific arguments of the polylogarithms in the six-point case are consistent with a recent all-orders analysis of the Landau equations~\cite{Prlina:2018ukf}.

In this article, we will focus on a particular class of DCI integrals inside the space of Steinmann hexagon functions. These integrals have a `double pentaladder' (hereafter just `pentaladder') topology, meaning they take the form of a ladder integral capped on each end by a pentagon with three external massless legs, for a total of six massless legs.  Starting at two loops, there are two integrals with this topology, denoted by $\Omega^{(L)}$ and $\tilde\Omega^{(L)}$, corresponding to two inequivalent numerator factors that render the pentagon integration infrared finite. These integrals constitute the most nontrivial part of the amplitude at two loops, and contribute to the amplitude at all loop orders~\cite{ArkaniHamed:2010kv}. Moreover, members of these classes of integrals are known to be related to each other at adjacent loop orders by a pair of second-order differential equations~\cite{Drummond:2010cz, Dixon:2011nj}.

Armed with these differential equations, we consider finite-coupling versions of $\Omega^{(L)}$ and $\tilde\Omega^{(L)}$ by summing over the loop order weighted by $(-g^2)^L$, as was done previously for a related box ladder integral~\cite{Broadhurst:2010ds}.  While these quantities are {\it not} the full finite-coupling six-point amplitude, they do constitute well-defined contributions to it that sum up an infinite class of Feynman integrals.  By exploiting the symmetries that preserve the dual coordinates on each side of their ladders, variables can be found that simplify the differential equations these integrals obey.  Remarkably, after performing a separation of variables, we obtain compact representations of the finite-coupling versions of $\Omega^{(L)}$ and $\tilde\Omega^{(L)}$ in terms of a single Mellin integral over products of hypergeometric functions. These representations are valid for any value of the coupling. Factoring the second-order differential operators into first-order operators, we are led to consider two additional classes of integrals, $\OL^{(L)}$ and $\WL^{(L)}$, that inherit this finite-coupling description. 
In order to generate more Steinmann hexagon functions, we go on to consider the enveloping space of
polylogarithmic functions that is generated by taking all possible derivatives of these integrals at arbitrarily high loop order.
We refer to this space of functions as the $\Omega$ space. It is graded by an integer weight, where for example $\Omega^{(L)}$ has weight $2L$.

Surprisingly, the nontrivial part of the $\Omega$ space is entirely encoded in the discontinuity of these integrals with respect to the channel carrying momentum along the ladder. After taking this discontinuity, the finite-coupling representation of each integral can be rewritten as a contour integral over a branch cut that collapses to a pole in the weak coupling expansion.  Perturbatively, the dependence on the kinematic variables reduces to powers of logarithms in one variable and single-valued harmonic polylogarithms (SVHPLs)~\cite{BrownSVHPLs} in the other two variables.  This simplicity allows us to recursively construct the function space corresponding to this discontinuity to arbitrary weight.  Promoting this space to the full $\Omega$ space also turns out to be incredibly simple, since the kernel of the discontinuity operation within that subspace
contains only two functions at each weight.

Using similar methods, we also resum the pentabox ladder integrals, which are capped by a pentagon on one end of the ladder and an off-shell box on the other end. These integrals contribute to seven- and higher-particle amplitudes in planar ${\cal N} = 4$ SYM.  There is an analogous enveloping space of polylogarithmic functions associated with these integrals, which can be easily constructed by taking a kinematic limit of the $\Omega$ space.

These new finite-coupling representations give us formidable control over the original $L$-loop integrals. In various kinematical limits, they lead to explicit formulae for the integrals to high loop orders. They also give us a handle on the structure of the $\Omega$ space, which (as a space of multiple polylogarithms) is endowed with a Hopf algebra and an associated coaction~\cite{Gonch2,Gonch3,FBThesis,Goncharov:2010jf,Brown:2011ik,Duhr:2011zq,Duhr:2012fh}.  In particular, the relevant discontinuities of the $\Omega^{(L)}$, $\tilde\Omega^{(L)}$, $\OL^{(L)}$, and $\WL^{(L)}$ integrals are related to each other by first-order differential operators, and this system of differential equations is encoded in the coaction. The coaction can thus be realized as a $4\times4$ matrix that acts on the vector of the discontinuities of these integrals. (The coaction on the integrals themselves maps to a slightly larger space of functions, and must be described by a larger matrix.)

The coaction on the discontinuity of these integrals can also be defined at finite coupling (that is, nonperturbatively) in the form of a matrix product of path-ordered exponentials.  By construction, this nonperturbative coaction satisfies a coaction principle~\cite{Schnetz:2013hqa,Brown:2015fyf,Panzer:2016snt}, meaning that the first entry of the coaction always maps to the original space of discontinuity functions, while the second entry can map to a larger space, in general.  We expect that this structure can be lifted to the full $\Omega$ space.  A similar coaction principle also seems to be at work in perturbative string theory~\cite{Schlotterer:2012ny,Drummond:2013vz}, $\phi^4$ theory~\cite{Panzer:2016snt}, QED~\cite{Schnetz:2017bko}, and the full space of Steinmann hexagon functions (where data currently exist through six loops)~\cite{Caron-Huot:2016owq,Caron-Huot:six_loops}.  The finite-coupling structure of the $\Omega$ space lends weight to the conjecture that ${\cal H}$ is endowed with a similar structure to all orders. In many ways, the $\Omega$ space thus serves as an instructive toy model for the full space of hexagon functions, as well as for quantum field theory more generally. 

In studying these integrals, we hope to inaugurate a new approach to Feynman integrals that goes beyond order-by-order progress in perturbation theory. The $\Omega^{(L)}$ and $\tilde\Omega^{(L)}$ integrals can now be described analytically to any order, as well as at finite coupling --- the already substantial all-orders understanding of this class of integrands is now complemented by a thorough understanding of the type of functions to which they integrate.  Similar types of functions can be expected to appear in planar ${\cal N}=4$ more generally, at least in the MHV and NMHV sectors.  We hope that other infinite families of integrals can be identified and characterized in a similar manner, eventually extending such an all-orders description to scattering amplitudes themselves. 

The remainder of this paper is organized as follows. In section~\ref{sec:finite_DCI_integrals} we define the $\Omega^{(L)}$ and $\tilde\Omega^{(L)}$ integrals.  We also define the related pentabox and box ladder integrals, for which the pentagon at one or both ends of the ladder is replaced by an off-shell box. We then introduce the differential equations these integrals satisfy. In section~\ref{sec:resummed_result} we leverage the symmetries of the $\Omega^{(L)}$ integrals, as well as their single-valuedness in the region where all cross ratios are close to 1, to resum them into a one-fold Mellin integral over hypergeometric functions. We do the same for $\tilde\Omega^{(L)}$, and introduce a family of related functions. In section~\ref{sec:sumsandpolys} we show that the finite-coupling representations of these integrals may be equivalently recast as an infinite series, corresponding to the Taylor expansion around a particular kinematic limit. By further expanding the coefficients of this series at weak coupling, we may resum it away from the limit for certain two-dimensional slices of the space of kinematics, in terms of multiple polylogarithms. Then, in section \ref{sec:strong_coupling} we analyze our integrals at strong coupling, finding evidence that they become exponentially suppressed for a large chunk of the Euclidean region.

Section~\ref{sec:spacesect} describes the $\Omega$ space of functions appearing in the coaction of the basic integrals. It showcases a space of functions relevant to the six-point scattering amplitude that can be constructed explicitly to all loop orders. First we consider the discontinuity of the functions with respect to the channel carrying momentum along the ladder.  This discontinuity is simpler to analyze, yet it contains nearly all the information about the full space. In particular, the discontinuity space can be efficiently reconstructed from its coaction, which we formulate nonperturbatively.  We conclude in section~\ref{ConclusionsSection} with a discussion of these results and possible directions for future work.  

This paper includes three appendices:  Appendix~\ref{hexvarappendix} collects relations between different sets of kinematic variables; appendix~\ref{ext_steinmann} describes some ``extended Steinmann relations'' that have been found in the full space of hexagon functions ${\cal H}$; and appendix~\ref{appendix:coproduct_relations} describes analogous relations for the spaces $\Omega$ and $\Omega_c$, as well as coproduct relations between the various integrals, and how a curious ``double coproduct'' operator acts on the $\Omega$ space. We also provide three ancillary files.  Two of them,
\texttt{omega1vwL0-8.m} and \texttt{omegauv0L0-8.m},
give the integral $\Omega^{(L)}$ on the surfaces $u=1$ and $w=0$,
respectively, through eight loops in terms of multiple polylogarithms.
The third, \texttt{omegacdiscwt0-12.m}, gives the $c$-discontinuity
of all the functions in the $\Omega$ space through weight 12.


\section{Finite Dual Conformal Invariant Integrals}
\label{sec:finite_DCI_integrals}

\subsection{Dual conformal symmetry}

In addition to the superconformal symmetry that follows from its Lagrangian formulation, ${\cal N}=4$ SYM develops a dual conformal symmetry in the planar limit~\cite{Drummond:2006rz,Bern:2006ew,Bern:2007ct,Alday:2007hr,Drummond:2008vq}. This new symmetry is associated with conformal transformations acting on the dual (or region) coordinates $x_i^{\alpha \dot \alpha}$, defined via
\be
p_i^{\alpha \dot \alpha} = \lambda_i^\alpha \tilde \lambda_i^{\dot \alpha} = x_i^{\alpha \dot \alpha} - x_{i+1}^{\alpha \dot \alpha} \, ,
\ee
where $p_i^{\alpha \dot \alpha}$ is the momentum of the $i^{\text{th}}$ scattering particle, and $x_{n+1}^{\alpha \dot \alpha} \equiv x_1^{\alpha \dot \alpha}$. Planarity implies that these coordinates can only appear in integrals via the squared differences
\be
x_{ij}^2 \equiv (x_i- x_j)^2 = \det(x_i^{\alpha \dot \alpha} - x_j^{\alpha \dot \alpha}),
\label{eq:squared_differences}
\ee
where $x_{i,i+1}^2 = 0$ when leg $i$ is massless.

The planar loop integrand also depends on dual coordinates $x_r$, $x_s$, etc., associated with the interior region of each loop.  After dividing out by the tree-level MHV superamplitude, the loop integrand multiplied by the integration measure becomes dual conformal invariant~\cite{Drummond:2008vq,ArkaniHamed:2010gh}.
In particular, such an object must be invariant under the dual conformal inversion operator $I$,
\be
I[x_i^{\alpha \dot \alpha}] = \frac{x_i^{\alpha \dot \alpha}}{x_i^2} \qquad
\Rightarrow \qquad I[x_{ij}^2] = \frac{x_{ij}^2}{x_i^2 x_j^2} \,, \quad
I[d^4x_r] = \frac{d^4x_r}{(x_r^2)^4} \,.
\ee
As a consequence, external dual coordinates should appear the same number of times in the numerator and denominator of the integrand, while the dual loop coordinates should appear four more times in the denominator.

For example, the two-loop pentaladder integrals can be written in terms of the integral
\be
{\cal I}_{dpl}^{(2)}\ \propto\
 \int \frac{d^4 x_{r}}{i \pi^2}\frac{d^4x_{s}}{i \pi^2}
\frac{x_{ar}^2 x_{bs}^2}{(x_{1r}^2 x_{2r}^2 x_{3r}^2 x_{4r}^2) x_{rs}^2 (x_{4s}^2 x_{5s}^2 x_{6s}^2 x_{1s}^2)}  \,, \label{eq:I_dp}
\ee
where, in addition to dual coordinates associated with each loop, we have introduced a pair of points $x_a^{\alpha \dot \alpha}$ and $x_b^{\alpha \dot \alpha}$ that solve the null-separation conditions $x_{a1}^2 = x_{a2}^2 = x_{a3}^2 = x_{a4}^2 = 0$ and $x_{b4}^2 = x_{b5}^2 = x_{b6}^2 = x_{b1}^2 = 0$~\cite{Drummond:2010cz}. This choice suppresses the integrand in each of the limits where the denominator vanishes, rendering the integral infrared (IR) finite.

These conditions each admit two parity-conjugate solutions, namely
\begin{align}
x_{a^1}^{\alpha \dot \alpha} &= \frac{\lambda^\alpha_1 \left( \lambda_{3 \beta}\ x_3^{\beta \dot \alpha} \right) - \lambda^\alpha_3 \left(\lambda_{1 \beta} \ x_1^{\beta \dot \alpha} \right)}{\langle 1 3 \rangle}, 
\quad 
&x_{a^2}^{\alpha \dot \alpha} &= \frac{\left(x_3^{\alpha \dot \beta} \ \tilde \lambda_{3 \dot \beta} \right) \tilde \lambda^{\dot \alpha}_1 - \left( x_1^{\alpha \dot \beta} \ \tilde \lambda_{1 \dot \beta} \right) \tilde \lambda^{\dot \alpha}_3}{[ 1 3 ]}, \label{eq:x_a} \\
x_{b^1}^{\alpha \dot \alpha} &= \frac{\lambda^\alpha_4 \left( \lambda_{6 \beta}\ x_6^{\beta \dot \alpha} \right) - \lambda^\alpha_6 \left(\lambda_{4 \beta} \ x_4^{\beta \dot \alpha} \right)}{\langle 4 6 \rangle}, 
\quad 
&x_{b^2}^{\alpha \dot \alpha} &= \frac{\left(x_6^{\alpha \dot \beta} \ \tilde \lambda_{6 \dot \beta} \right) \tilde \lambda^{\dot \alpha}_4 - \left(x_4^{\alpha \dot \beta}\ \tilde \lambda_{4 \dot \beta} \right) \tilde \lambda^{\dot \alpha}_6}{[ 4 6 ]}, \label{eq:x_b}
\end{align}
where $\langle i j \rangle \equiv \epsilon_{\alpha \beta}\lambda^\alpha_i \lambda^\beta_j$ and $[ i j ] \equiv \epsilon_{\dot \alpha \dot \beta} \tilde \lambda^{\dot \alpha}_i \tilde \lambda^{\dot \beta}_j$. These solutions seem to give us four possible choices of numerator in the pentaladder integral~\eqref{eq:I_dp}. However, only two of these choices give rise to different integrals, because replacing both $x_a^{\alpha \dot \alpha}$ and $x_b^{\alpha \dot \alpha}$ with their parity conjugates gives rise to integrals that differ only by terms that vanish after integration~\cite{Drummond:2010cz}.

Choosing pairs $(x_{a^1},x_{b^1})$ or $(x_{a^2},x_{b^2})$ that are related by the cyclic shift $i \rightarrow i+3$ gives rise to the integral $\Omega^{(2)}$ (up to a kinematic prefactor required to make it DCI, given below).
On the other hand, choosing the pair $(x_{a^1},x_{b^2})$ gives rise to the integral $\tilde\Omega^{(2)}$ (again, up to a kinematic prefactor),
which is parity conjugate to the integral $(x_{a^2},x_{b^1})$.

These integrals can also be expressed in terms of momentum twistors~\cite{Hodges:2009hk,Mason:2009qx}
\be
Z^R_i = (\lambda_i^\alpha, x_i^{\beta \dot \alpha} \lambda_{i \beta}) \, ,
\ee
where $R = (\alpha, \dot \alpha)$ is a combined $SU(2,2)$ index. Momentum twistors live in the projective space $\mathbb{CP}^3$, as they are invariant under the independent little group rescalings, $Z_i \rightarrow t_i Z_i$ for each $i$. They are related to the squared differences defined in \eqn{eq:squared_differences} by
\be
x_{ij}^2 = \frac{\langle i-1, i, j-1, j \rangle}{\langle i-1,i \rangle \langle j-1, j \rangle} \, , \label{eq:squared_diff_to_momentum_twistors}
\ee
where the four-bracket $\langle ijkl \rangle \equiv \langle Z_i Z_j Z_k Z_l \rangle = \epsilon_{RSTU} Z_i^R Z_j^S Z_k^T Z_l^U$ is invariant under $SU(2,2)$, but is not projectively invariant.  The spinor products $\langle i j \rangle$ are not DCI, but cancel out in projectively-invariant, DCI ratios. Using~\eqn{eq:squared_diff_to_momentum_twistors}, the expression for $\Omega^{(2)}$ can be written in terms of momentum twistors as
\begin{align} \label{eq:omega_2}
\Omega^{(2)} &=\int \frac{d^4Z_{AB}}{i \pi^2}\frac{d^4Z_{CD}}{i \pi^2} 
\frac{\l AB13\r }{\big(\l AB61\r\l AB12\r\l AB23\r\l AB34\r\big)} \nonumber \\
&\hspace{3cm} \times \frac{\l CD46\r \l 1256 \r \l 2345 \r \l 6134\r}{\l ABCD\r \big(\l CD34\r\l CD45\r \l CD56\r \l CD61\r\big)}, 
\end{align}
which has been made projectively invariant and DCI by the inclusion of the kinematic factor $\l 1256 \r \l 2345 \r \l 6134\r$. The modified index structure of the integration variables, from single dual indices to pairs of momentum twistor labels, encodes the fact that points in dual space map to lines in momentum twistor space. We have additionally used the replacements $x_a \rightarrow Z_1 Z_3, x_b \rightarrow Z_4 Z_6$, which selects out $x_{a^1}^{\alpha \dot \alpha}$ and $x_{b^1}^{\alpha \dot \alpha}$ from
\eqns{eq:x_a}{eq:x_b}. Notice that this integral is invariant under the dihedral transformation that exchanges momentum twistors (legs) $1 \leftrightarrow 3$ and $4 \leftrightarrow 6$ while leaving $2$ and $5$ invariant.

To write $\tilde\Omega^{(2)}$ in the language of momentum twistors, we use the fact that parity maps $Z_i$ to the ray orthogonal to $Z_{i-1}$, $Z_i$, and $Z_{i+1}$. In particular, mapping $Z_i$ and $Z_j$ to their parity conjugates sends the four-bracket $\langle Z_i Z_j Z_k Z_l \rangle \rightarrow \langle (i-1 i  i+1) \cap (j-1 j j+1) Z_k Z_l \rangle$, where $(i-1 i  i+1) \cap (j-1 j j+1)$ denotes the intersection of the hyperplanes spanned by $\{ Z_{i-1}, Z_i, Z_{i+1} \}$, and $\{Z_{j-1}, Z_j, Z_{j+1}\}$, respectively. Using this map to send $x_{b^1}^{\alpha \dot \alpha}$ to $x_{b^2}^{\alpha \dot \alpha}$ in \eqn{eq:omega_2}, we arrive at the expression
\begin{align} \label{eq:omegat_2}
\tilde{\Omega}^{(2)} &=\int \frac{d^4Z_{AB}}{i \pi^2}\frac{d^4Z_{CD}}{i \pi^2} \frac{ \l AB13 \r}{\big(\l AB61\r\l AB12\r\l AB23\r\l AB34\r\big)} \nonumber \\
&\hspace{3cm} \times \frac{\big(\l D345\r \l C561\r -  \l C345\r \l D561\r \big) \l 1246 \r \l 2346 \r}{\l ABCD\r \big(\l CD34\r\l CD45\r \l CD56\r \l CD61\r\big)},
\end{align}
where we have made use of the identity
\be
\l i j (k l m) \cap (n o p) \r = \l i k l m \r \l j n o p \r -  \l j k l m \r \l i n o p \r \,,
\ee
and have replaced the previous kinematic factor by $\l 1246 \r \l 2346 \r$. (We have additionally multiplied by an overall minus sign to stay consistent with the definition in the literature~\cite{Dixon:2011nj}.) Unlike $\Omega^{(2)}$, which is parity even, $\tilde{\Omega}^{(2)}$ has both a parity even and parity odd part. Like $\Omega^{(2)}$, it is symmetric under the dihedral transformation $1 \leftrightarrow 3, 4 \leftrightarrow 6$.


\subsection{The Box Ladder, Pentabox Ladder, and (Double) Pentaladder Integrals}
\label{sec:integral_definitions}

\begin{figure}[t]
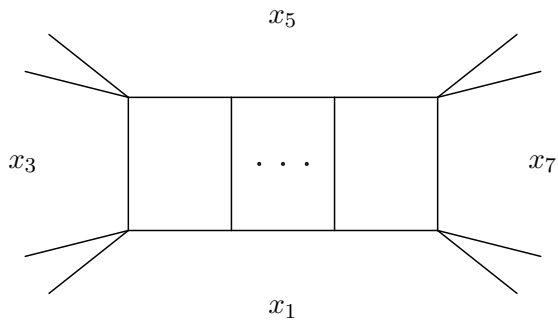

\begin{center}
\begin{fmfchar*}(300,140)
	\fmfset{dot_size}{0.18mm}
	 %
	\fmfforce{(.305w,.68h)}{p3}
	\fmfforce{(.435w,.68h)}{p4}
	\fmfforce{(.565w,.68h)}{p5}
	\fmfforce{(.695w,.68h)}{p6}
	\fmfforce{(.695w,.32h)}{p10}
	\fmfforce{(.565w,.32h)}{p11}
	\fmfforce{(.435w,.32h)}{p12}
	\fmfforce{(.305w,.32h)}{p13}
	%
	\fmfforce{(.47w,.5h)}{d1}
	\fmfforce{(.5w,.5h)}{d2}
	\fmfforce{(.53w,.5h)}{d3}
	%
	\fmfforce{(.175w,.75h)}{e1}
	\fmfforce{(.205w,.85h)}{e2}
	\fmfforce{(.795w,.85h)}{e3}
	\fmfforce{(.825w,.75h)}{e4}
	\fmfforce{(.825w,.25h)}{e5}
	\fmfforce{(.795w,.15h)}{e6}
	\fmfforce{(.205w,.15h)}{e7}
	\fmfforce{(.175w,.25h)}{e8}
	%
	\fmfforce{(.5w,.15h)}{l1}
	\fmfforce{(.2w,.5h)}{l3}
	\fmfforce{(.5w,.85h)}{l5}
	\fmfforce{(.8w,.5h)}{l7}
	%
	\fmf{plain, width=.2mm}{p3,p4}
	\fmf{plain, width=.2mm}{p4,p5}
	\fmf{plain, width=.2mm}{p5,p6}
	\fmf{plain, width=.2mm}{p6,p10}
	\fmf{plain, width=.2mm}{p10,p11}
	\fmf{plain, width=.2mm}{p11,p12}
	\fmf{plain, width=.2mm}{p12,p13}
	\fmf{plain, width=.2mm}{p13,p3}
	%
	\fmf{plain, width=.2mm}{p4,p12}
	\fmf{plain, width=.2mm}{p5,p11}
	%
	\fmf{plain, width=.2mm}{p3,e1}
	\fmf{plain, width=.2mm}{p3,e2}
	\fmf{plain, width=.2mm}{p6,e3}
	\fmf{plain, width=.2mm}{p6,e4}
	\fmf{plain, width=.2mm}{p10,e5}
	\fmf{plain, width=.2mm}{p10,e6}
	\fmf{plain, width=.2mm}{p13,e7}
	\fmf{plain, width=.2mm}{p13,e8}
	%
	\fmfdot{d1}
	\fmfdot{d2}
	\fmfdot{d3}
	%
	\fmfv{label={$x_1$}, label.dist=.1cm}{l1}
	\fmfv{label={$x_3$}, label.dist=.1cm}{l3}
	\fmfv{label={$x_5$}, label.dist=.1cm}{l5}
	\fmfv{label={$x_7$}, label.dist=.1cm}{l7}
\end{fmfchar*}
\end{center}
\caption{The eight- or higher-point $L$-loop ladder integral, labelled by dual coordinates.}
\label{fig:ladder_L}
\end{figure}

Before introducing the $L$-loop generalizations of $\Omega^{(2)}$ and $\tilde\Omega^{(2)}$, let us first consider the simpler `box ladder' integrals shown in~\fig{fig:ladder_L}.  The box ladder integrals involve only four dual coordinates, but none are null separated, so the first all-massless scattering amplitude to which they could contribute would be an eight-point amplitude.

The representative of this class at one loop is just the DCI four-mass box integral,
\be
{\cal I}_l^{(1)}(x_1, x_3, x_5, x_7) = \int \frac{d^4 x_r}{i \pi^2} \frac{x_{15}^2 x_{37}^2}{x_{1r}^2 x_{3r}^2 x_{5r}^2 x_{7r}^2}\, .
\label{eq:4mbox}
\ee
The $L$-loop integral can be defined iteratively by integrating over the appropriate dual coordinate in the $(L-1)$-loop integral,
\be
{\cal I}_l^{(L)}(x_1, x_3, x_5, x_7) = \int \frac{d^4 x_r}{i \pi^2} \frac{x_{15}^2 x_{37}^2}{x_{1r}^2 x_{3r}^2 x_{5r}^2 x_{7r}^2} {\cal I}_l^{(L-1)}(x_1, x_3, x_5, x_r)\,,
\ee
weighted by the appropriate propagator factors.

The box ladder integrals depend on only two cross ratios, conventionally expressed in terms of the variables $z$ and $\bar z$ defined by
\be
\frac{x_{13}^2 x_{57}^2}{x_{15}^2 x_{37}^2} = \frac{1}{(1-z)(1-\bar z)}\,,\quad 
\frac{x_{17}^2 x_{35}^2}{x_{15}^2 x_{37}^2} = \frac{z \bar z}{(1-z)(1-\bar z)}\,,\label{eq:boxvars} 
\ee
where $\bar z = z^*$ on the Euclidean sheet where the cross ratios are real and positive.

In general, for a sequence of $L$-loop ladder integrals ${\cal I}^{(L)}$,
we define the finite-coupling, or resummed, version by
\be
{\cal I}(g^2) = \sum_{L=0}^\infty \,(-g^{2})^L\,{\cal I}^{(L)} \,,
\label{definefinitecoupling}
\ee
i.e.~we just drop the $(L)$ superscript.
Typically the integral can be normalized so that ${\cal I}^{(L)}$ is
a pure function, that is, an iterated integral with no rational prefactor,
for $L\geq1$, while the tree quantity ${\cal I}^{(0)}$ is rational.

The box ladder integrals have long been known to all loop orders~\cite{Usyukina:1993ch,Isaev:2003tk}, and can be written as~\cite{Usyukina:1993ch,Fleury:2016ykk,Basso:2017jwq}
\be
{\cal I}_l^{(L)}(x_1, x_3, x_5, x_7)
= \frac{(1-z)(1-\zb)}{z- \zb} \ f^{(L)}(z, \bar z), \label{eq:normalization boxes}
\ee
where
\be
f^{(L)}(z, \bar z) = \sum_{r=0}^L \frac{(-1)^{r} (2L-r)!}{r! (L-r)! L!} \ln^r(z \bar z) \left( \text{Li}_{2L-r}(z) - \text{Li}_{2L-r}(\bar z) \right).
\label{boxladderL}
\ee
This class of integrals has been evaluated at finite coupling, i.e.~resummed to all orders~\cite{Broadhurst:2010ds}:
\be
f(z,\zb) = \sum_{L=0}^\infty (-g^2)^L f^{(L)}(z,\zb)
= \int_{2g}^\infty \frac{\zeta d\zeta}{\sqrt{\zeta^2-4g^2}}\, 2\cos\left(\frac12\sqrt{\zeta^2-4g^2} \ln\frac{1}{z \bar z}\right) \frac{\sinh[(\pi-\phi)\zeta]}{i\sinh(\pi \zeta)}\,.\label{eq:finitebox}
\ee
Here we have changed normalization by multiplying the result given in eq.~(21) of~\cite{Broadhurst:2010ds} by $t\mu/i$, and changed integration variables from $z$ to $\zeta$ to avoid confusion with our variables $z$ and $\bar z$ (in our variables, $\phi=\textrm{arg}\,z$). Finally, we have taken $\kappa^2\to 4g^2$ in order to match our coupling and normalization conventions. This result will provide a cross-check of the method used to obtain our main finite-coupling results in section~\ref{sec:resummed_result}.


\begin{figure}[t]
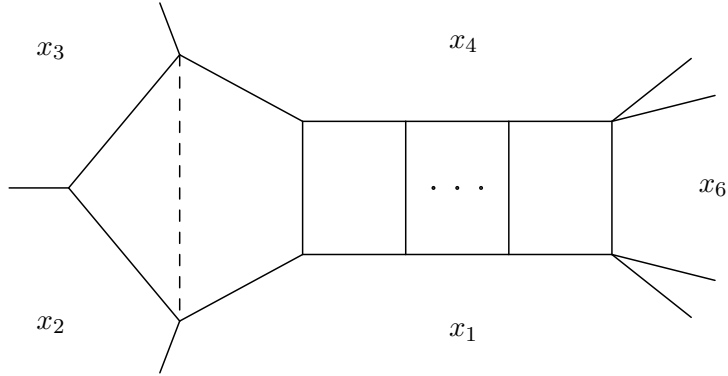

\begin{center}
\begin{fmfchar*}(300,140)
	\fmfset{dot_size}{0.18mm}
	 %
	\fmfforce{(.115w,.5h)}{p1}
	\fmfforce{(.255w,.86h)}{p2}
	\fmfforce{(.41w,.68h)}{p3}
	\fmfforce{(.54w,.68h)}{p4}
	\fmfforce{(.67w,.68h)}{p5}
	\fmfforce{(.8w,.68h)}{p7}
	\fmfforce{(.8w,.32h)}{p9}
	\fmfforce{(.67w,.32h)}{p11}
	\fmfforce{(.54w,.32h)}{p12}
	\fmfforce{(.41w,.32h)}{p13}
	\fmfforce{(.255w,.14h)}{p14}
	%
	\fmfforce{(.575w,.5h)}{d1}
	\fmfforce{(.605w,.5h)}{d2}
	\fmfforce{(.635w,.5h)}{d3}
	%
	\fmfforce{(.04w,.5h)}{e1}
	\fmfforce{(.23w,1.0h)}{e2}
	\fmfforce{(0.9w,0.85h)}{e3}
	\fmfforce{(0.93w,0.75h)}{e4}
	\fmfforce{(0.93w,0.25h)}{e5}
	\fmfforce{(0.9w,0.15h)}{e6}
	\fmfforce{(.23w,0.0h)}{e7}
	%
	\fmfforce{(.12w,.84h)}{l1}
	\fmfforce{(.59w,.85h)}{l2}
	\fmfforce{(0.9w,.5h)}{l34}
	\fmfforce{(.59w,.15h)}{l5}
	\fmfforce{(.12w,.16h)}{l6}
	%
	\fmf{plain, width=.2mm}{p1,p2}
	\fmf{plain, width=.2mm}{p2,p3}
	\fmf{plain, width=.2mm}{p3,p4}
	\fmf{plain, width=.2mm}{p4,p5}
	\fmf{plain, width=.2mm}{p5,p7}
	\fmf{plain, width=.2mm}{p7,p9}
	\fmf{plain, width=.2mm}{p9,p11}
	\fmf{plain, width=.2mm}{p11,p12}
	\fmf{plain, width=.2mm}{p12,p13}
	\fmf{plain, width=.2mm}{p13,p14}
	\fmf{plain, width=.2mm}{p14,p1}
	%
	\fmf{dashes, width=.2mm}{p2,p14}
	%
	\fmf{plain, width=.2mm}{p3,p13}
	\fmf{plain, width=.2mm}{p4,p12}
	\fmf{plain, width=.2mm}{p5,p11}
	%
	\fmf{plain, width=.2mm}{p1,e1}
	\fmf{plain, width=.2mm}{p2,e2}
	\fmf{plain, width=.2mm}{p7,e3}
	\fmf{plain, width=.2mm}{p7,e4}
	\fmf{plain, width=.2mm}{p9,e5}
	\fmf{plain, width=.2mm}{p9,e6}
	\fmf{plain, width=.2mm}{p14,e7}
	%
	\fmfdot{d1}
	\fmfdot{d2}
	\fmfdot{d3}
	%
	\fmfv{label={$x_3$}, label.dist=.1cm}{l1}
	\fmfv{label={$x_4$}, label.dist=.1cm}{l2}
	\fmfv{label={$x_6$}, label.dist=.1cm}{l34}
	\fmfv{label={$x_1$}, label.dist=.1cm}{l5}
	\fmfv{label={$x_2$}, label.dist=.1cm}{l6}
\end{fmfchar*}
\end{center}
\caption{The seven- or higher-point pentabox ladder integral, labelled by dual coordinates, where the ladder is formed out of $L-1$ loops. The dashed line represents a numerator factor that renders the integral DCI.}
\label{fig:pentaboxladder_L}
\end{figure}

We can promote the ladder integrals to the pentabox integrals shown in~\fig{fig:pentaboxladder_L} by attaching a pentagon to the end of one of the ladder integrals. This is done by carrying out a single integration on the ladder integral of the form
\be
{\cal I}_{pl}^{(L)}(x_1,x_2,x_3,x_4,x_6) = \frac{x_{14}^2 x_{26}^2 x_{36}^2 }{x_{6a}^2}\int \frac{d^4 x_r}{i \pi^2} \frac{x_{ar}^2}{x_{1r}^2 x_{2r}^2 x_{3r}^2 x_{4r}^2 x_{6r}^2} {\cal I}_{l}^{(L-1)}(x_1,x_r,x_4,x_6) \,,
\ee
where the point $x_a$ should be chosen to be null-separated from the dual variables $x_1$, $x_2$, $x_3$, and $x_4$.  Choosing $x_{a^1}$ or $x_{a^2}$ in \eqn{eq:x_a} gives the same result, i.e.~this integral is parity-even.  The pentabox ladder integrals involve five dual coordinates.  Since two pairs of coordinates are not null separated, the first all-massless scattering amplitude they can appear in is a seven-point amplitude.

The pentabox ladder integrals can be alternatively defined by attaching $L-1$ boxes to the one-loop pentagon integral,
\be
{\cal I}_{pl}^{(1)}(x_1,x_2,x_3,x_4,x_6) = \frac{x_{14}^2 x_{26}^2 x_{36}^2 }{x_{6a}^2} \int \frac{d^4 x_r}{i \pi^2} \frac{x_{ar}^2}{x_{1r}^2 x_{2r}^2 x_{3r}^2 x_{4r}^2 x_{6r}^2},
\label{eq:pent1}
\ee
where the boxes are added iteratively by the integration
\be
{\cal I}_{pl}^{(L)}(x_1,x_2,x_3,x_4,x_6) = \frac{x_{14}^2 x_{26}^2 x_{36}^2 }{x_{6a}^2}\int \frac{d^4 x_r}{i \pi^2} \frac{x_{ar}^2 }{x_{1r}^2 x_{2r}^2 x_{3r}^2 x_{4r}^2 x_{6r}^2} {\cal I}_{pl}^{(L-1)}(x_1,x_2,x_3,x_4,x_r) \, .
\ee
The pentabox ladder integrals depend on the cross-ratios
\be
u = \frac{x_{16}^2 x_{24}^2}{x_{26}^2 x_{14}^2}  \,, \qquad
v = \frac{x_{46}^2 x_{13}^2}{x_{36}^2 x_{14}^2} \,.
\ee

Our main interest in this paper is in the (double) pentaladder integrals, which involve six dual coordinates, all null separated from their neighbors, so that these integrals will appear in all-massless six-point amplitudes.

\begin{figure}[t]
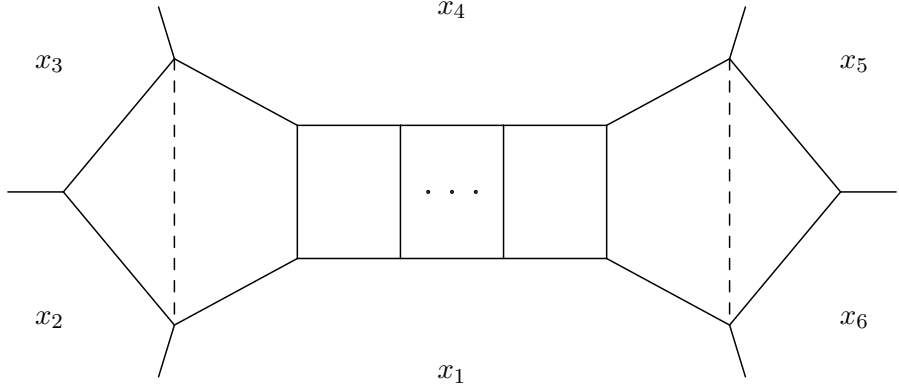

\begin{center}
\begin{fmfchar*}(300,140)
	\fmfset{dot_size}{0.18mm}
	 %
	\fmfforce{(.01w,.5h)}{p1}
	\fmfforce{(.15w,.86h)}{p2}
	\fmfforce{(.305w,.68h)}{p3}
	\fmfforce{(.435w,.68h)}{p4}
	\fmfforce{(.565w,.68h)}{p5}
	\fmfforce{(.695w,.68h)}{p6}
	\fmfforce{(.85w,.86h)}{p7}
	\fmfforce{(.99w,.5h)}{p8}
	\fmfforce{(.85w,.14h)}{p9}
	\fmfforce{(.695w,.32h)}{p10}
	\fmfforce{(.565w,.32h)}{p11}
	\fmfforce{(.435w,.32h)}{p12}
	\fmfforce{(.305w,.32h)}{p13}
	\fmfforce{(.15w,.14h)}{p14}
	%
	\fmfforce{(.47w,.5h)}{d1}
	\fmfforce{(.5w,.5h)}{d2}
	\fmfforce{(.53w,.5h)}{d3}
	%
	\fmfforce{(-.06w,.5h)}{e1}
	\fmfforce{(.13w,1.0h)}{e2}
	\fmfforce{(.87w,1.0h)}{e3}
	\fmfforce{(1.06w,.5h)}{e4}
	\fmfforce{(.87w,0.0h)}{e5}
	\fmfforce{(.13w,0.0h)}{e6}
	%
	\fmfforce{(.02w,.84h)}{l1}
	\fmfforce{(.5w,.95h)}{l2}
	\fmfforce{(.98w,.84h)}{l3}
	\fmfforce{(.98w,.16h)}{l4}
	\fmfforce{(.5w,.05h)}{l5}
	\fmfforce{(.02w,.16h)}{l6}
	%
	\fmf{plain, width=.2mm}{p1,p2}
	\fmf{plain, width=.2mm}{p2,p3}
	\fmf{plain, width=.2mm}{p3,p4}
	\fmf{plain, width=.2mm}{p4,p5}
	\fmf{plain, width=.2mm}{p5,p6}
	\fmf{plain, width=.2mm}{p6,p7}
	\fmf{plain, width=.2mm}{p7,p8}
	\fmf{plain, width=.2mm}{p8,p9}
	\fmf{plain, width=.2mm}{p9,p10}
	\fmf{plain, width=.2mm}{p10,p11}
	\fmf{plain, width=.2mm}{p11,p12}
	\fmf{plain, width=.2mm}{p12,p13}
	\fmf{plain, width=.2mm}{p13,p14}
	\fmf{plain, width=.2mm}{p14,p1}
	%
	\fmf{dashes, width=.2mm}{p2,p14}
	\fmf{dashes, width=.2mm}{p7,p9}
	%
	\fmf{plain, width=.2mm}{p3,p13}
	\fmf{plain, width=.2mm}{p4,p12}
	\fmf{plain, width=.2mm}{p5,p11}
	\fmf{plain, width=.2mm}{p6,p10}
	%
	\fmf{plain, width=.2mm}{p1,e1}
	\fmf{plain, width=.2mm}{p2,e2}
	\fmf{plain, width=.2mm}{p7,e3}
	\fmf{plain, width=.2mm}{p8,e4}
	\fmf{plain, width=.2mm}{p9,e5}
	\fmf{plain, width=.2mm}{p14,e6}
	%
	\fmfdot{d1}
	\fmfdot{d2}
	\fmfdot{d3}
	%
	\fmfv{label={$x_3$}, label.dist=.1cm}{l1}
	\fmfv{label={$x_4$}, label.dist=.1cm}{l2}
	\fmfv{label={$x_5$}, label.dist=.1cm}{l3}
	\fmfv{label={$x_6$}, label.dist=.1cm}{l4}
	\fmfv{label={$x_1$}, label.dist=.1cm}{l5}
	\fmfv{label={$x_2$}, label.dist=.1cm}{l6}
\end{fmfchar*}
\end{center}
\caption{The six-point integral $\Omega^{(L)}$, labelled by dual coordinates, where the ladder is formed out of $L-2$ loops. The dashed lines represent numerator factors that render the integral DCI.}
\label{fig:omega_L}
\end{figure}

There are two classes of pentaladder integrals that can be defined, corresponding to the two inequivalent numerator choices highlighted in the last section. The diagram for the first class of integrals, $\Omega^{(L)}$, is shown in~\fig{fig:omega_L}. The dashed lines in this diagram indicate the numerator factors $x_{a^1}^{\alpha \dot \alpha}$ and $x_{b^1}^{\alpha \dot \alpha}$, although we could have equivalently chosen $x_{a^2}^{\alpha \dot \alpha}$ and $x_{b^2}^{\alpha \dot \alpha}$ (which, in our convention, would have swapped these dashed lines for wavy lines). The diagram for $\tilde{\Omega}^{(L)}$ differs by the exchange of just one of these numerator factors for its parity conjugate---or, graphically, by the exchange of one of the dashed lines for a wavy line. 

These pentaladder integrals can be most easily defined in momentum twistor space, as repeated insertions of a box into $\Omega^{(2)}$ and $\tilde{\Omega}^{(2)}$. For example, the three loop integrals may be obtained from \eqns{eq:omega_2}{eq:omegat_2} by the replacement
\be
\begin{array}{cc}
\Omega^{(2)}&\to\Omega^{(3)}\\
\tilde\Omega^{(2)}&\to\tilde\Omega^{(3)}\\
\end{array}: \quad \frac{1}{\l ABCD\r}\to\int \frac{d^4Z_{EF}}{i\pi^2}\frac{\l6134\r}{\l ABEF \r \l EF61 \r \l EF34 \r \l EFCD \r}\,,
\ee
with an obvious generalization to higher loops.

Six-point DCI integrals can in general depend on the three cross ratios
\be\label{uvw_def}
u = \frac{x_{13}^2\,x_{46}^2}{x_{14}^2\,x_{36}^2}\,, 
\qquad v = \frac{x_{24}^2\,x_{51}^2}{x_{25}^2\,x_{41}^2}\,, \qquad
w = \frac{x_{35}^2\,x_{62}^2}{x_{36}^2\,x_{52}^2}\,.
\ee
As at two loops, $\Omega^{(L)}$ and $\tilde\Omega^{(L)}$ are both symmetric under the simultaneous exchange of legs $1 \leftrightarrow 3$ and $4 \leftrightarrow 6$.  This transformation exchanges $u$ and $v$, but we will have to be careful about signs when transforming the parity-odd part of $\tilde\Omega^{(L)}$.  The distinction between these integrals in terms of their numerator factors follows the same rule as at two loops---namely, $\Omega^{(L)}$ picks out pairs of points $x_{a}$ and $x_{b}$ that are related by the cyclic shift $i \rightarrow i+3$ in their dual indices, while for $\tilde\Omega^{(L)}$ these points are related by $i \rightarrow i+3$ plus parity conjugation.

\subsection{Climbing the ladders with differential equations}

These ladder integrals all share one crucial attribute: they satisfy a set of differential equations that relate adjacent loop orders~\cite{Drummond:2010cz}. For the box ladders, this differential equation is quite simple to write down, and has been known for some time:
\be
 \left[z\del_z \zb\del_{\zb} - g^2\right] f(z,\zb,g^2)=0\,.
 \label{eq:diffbox}
\ee
We have rearranged the traditional presentation of this relation, which relates $f^{(L)}$ to $f^{(L-1)}$, to the all-loop expression $f(z,\zb,g^2)$ in order to emphasize that it really is valid for finite coupling.

A similar, if slightly more complicated, differential equation applies to the pentabox ladders. We define the quantity
\be\label{eq:PsiDef}
\Psi^{(L)}(u,v) = (1-u-v) \, {\cal I}_{pl}^{(L)}(x_1,x_2,x_3,x_4,x_6)\,,
\ee
and its finite-coupling version $\Psi(u,v,g^2)$ via
\eqn{definefinitecoupling}.
Then $\Psi(u,v,g^2)$ obeys the differential equation
\be
\left[(1-u-v)uv\del_u\del_v  + g^2 \right]\Psi(u,v,g^2) = 0.
\label{PsiDiffeq}
\ee

Each of the (double) pentaladders also satisfies a differential equation,
which we first give at fixed loop order $L$.  The differential
equation relating $\Omega^{(L)}$ and $\Omega^{(L-1)}$ was derived in momentum-twistor space, and
reads~\cite{Drummond:2010cz}
\be
\frac{\l 1234\r \l 2345\r}{\l 6134\r} Z_1{\cdot}\frac{\del}{\del Z_2}
\left( \frac{1}{\l 2345\r}  Z_6{\cdot}\frac{\del}{\del Z_1} \Omega^{(L)}
\right)
= - \Omega^{(L-1)}\,.
\label{Omdiffeq}
\ee
The corresponding relation between $\tilde\Omega^{(L)}$ and $\tilde\Omega^{(L-1)}$ is
\be
\frac{\l 1234\r \l 2346\r}{\l 6134\r} Z_1 \cdot \frac{\del}{\del Z_2}
\left( \frac{1}{\langle 2346 \rangle} Z_6 \cdot \frac{\del}{\del Z_1}
\tilde \Omega^{(L)} \right) = - \tilde \Omega^{(L-1)}\,.
\label{Omtdiffeq}
\ee
This equation was given in ref.~\cite{Dixon:2011nj} for $L=2$,
but the same derivation holds for any $L>2$, except that the sign of the
right-hand side needs to be flipped;
we also redefine $\tilde\Omega^{(1)}\to -\tilde\Omega^{(1)}$
with respect to ref.~\cite{Dixon:2011nj}.  Note that $\Omega^{(1)}$ and $\tilde\Omega^{(1)}$ are one-loop hexagon integrals with double numerator insertions~\cite{ArkaniHamed:2010gh,Dixon:2011nj}.

We define the finite-coupling versions of
$\Omega^{(L)}$ and $\tilde\Omega^{(L)}$ by,
\be
\Omega\ =\ \sum_L \,(-g^{2})^L\,\Omega^{(L)} \,, \qquad
\tilde\Omega\ =\ \sum_L \,(-g^{2})^L\,\tilde\Omega^{(L)}\,.
\label{Omfinitecouplingdef}
\ee
The finite-coupling analogs of the above differential equations will be
discussed in section~\ref{sec:resummed_result}, after we introduce
some new kinematic variables which dramatically simplify them.


\section{Ladders at Finite Coupling} 
\label{sec:resummed_result}

\subsection{Separated form of the differential equations}

The ladder in the $\Omega^{(L)}$ and $\tilde\Omega^{(L)}$ integrals is framed by the dual coordinates $x_1$ and $x_4$, as can be seen in \fig{fig:omega_L}. We will exploit the symmetries that preserve these two points in order to write a finite-coupling expression for these integrals.
The same technique will also be applied to other systems containing the same ladder.

Thanks to dual conformal symmetry we can put $x_1$ and $x_4$ at zero and infinity, respectively.
It is then easy to see that the symmetry preserving their location
consists of SO(4) rotations and scale transformations.
Writing the SO(4) algebra as a product of two SU(2)'s, this indicates that the ladders are controlled by a
SU(2)${}_L\times$SU(2)${}_R\times$GL(1) symmetry.
The idea will be to find variables which transform as simply as possible under each factor of this group.

Let us first parametrize the hexagon kinematics explicitly in this frame using the embedding formalism.
Each dual coordinate $x_i^\mu \equiv \sigma^\mu_{\alpha \dot \alpha} x_i^{\alpha \dot \alpha}$ (where $\sigma$ are the usual Pauli matrices)
is encoded as a null six-vector $X_i \equiv (x_i^\mu,X_i^+,X_i^-)$,
with respect to the metric $X_i\cdot X_j \equiv X_i^+ X_j^- + X_i^- X_j^+ - x_i\cdot x_j$:
\be
\def\hs{{\hspace{3mm}}} X_i = \left(\begin{array}{c@\hs c@\hs c@\hs c@\hs c@\hs c}
 0& p_2^\mu&-p_4^\mu&0&p_5^\mu&-p_1^\mu\\
 0&0&-p_2{\cdot}p_4&1&-p_1{\cdot}p_5&0\\
 1&1&0&0&0&1\end{array}\right) \,,
\label{XarrayOmega}
\ee
where $i$ labels the columns of the matrix.
Here we have put the points $X_1$ and $X_4$ at 0 and $\infty$, respectively. 
Because the six external momenta in the ladder integral are massless,
in addition to $X_i^2 = 0$ for each $i$, we also have that $X_i\cdot X_{i+1}=0$.
This forces many components to vanish, and implies in addition that $p_i^2=0$.
Note that the $p_i$ are related to, but not equal to, the momenta in the original frame.

The ladder integrals depend only on the cross ratios (\ref{uvw_def}),
which now evaluate to
\bea
 u &\equiv& \frac{X_1\cdot X_3 \, X_4\cdot X_6}{X_1\cdot X_4 \, X_3\cdot X_6}
 = \frac{p_2\cdot p_4}{(p_1+p_2)\cdot p_4} \,, \qquad
 v \equiv \frac{X_2\cdot X_4 \, X_5\cdot X_1}{X_2\cdot X_5 \, X_4\cdot X_1}
   = \frac{p_1\cdot p_5}{(p_1+p_2)\cdot p_5} \,, \nonumber\\
 w &\equiv& \frac{X_3\cdot X_5 \, X_6\cdot X_2}{X_3\cdot X_6 \, X_5\cdot X_2}
   = \frac{p_1\cdot p_2\, p_4\cdot p_5}
          {(p_1+p_2)\cdot p_4 \,(p_1+p_2)\cdot p_5} \,.
\label{uvwintermsofp}
\eea
Notice that these variables
are invariant under separate rescalings of $p_4$ and $p_5$,
and are also invariant under a common rescaling of $p_1$ and $p_2$.

Let us now look at the action
of the SU(2)${}_L\times$SU(2)${}_R\times$GL(1) symmetry on one endpoint of the ladder, that is, on $X_2$ and $X_3$, which according to
\eqn{XarrayOmega} depend on $p_2$ and $p_4$.
The GL(1) scale transformations act simply by rescaling $p_2$.
Using the spinor helicity factorization for null vectors,
$p_i=\lambda_{i\alpha} \tilde\lambda_{i\dot\alpha}$, the first SU(2)
factor acts on holomorphic spinors $\lambda_{2\alpha}$ and $\lambda_{4\alpha}$,
and the second on anti-holomorphic spinors $\tilde\lambda_{2\dot\alpha}$ and $\tilde\lambda_{4\dot\alpha}$.
This motivates introducing the following variables:
\be
x\equiv \frac{\l14\r\l25\r}{\l15\r\l24\r} \,, \qquad
y\equiv \frac{[14][25]}{[15][24]} \,, \qquad
z\equiv \frac{p_2\cdot p_4\, p_2\cdot p_5}
           {p_1\cdot p_4 \, p_1\cdot p_5} \,,
\label{initialxyzdef}
\ee
with $\l14\r = \epsilon_{\alpha\beta}\lambda_{1}^\alpha\lambda_{4}^\beta$,
$[14] = -\epsilon_{\dot\alpha\dot\beta}
 \tilde\lambda_{1}^{\dot\alpha}\tilde\lambda_{4}^{\dot\beta}$, so that $\l14\r [41] = 2 p_1\cdot p_4$, and similarly
for the other spinor products.
Each of $x$ and $y$ is invariant under one of the SU(2)'s but not the other,
and scale transformations act only on $z$.

To find the change of variables between $(u,v,w)$ and $(x,y,z)$, first note from eq.~(\ref{uvwintermsofp}) that
\be
\frac{1-u}{u} = \frac{p_1\cdot p_4}{p_2\cdot p_4} \,, \qquad
\frac{1-v}{v} = \frac{p_2\cdot p_5}{p_1\cdot p_5} \,,
\label{mumvintermsofvp}
\ee
which readily yields
\be
xy = \frac{(1-u)(1-v)}{uv}, \qquad z = \frac{u(1-v)}{v(1-u)} \,.
\label{zasuvw}
\ee
Using the Schouten identity in \eqn{initialxyzdef} one similarly finds
\be
 (1-x)(1-y) = \frac{w}{uv}\,.  \label{omxomy}
\ee
Solving \eqns{zasuvw}{omxomy} for $x$ and $y$ in terms of $u,v,w$, we get
\be
x = 1+\frac{1-u-v-w+\sqrt{\Delta}}{2uv}\,, \qquad\quad
y = 1+\frac{1-u-v-w-\sqrt{\Delta}}{2uv}\,, \label{definexy}
\ee
with $\Delta=(1-u-v-w)^2-4uvw$ as usual.  The choice of sign in front of
$\sqrt{\Delta}$ in \eqn{definexy} is somewhat arbitrary; parity flips
this sign and exchanges $x \lr y$.

The differential equations~\eqref{Omdiffeq} and \eqref{Omtdiffeq}
are expressed in terms of momentum twistors, so it is useful
to write $x,y,z$ in terms of ratios of momentum-twistor four-brackets.
In appendix~\ref{hexvarappendix} we recall the momentum-twistor
representations of the cross ratios $u,v,w$, and also of the variables
$y_u,y_v,y_w$.  Both $u,v,w$ and $x,y,z$ are rational functions
of $y_u,y_v,y_w$; see eqs.~\eqref{u_from_y} and \eqref{defx}--\eqref{defz}.
Using \eqns{u_from_y}{momtwistorreps} in \eqn{definexy},
the momentum-twistor representations of $x,y,z$ are
\be
x = \frac{\l 1246\r \l 1356\r}{\l 1236\r \l 1456\r} \,, \qquad
y = \frac{\l 1345\r \l 2346\r}{\l 1234\r \l 3456\r} \,, \qquad
z = \frac{\l 1236\r \l 1246\r \l 1345\r \l 3456\r}
         {\l 1234\r \l 1356\r \l 1456\r \l 2346\r} \,.
\label{xyz_yi_momtw}
\ee
In this representation, it is easy to show that the dihedral flip $Z_i \lr Z_{4-i}$ that leaves $\Omega^{(L)}$ and $\Omt^{(L)}$ invariant transforms the above variables as $x\leftrightarrow y$ and $z\to1/z$. Also notice that under the cyclic transformation $Z_i \to Z_{i+3}$,
$x$ is exchanged with $y$, while $z$ is left invariant, allowing us
to identify this transformation with parity.  This transformation
also sends the dual coordinates $x_i \to x_{i+3}$.
Inspecting \fig{fig:omega_L}, we see that $\Omega^{(L)}$ is
invariant under parity because the left and right numerator
factors transform into each other under $i\to i+3$. In contrast,
$\Omt^{(L)}$ has both parity-even and parity-odd parts.

The $x,y,z$ variables simplify the momentum-twistor differential
operators appearing in \eqns{Omdiffeq}{Omtdiffeq}.  Using the Schouten
identity for four-brackets and the chain rule, we have
\bea
Z_6 \cdot \frac{\del}{\del Z_1} &=&  \frac{\l1346\r\l2345\r}{\l 1234\r\l1345\r}\left( y\partial_y + z\partial_z\right),\\
 Z_1 \cdot \frac{\del}{\del Z_2} &=&  \frac{\l1346\r}{\l 2346\r}\left( y\partial_y - z\partial_z\right)\,.
\eea
Using these relations, the differential equation \eqref{Omdiffeq}
for $\Omega$ becomes
\be
 - \frac{\l2345\r}{\l1345\r}  \frac{\l1346\r}{\l2346\r}
  (y\del_y - z\del_z) (y\del_y + z\del_z) \Omega^{(L)} = -\Omega^{(L-1)} \label{Omdiffop}
\ee
which can be expressed in terms of $y$ and $z$ only as
\be
\frac{1-y}{y} \Bigl[ (y\del_y)^2 - (z\del_z)^2 \Bigr] \Omega^{(L)}
 =  - \Omega^{(L-1)} \,. \label{Omdiffy}
\ee
At finite coupling, using the definition~\eqref{Omfinitecouplingdef},
the $\Omega$ ladders thus satisfy the equations:
\bea
\biggl[ \frac{1-y}{y} \Bigl( (y\del_y)^2 - (z\del_z)^2 \Bigr)
      - g^2 \biggr] \Omega(x,y,z,g^2)
 &=& 0 \,, \label{Omdiffyfin}\\
\biggl[ \frac{1-x}{x} \Bigl( (x\del_x)^2 - (z\del_z)^2 \Bigr)
      - g^2 \biggr] \Omega(x,y,z,g^2)
 &=& 0 \,, \label{Omdiffxfin}
\eea
where we have also used the fact that $\Omega(x,y,z,g^2)$ is even
under parity, $x\lr y$, to add the second equation.
This form of the equations will be very convenient because they now take on a separated form,
thanks to switching to the $x,y,z$ variables.

The corresponding differential equations for $\Omt(x,y,z,g^2)$,
derived in a similar way from \eqn{Omtdiffeq}, are
\bea
\biggl[ \frac{1-y}{y} \Bigl( (y\del_y)^2 - (z\del_z)^2 \Bigr)
   - \frac{1}{y} (y\del_y + z\del_z ) - g^2 \biggr] \Omt(x,y,z,g^2)
&= 0 \,, \label{Omtdiffyfin}\\
\biggl[ \frac{1-x}{x} \Bigl( (x\del_x)^2 - (z\del_z)^2 \Bigr)
  - \frac{1}{x} (x\del_x - z\del_z ) - g^2 \biggr] \Omt(x,y,z,g^2)
&= 0 \,. \label{Omtdiffxfin}
\eea
In particular, expressing \eqref{Omtdiffeq} in terms of the $x,y,z$ variables leads to \eqref{Omtdiffyfin}, whereas \eqref{Omtdiffxfin} follows from the latter by a flip transformation, as discussed under \eqref{xyz_yi_momtw}. Notice from the momentum-twistor forms~\eqref{Omdiffeq}
and \eqref{Omtdiffeq} that the second-order part of the differential operator
is exactly the same for $\Omega$ and $\Omt$, because
the $\l2345\r$ and $\l2346\r$ factors cancel in the second-order terms. 
The extra linear term for $\Omt$ arises when the operator
$Z_1 \cdot \del/\del Z_2$ acts on the factor $\l2345\r/\l2346\r$.


\subsection{Pentaladders}
\label{sec:doublepentasubsection}

We now turn to the solution of the differential equations~\eqref{Omdiffyfin}
and \eqref{Omdiffxfin} for $\Omega$, and~\eqref{Omtdiffyfin}
and \eqref{Omtdiffxfin} for $\Omt$.
We begin by diagonalizing $z\partial_z$ using a Mellin representation.
(A related double Mellin representation of the box ladder integral has been
obtained using integrability~\cite{Fleury:2016ykk}.)
We seek separated solutions for $\Omega(x,y,z,g^2)$ of the form
\be
z^{i\nu/2} F_\nu(x,y).
\label{OmzFparam}
\ee
\Eqn{Omdiffxfin} then gives
\be
\Bigl[ (1-x)(x\del_x)^2 +\tfrac14(1-x)\nu^2-x g^2 \Bigr] F_\nu (x,y) = 0,
\label{Omdiffeqxy}
\ee
while \eqn{Omdiffyfin} gives the identical equation for $F$ in $y$.
The four independent solutions to the pair of differential
equations can be labeled by the signs of $\nu$:
\be
F_{\pm\nu}^{j(\nu)}(x)F_{\pm\nu}^{j(\nu)}(y),
\qquad j(\nu) \equiv i\sqrt{\nu^2+4g^2}\,,
\label{eq:general solution}
\ee
where $F_\nu^j$ are hypergeometric functions, normalized to $F_\nu^j(1)=1$:
\be
 F_{\nu}^{j}(x) \equiv \frac{\Gamma(1+\frac{i\nu+j}{2})\Gamma(1+\frac{i\nu-j}{2})}{\Gamma(1+i\nu)}\,x^{i\nu/2}\,{}_2F_1(\tfrac{i\nu+j}{2},\tfrac{i\nu-j}{2},1+i\nu,x).
 \label{eq:defF}
\ee
Below when discussing the box ladders we will find that $j(\nu)-1$ is the SO(4) spin, which suggests viewing the differential equations~\eqref{Omdiffyfin} and \eqref{Omdiffxfin} intuitively as two relations among the three Casimir invariants of SU(2)${}_L\times$SU(2)${}_R\times$GL(1).

To find the physically relevant combination of the solutions (\ref{eq:general solution}),
we impose the fact that $\Omega$ must be smooth in the entire positive octant $u,v,w>0$.
In particular, consider the neighborhood of the point $(u,v,w)=(1,1,1)$,
where $x$ and $y$ are both small. The function should admit a regular Taylor series expansion.
However only the combination $x y$ is regular: $x/y$ depends in a complicated way on the angle of approach.
From the behavior of \eqn{eq:defF} near the origin,
we see that requiring the leading term in this limit
to be a power of $xy$ leaves only two acceptable solutions:
$F_{+\nu}^{j(\nu)}(x)F_{+\nu}^{j(\nu)}(y)$
and $F_{-\nu}^{j(\nu)}(x)F_{-\nu}^{j(\nu)}(y)$.

To get a further constraint, we note from \eqns{zasuvw}{omxomy} that:
\be
\frac{w}{uv} = (1-x)(1-y), \qquad (x-1)+(y-1) = \frac{1-u-v-w}{uv} \,,
\ee
which imply that $(1-x)$ and $(1-y)$ can both switch sign in the positive octant.
They can only switch simultaneously when $1-u-v-w$ switches sign.
However, the individual hypergeometric functions contain singular logarithms of the form $\ln(1-x)$
in their Taylor series.\footnote{%
These logarithms can be identified using the relations~\eqref{xdxF},
\eqref{xdxFpALT}, \eqref{Fpshiftindex} and \eqref{xdxFp}. While $F_{\nu}^{j}(x)$
is finite as $x\to1$, its derivative behaves like $-g^2 \, \ln(1-x)$.}
Smoothness in the positive octant thus requires that they combine into the regular combination $\ln[(1-x)(1-y)]=\ln\frac{w}{uv}$.
This singles out a unique linear combination of the above two functions.
We conclude that an integral representation of the following form must hold:
\be\label{intermed double penta}
\Omega(u,v,w,g^2)
= \int_{-\infty}^\infty d\nu \, c(\nu,g^2) \, z^{i\nu/2} \,
 \frac{F_{+\nu}^{j(\nu)}(x)F_{+\nu}^{j(\nu)}(y)-F_{-\nu}^{j(\nu)}(x)F_{-\nu}^{j(\nu)}(y)}{\sinh(\pi \nu)}\,,
\ee
where we have integrated over the dilatation eigenvalue $\nu$ with a yet undetermined coefficient $c(\nu,g^2)$.
The insertion of the explicit factor of $1/\sinh(\pi\nu)$ is motivated
by the following consideration:
Regularity of the Taylor series at $(u,v,w)=(1,1,1)$ implies that the singularities
of $\frac{c(\nu,g^2)}{\sinh(\pi\nu)}$ can be at most single poles at imaginary integers,
since in this limit $x,y\to 0$ and the integral can be done by residues (closing the contour below the real axis on the first term,
and above in the second term).  Thus $c(\nu,g^2)$ is an entire function.

To fix this function, we find a boundary condition at large $\nu$. At large $\nu$, the integral simplifies dramatically, because
one of the indices on the hypergeometric functions, $i\nu-j$, goes to zero, so 
using \eqn{zasuvw} we get simply
\be
z^{i\nu/2}\ F_{+\nu}^{j}(x)F_{+\nu}^{j}(y)
\to \left(\frac{u(1-v)}{v(1-u)}\right)^{i\nu/2}
  \ \left(\frac{(1-u)(1-v)}{uv}\right)^{i\nu/2}
= \left(\frac{1-v}{v}\right)^{i\nu} \,.
\label{largenuintegrand}
\ee
We see that for the above term, the large $\nu$ region dominates\footnote{Similarly, for the second term in \eqn{intermed double penta} the limit of large $\nu$ governs the $u\to +\infty$ limit. For our purposes, it will be sufficient to focus on the $v\to +\infty$ limit.}
the limit $v\to +\infty$. Choosing the phase, say,
\be
\log\left(\frac{1-v}{v}\right)\to \log\left(\frac{v-1}{v}\right)-i\pi  \,,
\ee
the integrand numerator acquires a factor $e^{\pi \nu}$,
canceling the same factor from the $\sinh(\pi\nu)$ in the denominator
and making it marginally convergent. Thus
\be
\lim_{v\to +\infty} \Omega(u,v,w,g^2)
 = \int_{\sim 1}^\infty d\nu\,2c(\nu,g^2) e^{-i\nu/v}\,.
\ee
If $c$ approaches a constant, we get linear growth in $v$ as $v\to\infty$.
Now, at tree level ($g^2=0$), we have that $\Omega(u,v,w,0)=1-u-v$.
(We didn't define $\Omega^{(0)}$ as an integral but it can be found by acting on the known
one-loop integral $\Omega^{(1)}$ with the differential operator~\eqref{Omdiffxfin}.)
Hence the tree-level quantity does grow linearly in the limit.
On the other hand, the loop corrections $\Omega^{(L)}$ with $L\geq 1$ grow at most logarithmically
in this limit, since they are uniform transcendental functions.
We conclude that the missing coefficient must be independent of the coupling, 
and equal to $c(\nu,g^2)=1/(2i)$.

Our final result is therefore:
\be
\boxed{
 \Omega(u,v,w,g^2) = \int_{-\infty}^\infty \frac{d\nu}{2i} \, z^{i\nu/2} \, 
 \frac{F_{+\nu}^{j(\nu)}(x)F_{+\nu}^{j(\nu)}(y)-F_{-\nu}^{j(\nu)}(x)F_{-\nu}^{j(\nu)}(y)}{\sinh(\pi \nu)}
      }
\label{result double penta} 
\ee
where $x,y,z$ are defined in \eqns{zasuvw}{definexy}, $j(\nu)=i\sqrt{\nu^2+4g^2}$, 
and the hypergeometric functions $F$ are defined in \eqn{eq:defF}.
This formula is the main result of this section.

As a simple check, at tree level, $g^2=0$, 
the hypergeometric functions become trivial and $F_{\nu}^{j(\nu)}(x)\to x^{i\nu/2}$.
From \eqn{uvwyfromxyz}, $\sqrt{xy/z}=(1-u)/u$
and $\sqrt{xyz}=(1-v)/v$, so we get
\be
\Omega(u,v,w,0) =  \int_{-\infty}^\infty \frac{d\nu}{2i \sinh(\pi\nu)} \left[ \left(\frac{1-v}{v}\right)^{i\nu}-\left(\frac{1-u}{u}\right)^{-i\nu}\right]\,.
\ee
Performing the integral by closing the contour in the lower half-plane and summing over residues at $\nu=-ik$, we reproduce the result $\Omega(u,v,w,0)=1-u-v$.
In general, it is straightforward to expand the integral \eqref{result double penta} around $(u,v,w)=(1,1,1)$, since $x$ and $y$ are both small and so only the residues at a finite number of poles contribute.
The residues give hypergeometric functions and only a finite number of terms
in their Taylor expansions are needed.
The expansion coefficients can therefore be obtained exactly in $g^2$.
At one loop, we have resummed the series around $(1,1,1)$,
finding as expected:
\be
 \Omega^{(1)}(u,v,w) \equiv \frac{-d}{dg^2} \Omega(u,v,w,g^2)\big|_{g^2=0}
 = \Li_2(1-u)+\Li_2(1-v)+\Li_2(1-w)+\ln u\ln v-2\zeta_2.
\ee
Section~\ref{sec:sumsandpolys} discusses how to perform
such expansions at higher loop orders.

The pentaladder integral $\Omt$ with mixed numerators
can be analyzed in an identical fashion.
The only change is that the solutions to the differential
equation \eqref{Omtdiffxfin} with the separated form~\eqref{OmzFparam}
are now $F^{j(\nu)\prime}_{+\nu}(x)$ and $F^{j(\nu)\prime}_{-\nu-2i}(x)$ with
\be
 F_{\nu}^{j\prime}(x) \equiv \frac{\Gamma(\frac{i\nu+j}{2})\Gamma(\frac{i\nu-j}{2})}{\Gamma(i\nu)}\,x^{i\nu/2}\,
 {}_2F_1(\tfrac{i\nu+j}{2},\tfrac{i\nu-j}{2},i\nu,x)
 \label{eq:defFprime} 
\ee
instead of \eqn{eq:defF}.  These functions are to be multiplied by the solutions
to~\eqref{Omtdiffyfin}, which are found (by letting $x\to y$ and $\nu\to-\nu$) to be $F^{j(\nu)\prime}_{-\nu}(y)$ and $F^{j(\nu)\prime}_{\nu-2i}(y)$.  As in the case of $\Omega$, imposing regularity in the positive octant fixes a unique combination of these solutions. It first requires the pairings $F_{+\nu}^{j(\nu)\prime}(x)F_{+\nu-2i}^{j(\nu)\prime}(y)$ and $F_{-\nu-2i}^{j(\nu)\prime}(x)F_{-\nu}^{j(\nu)\prime}(y)$, and then it imposes a relative minus sign between the two.

We also require that the large-$v$ limit
is saturated by its tree-level expression.
The one-loop result~\cite{ArkaniHamed:2010gh} is
\be
\Omt^{(1)}(u,v,w) = - \ln\left(\frac{u}{w}\right)\ \ln v
  + \frac{1-y_v}{1-y_u y_v} \ln\left(\frac{u}{v}\right)\ \ln w \,,
\label{Omt1}
\ee
normalized with the opposite sign as eq.~(E.4) of ref.~\cite{Dixon:2011nj}.
We note that this is not a pure transcendental function, although $\Omt^{(L)}$ is pure for $L\geq 2$.
The integral $\Omt^{(L)}$ has only been defined so far for $L\geq 1$ but we can define it at $L=0$
by applying the differential equation~\eqref{Omtdiffyfin}, which gives us,
after simplification using formulas in appendix \ref{hexvarappendix}:
\be
 \Omt^{(0)}(u,v,w) = -(1-u)(1-v) - v(1-u)\frac{1-y_v}{1-y_u} \left( 1+ \frac{\ln w}{1-w}\right)\,.
\ee
This expression has a strange form for a ``tree'' object, since it contains a logarithm.  However, hitting $\Omt^{(0)}$ once more with the differential operator gives zero, confirming that it is indeed the leading term of the $\Omt$ family.

In the limit of large $v$ (choosing the branch of the square root where $x\to 1$, corresponding to
$y_v\to 0$, $y_u\to\infty$ with $y_uy_v\to y$),
we have $\frac{1-y_v}{1-y_u}\to 0$.  Hence $\Omt^{(0)}$ is dominated by the first term, which grows linearly like $v(1-u)$.
The one-loop result~\eqref{Omt1} and the higher loop orders grow at most logarithmically, so the large-$v$ limit is again tree-level exact.
Matching the integral representation with this limit,
our final result for $\Omt$ is
\be
\boxed{
 \Omt(u,v,w,g^2) = g^2\int_{-\infty}^\infty \frac{d\nu}{2i} \, z^{i\nu/2} \, 
 \frac{F_{+\nu}^{j(\nu)\prime}(x)F_{+\nu-2i}^{j(\nu)\prime}(y)
 -F_{-\nu-2i}^{j(\nu)\prime}(x)F_{-\nu}^{j(\nu)\prime}(y)}{\sinh(\pi \nu)} \,.
 }
\label{result double penta 2}
\ee
We note the asymmetry between $x$ and $y$, which originates from the differential equations~\eqref{Omtdiffyfin}
and~\eqref{Omtdiffxfin}: 
$\Omt$ is not invariant under parity.

\subsection{Differential relations among pentaladder integrals}
\label{moredoublepentasubsection}

It turns out that the functions $\Omega$ and $\Omt$ are not independent,
but rather they appear as derivatives of one another. 
These relations can be understood as properties of
the  hypergeometric functions entering \eqns{result double penta}{result double penta 2}.

It will prove useful to first introduce other auxiliary integrals,
$\OL(u,v,w,g^2)$,
which we call the odd ladder because it
has odd parity and its perturbative coefficients $\OL^{(L)}$
all have odd weight,
and an even companion $\WL(u,v,w,g^2)$.
The odd ladder will be a generalization of the one-loop six-dimensional integral $\tilde\Phi_6$ studied in ref.~\cite{Dixon:2011ng}.
(See also ref.~\cite{DelDuca:2011ne}.)
The latter integral satisfies
\be
\sqrt{\Delta} \del_w \Omega^{(2)} = - \tilde\Phi_6 \,,
\label{tPhi6def}
\ee
and it is the first parity odd function in the space of hexagon functions.

By analogy with \eqn{tPhi6def}, we now define
\be
\OL^{(L-1)}\ \equiv\ \sqrt{\Delta} \del_w \Omega^{(L)}
               \ =\ ( - x \del_x + y \del_y ) \Omega^{(L)} \,,
\label{OdefL}
\ee
using \eqn{xdxminusydy} and identifying $\OL^{(1)} = - \tilde\Phi_6$.
The corresponding finite-coupling definitions of $\OL$ and its even companion $\WL$ are
\be
\OL\ \equiv\ \frac{1}{g^2} ( x \del_x - y \del_y ) \Omega \,,\qquad
\WL\ \equiv  ( x\del_x + y\del_y) \Omega\,.
\label{Odef}
\ee
This normalization of $\WL$ will prove convenient in
section~\ref{sec:spacesect}.

Given the finite-coupling solution for $\Omega$, \eqn{result double penta},
we can obtain one for the odd ladder integral simply by acting
with the differential operator on the products 
$F_{\pm\nu}^{j}(x)F_{\pm\nu}^{j}(y)$ in the integrand.
The hypergeometric function satisfies
\be
x\frac{d}{dx} F_{\nu}^{j(\nu)} (x)
= \frac{i\nu}{2} F_{\nu}^{j(\nu)}(x) + g^2 F_{\nu-2i}^{j(\nu)\prime}(x), \label{xdxF}
\ee
which can be verified, for example, using the hypergeometric function
series representation~\eqref{eq:2F1_def}.
Note that here we used that $-\nu^2-[j(\nu)]^2=4g^2$.

When we apply \eqn{xdxF} to \eqn{Odef}, using \eqn{result double penta},
the $i\nu/2$ terms cancel, and we are left with:
\bea
 \OL(u,v,w,g^2) &=& \int_{-\infty}^\infty \frac{d\nu}{2i}
 \frac{z^{i\nu/2}}{\sinh(\pi \nu)} \Bigl[
  F_{+\nu-2i}^{j(\nu)\prime}(x)F_{+\nu}^{j(\nu)}(y)-F_{+\nu}^{j(\nu)}(x)F_{+\nu-2i}^{j(\nu)\prime}(y)
\nonumber\\
&&\hskip2.8cm\null
 -F_{-\nu-2i}^{j(\nu)\prime}(x)F_{-\nu}^{j(\nu)}(y)+F_{-\nu}^{j(\nu)}(x)F_{-\nu-2i}^{j(\nu)\prime}(y) \Bigr]
\,. \label{resultOL}
\eea
Similarly,
\bea
 \WL(u,v,w,g^2) &=&
 g^2\int_{-\infty}^\infty \frac{d\nu}{2i}
 \frac{z^{i\nu/2}}{\sinh(\pi \nu)} \Bigl[
  F_{+\nu}^{j(\nu)\prime}(x)F_{+\nu}^{j(\nu)}(y)+F_{+\nu}^{j(\nu)}(x)F_{+\nu-2i}^{j(\nu)\prime}(y)
\nonumber\\
&&\hskip2.8cm\null
 -F_{-\nu-2i}^{j(\nu)\prime}(x)F_{-\nu}^{j(\nu)}(y)-F_{-\nu}^{j(\nu)}(x)F_{-\nu}^{j(\nu)\prime}(y) \Bigr]\,.
 \label{resultWL}
\eea

We are now in a position to discuss several first-order differential
relations among the integrals.
If we take the difference between \eqn{Omdiffyfin} and \eqn{Omdiffxfin}
so as to cancel the $z$ derivative, then the $x$ and $y$ derivatives
factorize as $(x\del_x)^2-(y\del_y)^2 = (x\del_x+y\del_y)(x\del_x-y\del_y)$,
which shows that $\Omega$ can also be written as a first derivative
of the odd ladder integral,
\be
\Omega = \frac{(1-x)(1-y)}{x-y} (x\del_x+y\del_y) \OL\,.
\label{OmfromOL}
\ee
This equation generalizes a relation found between $\Omega^{(1)}$
and a derivative of $\tilde\Phi_6$ in ref.~\cite{Dixon:2011ng}.

However, we also find empirically, to high orders in perturbation theory,
that the $x$ and $y$ derivatives of $\OL$ also contain
the parity-even part of $\Omt$, which we call $\Omte$,
and the $z$ derivative of $\OL$ generates the parity-odd part, $\Omto$,
where
\be
\Omt = \Omte + \Omto.
\label{OmteOmtosplit}
\ee
In particular, we find that
\bea
x \del_x \OL &=& - \Omte + \frac{x}{1-x} \Omega \,, \label{xdxOL}\\
y \del_y \OL &=&   \Omte - \frac{y}{1-y} \Omega \,, \label{ydyOL}
\eea
which gives back \eqn{OmfromOL} and also
\be
\Omte = \frac{xy}{x-y} \Bigl[ (1-x)\del_x + (1-y)\del_y \Bigr] \OL\,.
\label{OmtefromOL}
\ee
The parity-odd relation is just
\be
\Omto = - z \del_z \OL \,.
\label{OmtofromOL}
\ee
These two empirical relations combine to
\be
\Omt = \Biggl(\frac{xy}{x-y} \Bigl[ (1-x)\del_x + (1-y)\del_y \Bigr]- z\del_z\Biggr) \OL\,.  \label{eq: Omt from OL}
\ee
In appendix~\ref{ExtraCoproductRelationsSubAppendix}, we provide additional relations of a similar nature.

We would like to derive these differential relations from the finite-coupling integral representations.  In order to do so, it is useful to
consider the derivatives of
$F_{\nu-2i}^{j(\nu)\prime}(x)$ and $F_{\nu}^{j(\nu)\prime}(x)$.
If we apply the differential operator $(1-x)x \, d/dx$ to the left-
and right-hand sides of \eqn{xdxF}, we can use
the second-order differential equation~\eqref{Omdiffeqxy} satisfied
by $F_\nu^j$ to simplify the left-hand side, obtaining a formula
for the derivative of $F_{\nu-2i}^{j(\nu)\prime}(x)$:
\be
\frac{d}{dx} F_{\nu-2i}^{j(\nu)\prime}(x)
= -\frac{i\nu}{2x} F_{\nu-2i}^{j(\nu)\prime}(x)
+ \frac{1}{1-x} F_{\nu}^{j(\nu)}(x). \label{xdxFpALT}
\ee
Furthermore, only two out of the three functions $F_{+\nu-2i}^{j(\nu)\prime}(x)$,
$F_{+\nu}^{j(\nu)\prime}(x)$ and $F_{+\nu}^{j(\nu)}(x)$ can be linearly independent;
indeed, hypergeometric identities can be used to show that
\be
F_{+\nu-2i}^{j(\nu)\prime}(x)
= F_{+\nu}^{j(\nu)\prime}(x) - \frac{i\nu}{g^2} F_{+\nu}^{j(\nu)}(x).
\label{Fpshiftindex}
\ee
Combining \eqns{xdxFpALT}{Fpshiftindex}, we obtain an equation
for the derivative of $F_{\nu}^{j(\nu)\prime}(x)$:
\be
\frac{d}{dx} F_{\nu}^{j(\nu)\prime}(x)
= \frac{i\nu}{2x} F_{\nu}^{j(\nu)\prime}(x) + \frac{1}{1-x} F_{\nu}^{j(\nu)}(x).
\label{xdxFp}
\ee

The physical significance of the four functions we have introduced is
now clear: they form a complete
basis for products of $F$ and $F'$ with arguments $x$ and $y$ that are smooth in the Euclidean region.
More precisely, we can form the four-vector
\be V_i(\nu,g^2) = \left\{ \WL,\ \Omega,\ \Omt_{\rm e},\ \OL \right\}_\nu,
\quad i=1,2,3,4,
\label{Videf}
\ee
where the subscript $\nu$ means to focus on the $\nu$-integrand
(i.e.~to drop $\int d\nu/(2i)$ in each of
the integrals in \eqn{resultWL}, \eqn{result double penta}, \eqn{result double penta 2} and \eqn{resultOL}).
Because this basis is complete, all the $x$, $y$ and $z$ derivatives of the $V_i$ can be expressed as linear combinations of the $V_i$, allowing the total differential to be written in matrix form:
\be
d V_i(\nu,g^2) = (dM_{ij}(\nu,g^2)) V_j(\nu,g^2)\,.
\label{dVisMV}
\ee
Computing the derivatives,
we find the following explicit form for the matrix $M$:
\be
M= \left(\begin{array}{c@{\hspace{4mm}}c@{\hspace{4mm}}c@{\hspace{4mm}}c}
\frac{i\nu}{2} \ln z& -g^2 \ln c -\frac{\nu^2}{2}\ln(xy)& g^2\ln(xy) & 0 \\
\frac{1}{2}\ln(xy) & \frac{i\nu}{2}\ln z & 0 & \frac{g^2}{2} \ln \frac{x}{y} \\
-\frac{1}{2}\ln c & 0 & \frac{i\nu}{2}\ln z & \frac{g^2}{2}\ln \frac{1-x}{1-y} +\frac{\nu^2}{4}\ln\frac{x}{y} \\
0 & -\ln \frac{1-x}{1-y} & -\ln\frac{x}{y} & \frac{i\nu}{2}\ln z \end{array}\right) \label{matrix section3}
\ee
where we have abbreviated $c\equiv (1-x)(1-y)$.

The matrix $M$ provides us with first-order differential relations
between integrals.  Since the variable $\nu$ is to be integrated over,
to find these relations we should take combinations of $x$ and $y$
derivatives that are independent of $\nu$.
For example, the $x$ and $y$ derivatives of the second row of $M$ give us back the definitions \eqn{Odef},
whereas the $x$ and $y$ derivatives of the fourth row give the two nontrivial relations \eqn{OmfromOL} and \eqn{OmtefromOL}.
Finally, from the first and third row we find two additional first-order relations:
\bea
 \frac{(1-x)(1-y)}{x-y} \left( x\del_x - y\del_y\right) \WL &=& g^2\Omega,\\
2(1-x)(1-y)\left( x\del_x + y\del_y\right) \Omt_{\rm e} &=& (x+y-2xy) \WL - g^2 (x-y) \OL\,.
\eea
The first follows readily from the factorized form of the $\Omega$ differential equation.
Using eq.~(\ref{OmtofromOL}), the second-order \eqn{Omtdiffyfin} for $\Omt$ could also be readily rederived from the matrix $M$.

This discussion shows that $\Omega$ and $\Omt$ naturally fit inside a common system.
In section \ref{sec:spacesect} we will use the matrix $M$ to define an enlarged set of transcendental functions,
which will be closed under the action of taking any derivative.


\subsection{Pentabox ladders}

Consider now the similar integral in \fig{fig:pentaboxladder_L} where we chop off the pentagon on the right and replace it by a box.  The two light-like-separated dual coordinates of the right pentagon are replaced by a single point $q=x_{6}$ which is not light-like separated from $x_1$ or $x_4$.

At one loop, for example, the integral we consider is given in \eqn{eq:pent1},
or in terms of seven momentum twistors,
\be
 \Psi^{(1)} = \int \frac{d^4Z_{AB}}{\pi^2}
 \frac{\l AB13\r \l 56(712)\cap(234)\r}{\l AB71\r\l AB12\r\l AB23\r\l AB34\r\l AB 56\r}.
\ee
Using the embedding formalism as in \eqn{XarrayOmega}, and again putting the sides of the ladder at $0$ and $\infty$, we parametrize the external kinematics as
\be
\def\hs{{\hspace{3mm}}} X_i = \left(\begin{array}{c@\hs c@\hs c@\hs c@\hs c}
 0& p_2^\mu&-p_4^\mu&0&q^\mu\\
 0&0&-p_2{\cdot}p_4&1&q^2/2\\
 1&1&0&0&1\end{array}\right) \,,
\label{XarrayPsi}
\ee
where $q^2\neq 0$ reflects the external masses on one side of the ladder.
The two cross-ratios are now:
\be
 u\equiv \frac{x_{16}^2x_{24}^2}{x_{26}^2x_{14}^2}=\frac{q^2}{(q-p_2)^2}\,, \qquad v\equiv \frac{x_{46}^2x_{13}^2}{x_{36}^2x_{14}^2}=\frac{p_2\cdot p_4}{(p_2-q)\cdot p_4}\,.
\ee
Scale transformations act by rescaling $q$, which leaves invariant the variable
\be
\frac{(1-u)(1-v)}{uv} = \frac{2p_2\cdot q}{q^2} \frac{q\cdot p_4}{p_2\cdot p_4}
\,.
\ee
Therefore we make an ansatz for $\Psi$ as a sum of terms
\be
 z^{i\nu/2} \, F(x),
\qquad x\equiv \frac{(1-u)(1-v)}{uv} \,,
\quad z \equiv \frac{u(1-v)}{v(1-u)} \,.
\label{PsiAnsatz}
\ee
Using the chain rule, we can rewrite $u$ and $v$ partial derivatives
in terms of $x$ and $z$ derivatives,
\be
-(1-v) v \del_v = x \del_x + z \del_z \,, 
\qquad
-(1-u) u \del_u = x \del_x - z \del_z \,. 
\ee
Applying also the identity $(1-u-v)/[(1-u)(1-v)] = -(1-x)/x$,
the differential equation~\eqref{PsiDiffeq} for $\Psi(u,v,g^2)$
(defined via \eqn{eq:PsiDef}) becomes
\be
\biggl[ \frac{1-x}{x} \Bigl( (x\del_x)^2 - (z\del_z)^2 \Bigr)
      - g^2 \biggr] \Psi(x,z,g^2) = 0,
\ee
which has exactly the same form as \eqn{Omdiffxfin} for $\Omega(x,y,z,g^2)$!
Thus for the ansatz~\eqref{PsiAnsatz}
we get exactly the equation for $F(x)$ given in \eqn{Omdiffeqxy}; that is,
the partial wave decomposition of $\Psi$ gives a sum of $F^{j(\nu)}_{\pm\nu}$ terms.
As in the double pentagon case, here
we could again argue that the relative coefficient is fixed by analyticity around $u,v=1$ and $u+v=1$, up to an entire function,
itself fixed from the $v\to\infty$ limit.

In fact the equivalence of the two problems was already understood from the differential equation in ref.~\cite{Drummond:2010cz}:
we can get to the pentabox ladders by taking the $w\to 0$ limit of the result \eqref{result double penta}, where $y\to 1$ and thus $F_{\pm \nu}^j(y)\to1$. This gives immediately:
\be
 \Psi(u,v,g^2)= \Omega(u,v,0,g^2) = \int_{-\infty}^\infty \frac{d\nu}{2i} \, z^{i\nu/2} \, 
\frac{F_{+\nu}^{j(\nu)}(x)-F_{-\nu}^{j(\nu)}(x)}{\sinh(\pi \nu)}\,. 
\label{result penta}
\ee


\subsection{Box ladders}

We can conduct a similar analysis for the box ladders depicted in \fig{fig:ladder_L}. In this case a finite-coupling expression was already derived in ref.~\cite{Broadhurst:2010ds}.  We will show that we reproduce this expression, which we presented in our conventions in \eqn{eq:finitebox}.  We consider the ladder whose long sides are labelled by $x_1$ and $x_5$, as in \fig{fig:ladder_L}, so again we set these points to zero and infinity.  The data then are the ratio $x_3^2/x_7^2$ and the angle between $x_3$ and $x_7$, which are the norm and phase of the complex variable $z$ defined in \eqn{eq:boxvars}.  The conventional cross ratios $u$ and $v$ are defined by
\be
u=\frac{x_{13}^2x_{57}^2}{x_{15}^2x_{37}^2}=\frac{z \zb}{(1-z)(1-\zb)},\qquad
v=\frac{x_{35}^2x_{17}^2}{x_{15}^2x_{37}^2}=\frac{1}{(1-z)(1-\zb)}\,;
\ee
note that $u/v =z \zb$.  The sequence of ladders obeys the differential equation:
\be
 \left[z\del_z \zb\del_{\zb} - g^2\right] f(z,\zb,g^2)=0,
\ee
as previously presented in \eqn{eq:diffbox}. The one-loop case is
\be
 f^{(1)}(z,\zb) = 2(\Li_2(z)-\Li_2(\zb))-\ln(z\zb)\big(\Li_1(z)-\Li_1(\zb)\big).
\ee
The differential equation then requires
$f^{(0)}= (z-\zb)/[(1-z)(1-\zb)]$.
Again we make an ansatz for $f$ as a sum of terms of the form:
\be
 (z\zb)^{i\nu/2} F(\phi), \qquad e^{i\phi}\equiv \sqrt{z/\zb}\,;
\ee
that is, $\phi={\rm arg}\,z$ when $z$ is complex, and the differential equation becomes
\be
 (\nu^2-\del_\phi^2+4g^2)F(\phi)=0.
\ee
The general solution is a combination of
$F_{\pm j}(\phi)=e^{\pm i j \phi}$, with $j(\nu)= i\sqrt{\nu^2+4g^2}$ as before.
Because $j(\nu)$ is conjugate to $\phi$, the angle between $x_3$ and $x_7$,
it should be interpreted as SO(4) spin.\footnote{More precisely, comparing with the Gegenbauer polynomials with spin $\ell$, $C^{(1)}_\ell(\cos\phi) = \frac{\sin((\ell+1)\phi)}{\sin\phi}$, with the denominator corresponding to the $(z-\zb)$ factored out in \eqn{eq:normalization boxes}, we see that $j(\nu)-1$ should be identified with SO(4) spin.}

To find the correct combination of solutions $F_{\pm j}(\phi)$, we wish to impose that the loop corrections to $f$ vanish when $z=\zb$, as a consequence of the factor $(z-\zb)$ removed in \eqn{eq:normalization boxes}.
Working in the Euclidean region $\zb=z^*$, this occurs for two different values of the phase: $\phi=0,\pi$.
As we will see below, the proper interpretation of the vanishing at $\phi=0$ turns out to be subtle,
because there is a singularity at $z=\zb=1$. In fact the restriction to $\phi=0$ at tree level results in a nonvanishing distribution
supported at that point.  However, $f$ vanishes identically at $\phi=\pi$.
This means that for each value of $\nu$ only the combination of solutions $F_{\pm j}(\phi)$ that vanishes at $\phi=\pi$ contributes,
so that we can write:
\be
 f(z,\zb,g^2) = \int_{-\infty}^\infty d\nu \, (z\zb)^{i\nu/2} \,
 \frac{\sin[(\pi-\phi)j(\nu)]}{\sin(\pi j(\nu))} c(\nu,g^2).
\ee
The symmetries of the integral imply that $c(\nu,g^2)$ is an even function of $\nu$.
Now consider single-valuedness as $(z\zb)\to 0$. There the $\nu$-integral will be done by residues in the lower-half plane
and each residue should correspond to an integer spin $j$ in order to be single-valued.
This implies that the only singularities of $\frac{c(\nu,g^2)}{\sin(\pi j)}$ are single poles at integer spin,
and therefore $c(\nu,g^2)$ is an entire function.

To fix the asymptotics of $c(\nu,g^2)$ we consider the limit $\zb\to 1$ (with $z$ otherwise fixed).
We get a power divergence at tree level,
but only logs at loop level; in the above expression this divergence will come from large positive $\nu$, where $j\approx i\nu$:
\be
 \frac{1}{\zb-1} \sim \int_{\sim 1}^\infty d\nu\,c(\nu,g^2) e^{i\nu(\zb-1)}
 \quad \Rightarrow \quad
 c(\nu,g^2)\to \frac{1}{i}.
\ee
Because $c(\nu,g^2)$ is entire, this behavior determines it uniquely:
\be
 f(z,\zb,g^2) = \int_{-\infty}^\infty d\nu\,(z\zb)^{i\nu/2}
 \frac{\sin[(\pi-\phi)j(\nu)]}{i\sin(\pi j(\nu))}\,,\qquad j(\nu)=i\sqrt{\nu^2+4g^2}.  \label{result box ladder}
\ee
We can now understand the vanishing at $\phi=0$ more precisely.
The sine factors cancel in this limit and the integral produces a delta-function $\delta(|z|-1)$ that is independent of $g^2$.  This reproduces precisely the singular behavior of the tree-level function, and otherwise it vanishes for generic $|z|$.

We remark that \eqn{result box ladder} resembles an integrability-based representation of the box ladder integral~\cite{Fleury:2016ykk}, in which the variables analogous to $\nu$ are spectral parameters.

Let us compare with \eqn{eq:finitebox}.  Setting $\nu=\sqrt{\zeta^2-4g^2}$, $\zeta=-ij(\nu)$ in this integral, we get
\be
f(z,\zb)= \int_{0}^\infty d\nu\, 2\cos\left(\frac12\nu \ln\frac{u}{v}\right) \frac{\sin[(\pi-\phi)j]}{i\sin(\pi j)}\,,
\ee
which is precisely the same as \eqn{result box ladder}. We conclude that the method works for the box ladders as well, although it involves an additional subtlety because of the tree level singularity at $z,\zb\to1$.


\section{Sum Representation and Perturbative Evaluation as Polylogarithms}
\label{sec:sumsandpolys}

The integral representations presented in the previous section capture the $\Omega$ integrals fully at finite coupling. If one is interested in extracting numerical values, or in finding $\Omega^{(L)}$ at a particular loop order, it is useful to derive alternate representations in terms of infinite sums. In this section we will derive a representation of this sort, and use it to efficiently find polylogarithmic expressions for $\Omega^{(L)}$ in specific limits.


\subsection{Sum representation}
\label{sumsubsection}

In \eqn{result double penta}, we may change the sign of the integration
variable $\nu$ in the term containing $F_{-\nu}^j$, and rewrite the integral as
\be\label{eq:AllLoopOmegaP}
\Omega(u,v,w,g^2)=\cP\left( \int_{-\infty}^{+\infty}\frac{d\nu}{2i} 
(z^{i\nu/2}+z^{-i\nu/2})\frac{F_{+\nu}^j(x)F_{+\nu}^j(y)}{\sinh\pi \nu}\right)\,,
\ee
where $\cP$ denotes the Cauchy principal value, which is necessary because now the integrand has a pole on the integration contour, at $\nu=0$. This simply amounts to the prescription of including half the contribution of this pole to the integral.

Using the power series definition of the $_2F_1$ Gauss hypergeometric function\footnote{From the definition, it is evident that the function is symmetric in $a\leftrightarrow b$.},
\be\label{eq:2F1_def}
_2F_1(a,b,c,x)=\sum_{n=0}^\infty \frac{\Gamma(a+n)}{\Gamma(a)}\frac{\Gamma(b+n)}{\Gamma(b)}\frac{\Gamma(c)}{\Gamma(c+n)}\frac{x^n}{\Gamma(n+1)}\,,
\ee
we can deduce that $F_\nu^j(x)$ will produce poles when $(i\nu\pm j)/2+n=-k$, or equivalently when
\be
\nu=i \left(\frac{g^2}{k+n}+k+n\right)\,,\quad k\ge0\,,
\label{nupoles2F1}
\ee
namely only in the upper-half $\nu$-plane. We may thus choose to evaluate \eqn{eq:AllLoopOmegaP} by closing the contour in the lower-half plane, picking up poles at $\nu=-i k$ from the $\sinh(\pi\nu)$ factor in the denominator. Redefining 
\be
F_\nu^j(x)=x^{i\nu/2} \hat F_\nu^j(x)\,,
\ee
we thus arrive at the following series representation of the all-loop $\Omega$,
\be\label{eq:AllLoopOmegaSum}
\Omega(u,v,w,g^2)=-\sum_{k=1}^\infty [(-\sqrt{xyz})^k+(-\sqrt{xy/z})^{k}]\hat F_{-ik}^j(x)\hat F_{-ik}^j(y) - \hat F_{0}^{2ig}(x)\hat F_{0}^{2ig}(y)\,.
\ee
By virtue of \eqn{eq:2F1_def}, as well as the following argument transformation formula,
\be
_2F_1(a,b,c,x) = (1-x)^{c-a-b}\, _2F_1(c-a,c-b,c,x)\,,
\ee
we may express the functions $\hat F$ as
\begin{align}
\hat F_{-ik}^j(x)&=\frac{\Gamma(1+\frac{k+j}{2})\Gamma(1+\frac{k-j}{2})}{\Gamma(1+k)}+g^2 \sum_{n=1}^\infty \frac{\Gamma(\frac{k+j}{2}+n)\Gamma(\frac{k-j}{2}+n)}{\Gamma(1+k+n)\Gamma(n+1)}{x^n}\,,\label{eq:Fjk1}\\
\hat F_{-ik}^j(x)&=(1-x) \sum_{n=0}^\infty \frac{\Gamma(\frac{k+j}{2}+1+n)\Gamma(\frac{k-j}{2}+1+n)}{\Gamma(1+k+n)\Gamma(n+1)}x^n\,,\label{eq:Fjk2}
\end{align}
where $j=\sqrt{k^2-4g^2}$.
With the help of these formulas, and the sum representation~\eqref{eq:AllLoopOmegaSum}, we may easily obtain kinematic expansions of $\Omega$ around $x=y=0$ to the desired order.  Through \eqn{uvwyfromxyz}, these expansions are equivalent to expansions in $(u,v,w)$ around $(1,1,1)$.

In a similar way, we can expand the integrals $\OL$, $\Omto$, $\Omte$ and $\WL$
around $x=y=0$.  For this purpose, we also need the expansion of
$F_{\nu-2i}^{j(\nu)\prime}$.  We define
\be
x^{-i\nu/2} F^{j(\nu)\prime}_{\nu-2i}(x)\Bigr|_{\nu=-ik}
\equiv \hat F_{-ik}^{j\prime}(x)
= \sum_{n=1}^\infty \frac{\Gamma(\frac{k+j}{2}+n)\Gamma(\frac{k-j}{2}+n)}{\Gamma(1+k+n)\Gamma(n)}{x^n}\,.\label{eq:Fhatjk1}
\ee
Note that $\hat F_{-ik}^{j\prime}$ only differs from $\hat F_{-ik}^{j}$
by a factor of $n$ in the $n^{\rm th}$ term, and an overall factor of $1/g^2$.
The two series expansions are related by
\bea
x \frac{d}{dx} \hat F_{-ik}^{j}(x) &=& g^2 \hat F_{-ik}^{j\prime}(x),
\label{dFtoFp}\\
(1-x) \frac{d}{dx} \biggl[ x^k \, \hat F_{-ik}^{j\prime}(x) \biggr]
&=& x^k \, \hat F_{-ik}^{j}(x),
\label{dFptoF}
\eea
the latter result following from \eqn{eq:Fjk2}.

Then the series expansion of the all-orders odd ladder integral $\OL$ is:
\bea
\OL(u,v,w,g^2) &=& - \sum_{k=1}^\infty [(-\sqrt{xyz})^k+(-\sqrt{xy/z})^{k}]
\Bigl[ \hat F_{-ik}^{j\prime}(x)\hat F_{-ik}^j(y)
- \hat F_{-ik}^j(x)\hat F_{-ik}^{j\prime}(y) \Bigr]\nonumber\\
&&\null
 - \hat F_{0}^{2ig\,\prime}(x)\hat F_{0}^{2ig}(y)
 + \hat F_{0}^{2ig}(x)\hat F_{0}^{2ig\,\prime}(y)\,.
\label{eq:AllLoopOddladderSum}
\eea
The expansion of the odd part of $\Omt$
can be found by applying $-z\del_z$ to \eqn{eq:AllLoopOddladderSum}:
\be
\Omto(u,v,w,g^2) = \sum_{k=1}^\infty \frac{k}{2}
[ (-\sqrt{xyz})^k - (-\sqrt{xy/z})^{k} ]
\Bigl[ \hat F_{-ik}^{j\prime}(x)\hat F_{-ik}^j(y)
     - \hat F_{-ik}^j(x)\hat F_{-ik}^{j\prime}(y) \Bigr] \,.
\label{eq:AllLoopOmtoSum}
\ee
The expansion of the even part of $\Omt$ can be found by expanding
the all orders result~\eqref{result double penta 2} for $\Omt$ and then subtracting
off the odd part~\eqref{eq:AllLoopOmtoSum}.  The result is
\bea
\Omte(u,v,w,g^2) &=& - \sum_{k=1}^\infty [(-\sqrt{xyz})^k+(-\sqrt{xy/z})^{k}]
\biggl\{
\frac{k}{2} \Bigl[ \hat F_{-ik}^{j\prime}(x)\hat F_{-ik}^j(y)
                 + \hat F_{-ik}^j(x)\hat F_{-ik}^{j\prime}(y) \Bigr]\nonumber\\
&&\null\hskip1cm
+ g^2 \hat F_{-ik}^{j\prime}(x)\hat F_{-ik}^{j\prime}(y) \Biggr\}
 - g^2 \hat F_{0}^{2ig\,\prime}(x)\hat F_{0}^{2ig\,\prime}(y) \,.
\label{eq:AllLoopOmteSum}
\eea
Finally, the expansion of the even ladder integral $\WL$ is given by
\bea
\WL(u,v,w,g^2) &=& - g^2 \Biggl\{
\sum_{k=1}^\infty [(-\sqrt{xyz})^k+(-\sqrt{xy/z})^{k}]
\Bigl[ \hat F_{-ik}^{j\prime}(x)\hat F_{-ik}^j(y)
+ \hat F_{-ik}^j(x)\hat F_{-ik}^{j\prime}(y) \Bigr]\nonumber\\
&&\null\hskip1cm
 + \hat F_{0}^{2ig\,\prime}(x)\hat F_{0}^{2ig}(y) 
 + \hat F_{0}^{2ig}(x)\hat F_{0}^{2ig\,\prime}(y) \Biggr\}
\nonumber\\
&&\null
- \sum_{k=1}^\infty k [(-\sqrt{xyz})^k+(-\sqrt{xy/z})^{k}]
\hat F_{-ik}^j(x)\hat F_{-ik}^j(y) 
\,.
\label{eq:AllLoopEvenladderSum}
\eea
%


\subsection{Weak coupling expansion}
\label{weakcouplingsubsection}

Let us now discuss how to perform the weak coupling expansion of \eqn{eq:AllLoopOmegaSum}, as well as its kinematic resummation to multiple polylogarithms, at least in some limits.  We begin with the rightmost term in \eqn{eq:AllLoopOmegaSum}, coming from the $\nu=0$ residue where $j=2ig$. The $F_{0}^{2ig}$ functions appearing in this term are given by the all-order relation
\be
\hat F_{0}^{2ig}(x) = \frac{\pi g}{\sinh\pi g}\,\, _2F_1(ig,-ig,1,x)\,,
\label{F02ig}
\ee
where the factor involving $\Gamma$ functions in \eqn{eq:defF} was eliminated with the help of the reflection formula,
\be\label{eq:Gamma_reflection}
\Gamma(1-x)\Gamma(x)=\frac{\pi}{\sin \pi x}\,.
\ee
For its weak-coupling expansion, it will be more convenient to use the representation~\eqref{eq:Fjk1}, together with the identity
\be\label{eq:Gamma_expansion}
\Gamma(n+\epsilon)=\Gamma(1+\epsilon)\Gamma(n)\sum_{i=0}^\infty
\epsilon^i Z_{\fwboxL{27pt}{{\underbrace{1,\ldots,1}_{i\,\,\text{times}}}}}(n-1)\,,
\ee
where
\be
Z_{m_1,\ldots,m_j}(n)=\sum_{n\ge i_1>i_2>\ldots>i_j>0}
\frac{1}{i_1^{m_1} \, i_2^{m_2} \, \cdots \, i_j^{m_j}}\label{Zsum}
\ee
are Euler-Zagier sums, or particular values of $Z$-sums~\cite{Moch:2001zr}. 
Inserting \eqn{eq:Gamma_expansion} twice into the series expansion for the
$_2F_1$ in \eqn{F02ig}, we find
\be\label{eq:F0Expansion}
\hat F_{0}^{2ig}(x)= \frac{\pi g}{\sinh\pi g}
\left(1+g^2\sum_{n=1}^\infty\frac{x^n}{n^2}\sum_{j,k=0}^\infty 
(-1)^j (ig)^{j+k}Z_{\fwboxL{27pt}{{\underbrace{1,\ldots,1}_j}}}(n-1)Z_{\fwboxL{27pt}{{\underbrace{1,\ldots,1}_k}}}(n-1)\right)\,.
\ee
After reexpressing the product of Euler-Zagier sums as a linear combination thereof, with the help of the quasi-shuffle (also known as stuffle) algebra relations, for example
\be
Z_1(n)Z_1(n)=2Z_{1,1}(n)+Z_2(n)\,,
\ee
\eqn{eq:F0Expansion} may be immediately evaluated at any loop order in terms of harmonic polylogarithms (HPLs)~\cite{Remiddi:1999ew} with argument $x$,
\be
H_{m_1,m_2,\ldots,m_j}(x)=\sum_{n=1}^\infty\frac{x^{n}}{n^{m_1}}Z_{m_2,\ldots,m_j}(n-1).
\label{HPLfromZ}
\ee

Let us now look at the remaining terms in \eqn{eq:AllLoopOmegaSum}. Since $\hat F$ is symmetric in $j\leftrightarrow -j$, see for example \eqn{eq:Fjk1}, the choice of branch when we expand $j$ in the coupling is immaterial, and we may pick
\be\label{eq:jTok}
j(k)=k\sqrt{1-\frac{4g^2}{k^2}}=k\sum_{l=0}^\infty \binom{1/2}{l}\left(\frac{-4g^2}{k^2}\right)^l.
\ee
Separating the contribution that is small at weak coupling,
\be\label{eq:epsilonToj}
\epsilon(k,g^2)=\frac{j-k}{2}\,,
\ee
we may again use the identities \eqref{eq:Gamma_reflection} and \eqref{eq:Gamma_expansion}, this time for $\epsilon(k,g^2)$, in order to rewrite \eqn{eq:Fjk1} as
\bea
\hat F_{-ik}^j(x) &=& \frac{\pi \epsilon}{\sin \pi \epsilon}\Biggl[
\sum_{i=0}^\infty \epsilon^i Z_{\fwboxL{27pt}{{\underbrace{1,\ldots,1}_i}}}(k)
\label{eq:Fjk_Expansion}\\
&&\hskip0.8cm\null
+ g^2\sum_{n=1}^\infty \frac{x^n}{n(k+n)}\sum_{i,j=0}^\infty
(-1)^i \epsilon^{i+j} Z_{\fwboxL{27pt}{{\underbrace{1,\ldots,1}_i}}}(n-1)
                    Z_{\fwboxL{27pt}{{\underbrace{1,\ldots,1}_j}}}(k+n-1) \Biggr] \,.
\nonumber
\eea
This formula allows us to obtain the weak coupling expansion of $\hat F_{-ik}^j$ most efficiently, by first expanding in $\epsilon(k,g^2)$, and then in $g$ with the help of eqs.~\eqref{eq:jTok}--\eqref{eq:epsilonToj}. In this manner, it is evident that the most complicated sums in \eqn{eq:AllLoopOmegaSum} will always be of the form
\be
\sum_{k,n,m} \frac{(-r)^k}{k^p}\frac{x^n}{n(k+n)}Z_{1,\ldots,1}(n-1)Z_{1,\ldots,1}(k+n-1)\frac{y^m}{m(k+m)}Z_{1,\ldots,1}(m-1)Z_{1,\ldots,1}(k+m-1)\,,
\label{genericZZsum}
\ee
for $r=\sqrt{xyz}$ or $r=\sqrt{xy/z}$ and $p$ any positive integer.
The lengths of the strings of 1's in this expression are arbitrary.


\subsection{Kinematic resummation of 
\texorpdfstring{$\Omega^{(L)}(1,v,w)$}{Omega(L)(1,v,w)} and \texorpdfstring{$\Omega^{(L)}(u,v,0)$}{Omega(L)(u,v,0)}}

To our knowledge, no algorithm currently exists for directly evaluating these kinds of sums in terms of multiple polylogarithms. It would be very interesting to develop one based on our understanding of hexagon functions. However, it turns out that it is indeed possible to resum \eqn{eq:AllLoopOmegaSum} in the limit $y\to0$ and $z\to\infty$, with $x$ and $r=\sqrt{xyz}$ held fixed. Inspecting \eqn{uvwyfromxyz}, we see that this limit corresponds to the following two-dimensional subspace of hexagon kinematics,
\be\label{eq:1vwToxr}
u=1\,,\quad v=\frac{1}{1+r}\,,\quad w=\frac{1-x}{1+r}\,.
\ee
In this subspace, only the first term in \eqn{eq:Fjk_Expansion} survives in $\hat F_{-ik}^j(y=0)$.  We let $(-r)^k = x^k \times (-r/x)^k$, replace the summation variable $n$ with $n'=n+k$ in the other $\hat F_{-ik}^j(x)$ factor, and exchange the order of summation.  Then the most complicated sums take the form
\be\label{eq:Omega1vwSums}
\sum_{n'=1}^{\infty}\frac{x^{n'}}{n'}Z_{1,\ldots,1}(n'-1)\sum_{k=1}^{n'-1} \frac{(-r/x)^k}{k^p}Z_{1,\ldots,1}(k)\frac{1}{(n'-k)}Z_{1,\ldots,1}(n'-k-1)\,.
\ee
Crucially, the rightmost sum can be done with the help of algorithm B of ref.~\cite{Moch:2001zr}. This algorithm has been implemented in the \texttt{nestedsums} library \cite{Weinzierl:2002hv} within the \texttt{GiNaC} framework, and by interfacing it with \texttt{Mathematica} we are able to replace all sums in $k$ of the form~\eqref{eq:Omega1vwSums} with $Z$-sums with outer summation index $n'-1$ (possibly accompanied by rational factors $(r/x)^{n'}$). With the help of the quasi-shuffle algebra relations, we may rewrite their products with the leftmost $Z$-sum in \eqn{eq:Omega1vwSums} as linear combinations of $Z$-sums, similarly to what we did for $\hat F^{2ig}_0$ in \eqn{eq:F0Expansion}. Finally, we evaluate the remaining sum over $n'$ in terms of multiple polylogarithms with the help of their sum representation,
\be\label{LitoZ}
\text{Li}_{m_1,\ldots,m_j}(x_1,\ldots,x_j)=\sum_{i=1}^\infty \frac{x_1^{i}}{i^{m_1}}Z(i-1;m_2,\ldots,m_j;x_2,\ldots,x_j)\,,
\ee
where
\be
Z(n;m_1,\ldots,m_j;x_1,\ldots,x_j)
\equiv \sum_{n\ge i_1>i_2>\ldots>i_j>0}
\frac{x_1^{i_1}}{i_1^{m_1}}\ldots\frac{x_j^{i_j}}{i_j^{m_j}}\,.
\label{actualZsum}
\ee
Very similar techniques have been used to evaluate~\cite{Papathanasiou:2013uoa,Papathanasiou:2014yva} the leading, and part of the subleading, contribution to the hexagon Wilson loop OPE near the collinear limit, as well as to resum~\cite{Drummond:2015jea} all single-particle gluon bound states contributing to the double scaling limit, $v\to 0$ with $u,w$ fixed.

The systematic procedure we have described works in principle at any loop order, subject to limitations in computer power.  We have used it to obtain explicit expressions for $\Omega^{(L)}(1,v,w)$ through $L=8$ loops.  We quote here the first two loop orders,
\begin{align}
\Omega^{(1)}(1,v,w)=&-2 H_{1,1}(-r)+\text{Li}_{1,1}\left(-r,-\frac{x}{r}\right)
-2 H_2(-r)+H_2(x)-2\zeta_2\,,
\label{Om1_1vw}\\
\Omega^{(2)}(1,v,w)=&\ H_{2,2}(-r)-2 H_{3,1}(-r)-2 \left(H_{2,2}(-r)+2 H_{2,1,1}(-r)\right)-H_{2,2}(x)\nonumber\\
&+\text{Li}_{2,2}\left(-r,-\frac{x}{r}\right)+2 \text{Li}_{2,2}\left(x,-\frac{r}{x}\right)+\text{Li}_{3,1}\left(-r,-\frac{x}{r}\right)+\text{Li}_{3,1}\left(x,-\frac{r}{x}\right)\nonumber\\
&+2 \text{Li}_{2,1,1}\left(-r,1,-\frac{x}{r}\right)+\text{Li}_{2,1,1}\left(-r,-\frac{x}{r},-\frac{r}{x}\right)+2 \text{Li}_{2,1,1}\left(x,-\frac{r}{x},1\right)
\nonumber\\
&-\text{Li}_{2,1,1}\left(x,-\frac{r}{x},-\frac{x}{r}\right)+H_4(-r)+2\zeta_2\left(H_2(x)- H_2(-r)\right)-6\zeta_4\,, \label{Om2_1vw}
\end{align}
where
\be\label{eq:xrTovw}
r=\frac{1}{v}-1\,,\qquad x=1-\frac{w}{v}\,.
\ee
Results through eight loops are contained in the ancillary file
\texttt{omega1vwL0-8.m} provided with this paper.

In precisely the same fashion, we may also resum $\Omega^{(L)}(u,v,0)$, which, as discussed around \eqn{result penta}, is equivalent to the dual conformal pentabox ladder $\Psi^{(L)}(u,v)$ defined in \eqref{eq:PsiDef} and shown in \fig{fig:pentaboxladder_L}. Starting from the sum representation \eqref{eq:AllLoopOmegaSum}, as already mentioned the limit in question amounts to letting $y\to1$.  In this limit, the way $F_\nu^j$ are normalized implies that $\hat F_{-ik}^j(y)\to1$, also for $k=0$. Up to two loops we obtain
\begin{align}
\Psi^{(1)}(u,v)=&-H_{1,1}\left(-\sqrt{\frac{x}{z}}\right)-H_{1,1}\left(-\sqrt{x z}\right)+\text{Li}_{1,1}\left(-\sqrt{\frac{x}{z}},-\sqrt{x z}\right)\nonumber\\
&+\text{Li}_{1,1}\left(-\sqrt{x z},-\sqrt{\frac{x}{z}}\right)-\zeta _2-H_2\left(-\sqrt{\frac{x}{z}}\right)-H_2\left(-\sqrt{x z}\right)+H_2(x)\,,
\label{Psi1_uv}
\end{align}
\begin{align}
\Psi^{(2)}(u,v)=& -H_{2,1,1}\left(-\sqrt{\frac{x}{z}}\right)-H_{2,1,1}\left(-\sqrt{x z}\right)-H_{2,2}(x)+\text{Li}_{2,2}\left(x,-\sqrt{\frac{z}{x}}\right)\nonumber\\
&+\text{Li}_{2,2}\left(x,-\frac{1}{\sqrt{x z}}\right)+\text{Li}_{2,2}\left(-\sqrt{\frac{x}{z}},-\sqrt{x z}\right)+\text{Li}_{2,2}\left(-\sqrt{x z},-\sqrt{\frac{x}{z}}\right)\nonumber\\
&+\text{Li}_{2,1,1}\left(x,-\sqrt{\frac{z}{x}},1\right)-\text{Li}_{2,1,1}\left(x,-\sqrt{\frac{z}{x}},-\sqrt{\frac{x}{z}}\right)+\text{Li}_{2,1,1}\left(x,-\frac{1}{\sqrt{x z}},1\right)\nonumber\\
&-\text{Li}_{2,1,1}\left(x,-\frac{1}{\sqrt{x z}},-\sqrt{x z}\right)+\text{Li}_{2,1,1}\left(-\sqrt{\frac{x}{z}},1,-\sqrt{x z}\right)-\frac{7 \zeta _4}{4}\nonumber\\
&+\text{Li}_{2,1,1}\left(-\sqrt{x z},1,-\sqrt{\frac{x}{z}}\right)+\zeta _2 H_2(x)-\zeta _2 H_2\left(-\sqrt{\frac{x}{z}}\right)\nonumber\\
&-\zeta _2 H_2\left(-\sqrt{x z}\right)+H_4\left(-\sqrt{\frac{x}{z}}\right)+H_4\left(-\sqrt{x z}\right)\,,\label{Psi2_uv}
\end{align}
where
\be
x = \frac{(1-u) (1-v)}{u v}\,,\quad
z = \frac{u(1-v)}{v(1-u)}\,,\quad
\sqrt{\frac{x}{z}} = \frac{1-u}{u}\,,\quad
\sqrt{xz} = \frac{1-v}{v} \,.
\label{xzsqrts}
\ee
The polylog arguments are all rational in $u,v$ in this case.
Here as well we have carried out the computation up to $L=8$;
the resulting expressions may be found in the ancillary file
\texttt{omegauv0L0-8.m}.


\subsection{The line \texorpdfstring{$(1,1,w)$}{(1,1,w)}}
\label{Line11wsubsection}

The line $(u,v,w)=(1,1,w)$ corresponds to taking $y\to0$ at fixed $x$ and $z$,
with $w=1-x$ on this line.
Examining the series expansions found in section~\ref{sumsubsection},
we see that only the $k=0$ terms involving $\hat F^{2ig}_0$ and
$\hat F^{2ig\,\prime}_0$ survive, since $\sqrt{xyz} \to 0$
and $\sqrt{xy/z} \to 0$.  Also, $\hat F^{2ig\,\prime}_0(0) = 0$.
Therefore both the even and odd parts of $\Omt$ vanish on this line,
\be
\Omte(1,1,w) = \Omto(1,1,w) = 0.
\label{Omt11wvanish}
\ee
The vanishing of $\Omto$ is also a consequence of its antisymmetry
under $u\lr v$.

For $\Omega$, the $k=0$ term in \eqn{eq:AllLoopOmegaSum} gives,
using \eqn{F02ig},
\bea
\Omega(1,1,w,g^2) &=& - \hat F_{0}^{2ig}(0) \hat F_{0}^{2ig}(x) \nonumber\\
 &=& - \left( \frac{\pi g}{\sinh\pi g} \right)^2 \, _2F_1(ig,-ig,1,x)
\nonumber\\
&=& - \left( \frac{\pi g}{\sinh\pi g} \right)^2 \, 
\sum_{n=0}^\infty \frac{\prod_{k=0}^{n-1} ( k^2 + g^2 )}{(n!)^2} \, x^n
\nonumber\\
&=& - 1 + \sum_{L=1}^\infty (g^2)^L
\biggl[ - H_{\fwboxL{27pt}{{\underbrace{2,\ldots,2}_{L}}}}(x)
 + \sum_{m=1}^{L} (-1)^{m} (2-4m) \zeta_{2m} H_{\fwboxL{27pt}{{\underbrace{2,\ldots,2}_{L-m}}}}(x) \biggr]
\,,\nonumber\\
&~&  \label{Om_11w}
\eea
with $x=1-w$.  At $x=0$, the term in \eqn{Om_11w} with $m=L$ supplies
the value of $\Omega$ at $(u,v,w)=(1,1,1)$:
\be
\Omega(1,1,1,g^2) = - \left( \frac{\pi g}{\sinh\pi g} \right)^2
= - 1 + \sum_{L=1}^\infty (-g^2)^L (2-4L) \zeta_{2L} \,. \label{Om_111}
\ee
At $x=1$, we have
\bea
\Omega(1,1,0,g^2) &=&
- \left( \frac{\pi g}{\sinh\pi g} \right)^2 \, _2F_1(ig,-ig,1,1)
\nonumber\\
&=& - \frac{\pi g}{\sinh\pi g}
\ =\ - 1 - \sum_{L=1}^\infty (-g^2)^L \, \frac{2^{2L-1} - 1}{2^{2L-2}} \, \zeta_{2L}
\,.
\label{Om_110}
\eea

Note that the HPLs obey
\be
w \frac{d}{dw} \biggl[ (1-w) \frac{d}{dw} H_{\fwboxL{27pt}{{\underbrace{2,\ldots,2}_{N}}}}(1-w) \biggr]
= H_{\fwboxL{27pt}{{\underbrace{2,\ldots,2}_{N-1}}}}(1-w).
\label{HPL2ident}
\ee
Therefore $\Omega(1,1,w,g^2)$ satisfies the differential equation
\be
w \frac{d}{dw} \biggl[ (1-w) \frac{d}{dw} \Omega(1,1,w,g^2) \biggr]
= g^2 \, \Omega(1,1,w,g^2).
\label{Om11wDE}
\ee

The first few perturbative orders for $\Omega^{(L)}(1,1,w)$ are:
\bea
\Omega^{(0)}(1,1,w) &=& - 1\,, \label{On11w0}\\
\Omega^{(1)}(1,1,w) &=&  H_{2}(1-w) - 2 \zeta_2 \,, \label{On11w1}\\
\Omega^{(2)}(1,1,w) &=&  - H_{2,2}(1-w) + 2 \zeta_2 H_{2}(1-w)
- 6 \zeta_4 \,, \label{On11w2}\\
\Omega^{(3)}(1,1,w) &=&  H_{2,2,2}(1-w) - 2 \zeta_2 H_{2,2}(1-w)
+ 6 \zeta_4 H_{2}(1-w) - 10 \zeta_6 \,. \label{On11w3}
\eea
The one- and two-loop formulae
can be recovered from \eqns{Om1_1vw}{Om2_1vw} by letting $r\to0$,
which leaves only the HPLs with argument $x=1-w$.

Similarly, the odd ladder integral becomes
\bea
\OL(1,1,w,g^2) &=& - \hat F_{0}^{2ig}(0) \hat F_{0}^{2ig\,\prime}(x) \nonumber\\
 &=& \frac{1}{g^2} \, x \frac{d}{dx} \Omega(1,1,1-x,g^2)
\nonumber\\
&=& \sum_{L=1}^\infty (g^2)^{L-1}
\biggl[ - H_{\fwboxL{27pt}{{\underbrace{1,2,\ldots,2}_{L}}}}\hspace{.28cm} (x)
 + \sum_{m=1}^{L} (-1)^{m} (2-4m) \zeta_{2m} H_{\fwboxL{27pt}{{\underbrace{1,2,\ldots,2}_{L-m}}}}\hspace{.28cm} (x) \biggr]
\,,\nonumber\\
&=& - \frac{1}{g^2} (1-w) \frac{d}{dw} \Omega(1,1,w,g^2),
\label{OL_11w}
\eea
so it sits in the middle of the differential equation~\eqref{Om11wDE}.
On the line $(1,1,w)$, the even ladder integral is simply related
to the odd one at one higher loop:
\be
\WL(1,1,w,g^2) =  \, g^2 \, \OL(1,1,w,g^2).
\label{W11w}
\ee

As we will see in subsection \ref{sec:strong_coupling}, the $(1,1,w)$ limit offers us insight into the strong-coupling analysis of the integrals. In addition, we can study the radius of convergence in the $g$ plane of the perturbative expansion for $\Omega(1,1,w,g^2)$ using \eqn{Om_11w}.  The same arguments that lead to \eqn{nupoles2F1} show that the hypergeometric function $_2F_1(ig,-ig,1,x)$ has poles at $g=\pm i$, which are the poles in the $g$ plane closest to the origin.  They also match the location of the closest poles of the prefactor $\pi g/\sinh(\pi g)$.  Therefore the radius of convergence of the perturbative expansion of $\Omega(1,1,w,g^2)$ is unity for all $w$.  We can check this result at $w=1$ and $w=0$ by observing that the ratio of successive loop orders in \eqns{Om_111}{Om_110} goes to $-1$ as $L\to\infty$. 

We remark here that the resummed integral $\Omega(u,v,w,g^2)$ appears correctly weighted in the full BDS-like normalized MHV amplitude ${\cal E}(u,v,w,g^2)$, when $g^2$ is identified with the standard coupling parameter in planar ${\cal N}=4$ SYM, $g^2 = N_c g_{\rm YM}^2/(4\pi)^2$, where $g_{\rm YM}$ is the Yang-Mills coupling and the gauge group is $SU(N_c)$.  In order to establish that it is correctly weighted, one can use the ``rung rule'' for performing two-particle cuts in planar ${\cal N}=4$ SYM~\cite{Bern:1997nh,Bern:1998ug}. This rules provides the normalization of the $\Omega^{(L)}$ terms within the $L$-loop integrand, relative to the normalization at one lower loop.  Therefore it makes sense to compare the unit radius of convergence for $\Omega(1,1,w,g^2)$ with the radius of convergence for amplitudes.  The latter is not firmly established~\cite{Dixon:2014voa,Dixon:2015iva,Caron-Huot:2016owq}, but it appears to be closer to $1/16$, the value for the cusp anomalous dimension~\cite{Beisert:2006ez}, and much smaller than 1.  It would be interesting to use our finite-coupling representation~\eqref{result double penta} to investigate the perturbative radius of convergence of $\Omega(u,v,w,g^2)$ for more general kinematics than just the line $(1,1,w)$.


\subsection{The line \texorpdfstring{$(1,v,1)$}{(1,v,1)}}

Next we consider the line $(u,v,w)=(1,v,1)$.
From \eqns{eq:xrTovw}{uvwyfromxyz},
this corresponds to letting $y\to0$ with $x$ and $yz$ fixed, and then letting
$r=\sqrt{xyz}=-x$, where $v=1/(1-x)$, $x=1-1/v$.
Applying this substitution to the series representation of the
ladder integrals in eqs.~\eqref{eq:AllLoopOmegaSum},
\eqref{eq:AllLoopOddladderSum}--\eqref{eq:AllLoopEvenladderSum},
and using $F_{-ik}^{j\prime}(0)=0$, yields
\bea
\Omega(1,1/(1-x),1,g^2) &=& - \hat F_{0}^{2ig}(0) \hat F_{0}^{2ig}(x)
          - \sum_{k=1}^\infty x^k \hat F_{-ik}^j(0) \hat F_{-ik}^j(x) \,,
\label{Om_1v1}\\
\OL(1,1/(1-x),1,g^2) &=& - \hat F_{0}^{2ig}(0) \hat F_{0}^{2ig\,\prime}(x)
          - \sum_{k=1}^\infty x^k \hat F_{-ik}^j(0) \hat F_{-ik}^{j\prime}(x) \,,
\label{OL_1v1}\\
\Omte(1,1/(1-x),1,g^2) &=&
- \sum_{k=1}^\infty \frac{k}{2} x^k \hat F_{-ik}^j(0) \hat F_{-ik}^{j\prime}(x) \,,
\label{Omte_1v1}\\
\Omto(1,1/(1-x),1,g^2) &=& - \Omte(1,1/(1-x),1,g^2),
\label{Omto_1v1}\\
\WL(1,1/(1-x),1,g^2) &=& - g^2
\biggl[ \hat F_{0}^{2ig}(0) \hat F_{0}^{2ig\,\prime}(x)
     + \sum_{k=1}^\infty x^k \hat F_{-ik}^j(0) \hat F_{-ik}^{j\prime}(x) \biggr]
\nonumber\\
&&\null
- \sum_{k=1}^\infty k x^k \hat F_{-ik}^j(0) \hat F_{-ik}^j(x) \,.
\label{WL_1v1}
\eea
Hence $\Omt=\Omte+\Omto$ vanishes on the line $(1,v,1)$,
and using \eqns{dFtoFp}{dFptoF} we have
\bea
(1-x) \frac{d}{dx} \OL(x) &=& \Omega(x), \label{OLO1v1}\\
x \frac{d}{dx} \Omega(x) &=&  \WL(x). \label{OmW1v1}
\eea
So it is enough to specify the loop expansion of $\Omega(1,v,1)$ below.

Now $\Omega$, $\Omte$, $\OL$ and $\WL$
are symmetric under $u\lr v$ ($z\lr 1/z$),
while $\Omto$ is anti-symmetric.
Therefore, on the line $(u,1,1)$, with $u=1/(1-x)$, we can also
use the above formulas, except that the sign of $\Omto$ is reversed
so that in $\Omt$ it doubles $\Omte$ instead of cancelling it:
\bea
\Omto(1/(1-x),1,1,g^2) &=& \Omte(1/(1-x),1,1,g^2) = \Omte(1,1/(1-x),1,g^2),
\label{Omteorelation_u11}\\
\qquad \Omt(1/(1-x),1,1,g^2) &=& 2 \, \Omte(1,1/(1-x),1,g^2).
\label{Omt_on_u11}
\eea

On the line $(1,v,1)$, the first few orders of explicit results
for $\Omega^{(L)}$ are:
\bea
\Omega^{(1)}(1,v,1) &=&  -H_{2} -H_{1,1} - 2 \zeta_2 \,, \label{Om1v11}\\
\Omega^{(2)}(1,v,1) &=&  H_{4} + H_{2,2} - 6 \zeta_4 \,, \label{Om1v12}\\
\Omega^{(3)}(1,v,1) &=&  -2 H_{6}-H_{4,2}
-H_{3,3}-H_{2,4}- H_{2,2,2}
+ 6 \zeta_4 H_{2} - 10 \zeta_6 \,, \label{Om1v13}\\
\Omega^{(4)}(1,v,1) &=& 5 H_{8}
+2 (H_{6,2}+H_{5,3}+H_{4,4}+H_{3,5}+H_{2,6})
\nonumber\\  &&\hskip0cm\null
+H_{4,2,2}+H_{3,3,2}+H_{3,2,3}+H_{2,4,2}+H_{2,3,3}+H_{2,2,4}+H_{2,2,2,2}
\nonumber\\  &&\hskip0cm\null
-6 \zeta_4 (H_{4}+H_{2,2})
+10 \zeta_6 H_{2}-14 \zeta_8 \,, \label{Om1v14}\\
\Omega^{(5)}(1,v,1) &=& -14 H_{10}
-5 (H_{8,2}+H_{7,3}+H_{6,4}+H_{5,5}+H_{4,6}+H_{3,7}+H_{2,8})
\nonumber\\  &&\hskip0cm\null
-2 (H_{6,2,2}+H_{5,3,2}+H_{5,2,3}+H_{4,4,2}+H_{4,3,3}+H_{4,2,4}+H_{3,5,2}+H_{3,4,3}
\nonumber\\  &&\hskip0.3cm\null
   +H_{3,3,4}+H_{3,2,5}+H_{2,6,2}+H_{2,5,3}+H_{2,4,4}+H_{2,3,5}+H_{2,2,6})
\nonumber\\  &&\hskip0cm\null
-H_{4,2,2,2}-H_{3,3,2,2}-H_{3,2,3,2}-H_{3,2,2,3}-H_{2,4,2,2}-H_{2,3,3,2}
\nonumber\\  &&\hskip0cm\null
-H_{2,3,2,3}-H_{2,2,4,2}-H_{2,2,3,3}-H_{2,2,2,4}-H_{2,2,2,2,2}
\nonumber\\  &&\hskip0cm\null
+6 \zeta_4 (2 H_{6}+H_{4,2}+H_{3,3}+H_{2,4}+H_{2,2,2})
\nonumber\\  &&\hskip0cm\null
-10 \zeta_6 H_{2,2}+14 \zeta_8 H_{2}-18 \zeta_{10} \,, \label{Om1v15}
\eea
where $x=1-1/v$ is the implicit argument of $H_{\vec{w}} = H_{\vec{w}}(x)$.
These formulae can be obtained from the results obtained on $(1,v,w)$
by letting $r\to-x$, which collapses the multiple polylogarithms,
for example,
\be
\text{Li}_{2,1,1}\left(x,-\frac{r}{x},-\frac{x}{r}\right)
\to \text{Li}_{2,1,1}(x,1,1) = H_{2,1,1}(x),
\label{LitoH}
\ee
using \eqn{HPLfromZ}.

In the coefficients of the non-$\zeta$ terms
in eqs.~\eqref{Om1v11}--\eqref{Om1v15},
one can see the emergence of the Catalan numbers,
\be
C_n = \frac{(2n)!}{(n+1)!n!} = 2 \frac{(2n-1)!}{(n+1)!(n-1)!}
= 1,2,5,14,42,132,429,\ldots.
\label{Catalan}
\ee
Although it is not really apparent yet, the coefficients of the $\zeta_{2(k+1)}$
terms for $k>1$ are controlled by a $2k$-fold convolution of the
Catalan numbers.
Define
\be
C_{n,k} \equiv \frac{k}{2n-k} \binom{2n-k}{n} \,,
\label{Catalank}
\ee
which satisfies
\be
C_{n,k} = \sum_{i_1+i_2+\cdots+i_k=n} C_{i_1-1} C_{i_2-1} \ldots C_{i_k-1} \,,
\label{CatalankRecurse}
\ee
with $C_n\equiv0$ for $n<0$.  It also obeys
$C_{n+1,1} = C_{n+1,2} = C_{n}$.
Also note from eqs.~\eqref{Om1v12}--\eqref{Om1v15}
that the Catalan number required is related to the {\it depth},
i.e.~the length of the (compressed) HPL weight vector.
All weight vectors having the same depth appear
with the same coefficient, which is nonzero if all entries are $\geq2$.
(The only exception is at $L=1$, which contains $H_{1,1}$.)

We find that, for $L>1$, $\Omega^{(L)}(1,v,1)$ is given by
\be
\Omega^{(L)}(1,v,1) = (-1)^L \biggl[
X_0^{(L)} + \sum_{k=2}^L (2-4k) \zeta_{2k} X_k^{(L-k)} \biggr]
\,, \label{Om1v1L}
\ee
where the no-$\zeta$ term is
\be
X_0^{(L)} = H_{\fwboxL{27pt}{{\underbrace{2,2,\ldots,2}_{L}}}}\hspace{.28cm} (x)
  + \sum_{m=1}^{L-1} C_{L-m} \sum_{\vec{w} \in g_{m,2L}} H_{\vec{w}}(x)\,,
\label{XnoZeta}
\ee
and the $\zeta$ terms are
\bea
X_k^{(0)} &=& (-1)^k \,,
\label{XZeta0}\\
X_k^{(L)} &=& (-1)^k  \, H_{\fwboxL{27pt}{{\underbrace{2,2,\ldots,2}_{L}}}}\hspace{.28cm} (x)
  + \sum_{m=1}^{L-k+1} C_{L+k-1-m,2(k-1)} \sum_{\vec{w} \in g_{m,2L}} H_{\vec{w}}(x)
  \qquad L > 0 \,.
\label{XZeta}
\eea
Here $g_{m,n}$ is the set of weight vectors $\vec{w}=(w_1,w_2,\ldots,w_m)$
of depth $m$ and weight $\sum_{i=1}^m w_i = n$, with all $w_i \geq 2$.

Note that the first term of $X_0^{(L)}$ also appears in \eqn{Om_11w}
for $\Omega(1,1,w)$.
Also, using $C_{n+1,2}=C_{n}$ we see that
$X_2^{(L)} = X_0^{(L)}$; that is, the $\zeta_4$ terms are controlled
by $X_2^{(L-2)} = X_0^{(L-2)}$, which is exactly the same function
describing the no-$\zeta$ terms at two lower loops.
The $\zeta_6$ terms are the first to require
a true convolution of the Catalan numbers, i.e.~$C_{n,4}$.

We have checked \eqn{Om1v1L} exactly through six loops,
and through 13 loops via the series expansion~\eqref{Om_1v1}
to order $x^{10}$.  At 13 loops, the full answer contains
75,025 non-$\zeta$ terms, 10,946 $\zeta_4$ terms,
4,136 $\zeta_6$ terms, 1,351 $\zeta_8$ terms, 246 $\zeta_{10}$ terms,
13 $\zeta_{12}$ terms, and one each of the
$\zeta_{14},\zeta_{16},\ldots,\zeta_{26}$ terms.

Note that the non-Catalan term in $\Omega^{(L)}(1,v,1)$
is equal to the much simpler expression for
$\Omega^{(L)}(1,1,w)$, after identifying $1-x = w = 1/v$.
This is simply the $k=0$ term in \eqn{Om_1v1},
while the terms containing the Catalan numbers come from the $k>0$ terms.

\subsection{Strong-coupling behavior}
\label{sec:strong_coupling}

A remarkable feature of the finite-coupling $\Omega,\tilde \Omega$ integrals \eqref{result double penta} and \eqref{result double penta 2} is that we can evaluate them outside the radius of convergence of the weak-coupling region we used to derive them, 
all the way to strong coupling. Here we provide evidence that the functions become exponentially suppressed as $g\to\infty$ for a large subspace of the Euclidean domain. This is very similar to the observed strong-coupling behavior of the box ladder integrals for general kinematics~\cite{Broadhurst:2010ds}.

For simplicity, let us begin with the $(1,1,w)$ line, which we also analyzed in section \ref{Line11wsubsection}
where we focused on weak coupling. At $w=1$, the exponential suppression is clear from \eqn{Om_111}:
\be
\Omega(1,1,1,g^2)\ =\ - \left(\frac{\pi g}{\sinh\pi g}\right)^2
\ \sim\ - 4\pi^2 g^2 \exp(-2\pi g), \qquad  \hbox{as $g \to \infty$,}
\label{expsuppr111}
\ee
and similarly for $w=0$, from \eqn{Om_110}:
\be
\Omega(1,1,0,g^2)\ =\ - \frac{\pi g}{\sinh\pi g}
\ \sim\ - 2\pi g \exp(-\pi g),  \qquad \hbox{as $g \to \infty$.}
\label{expsuppr110}
\ee
Generally, going from $w=1$ to $w=0$ the absolute value of the function increases monotonically between these two limits.
To study in detail the behavior between the endpoints, the problem is reduced to the
asymptotic analysis of the (normalized) hypergeometric functions
$F^{2ig}_0(x)$ defined in \eqn{eq:defF}, which enters on the first line of \eqref{Om_11w}.
Fortunately, a very detailed saddle point analysis of this precise class of hypergeometric functions has been carried out
in ref.~\cite{doi:10.1142/S0219530514500389},
see in particular Theorems 3.1 and 3.2, which are valid in the region $x<0$ and $0<x<1$ respectively.
Focusing on the region $0<x<1$, we can write their result as:
\be\label{eq:Fstrong}
\lim_{g\to\infty} F_{\nu}^{j(\nu)}(x)= \sqrt{\pi g}\left(\tfrac{1}{x}-1\right)^{1/4}e^{-g \phi(x)}\,,
\qquad \phi(x)=\arccos(2x-1)\,,
\ee
which in fact holds for any fixed value of $|\nu|\ll g$, up to relative corrections of order $1/g$.
The angle $\phi$ varies continuously and monotonically from $\phi(0^+)=\pi$ to $\phi(1^-)=0$.
Including the second hypergeometric factor in \eqref{Om_11w}, which gives $F^{2ig}_0(0)= \frac{\pi g}{\sinh(\pi g)}$, we thus get:
\be
\Omega(1,1,w,g^2\rightarrow \infty) = -2(\pi g)^{3/2}\left(\tfrac{1}{x}-1\right)^{1/4} \times e^{-g(\phi(x)+\pi)}
\times\left( 1+ \mathcal{O}(1/g)\right)
\label{11w strong coupling}
\ee
where $x=1-w$.
The dependence of the exponent on $w$ could also be obtained simply by solving the hypergeometric differential equation at leading order in $g$.\footnote{We thank Bob Cahn for this observation.}
We observe that the exponent smoothly interpolates between \eqn{expsuppr111} and \eqn{expsuppr110}.
The prefactors do not quite go smoothly, but this
can be understood as a breakdown of the saddle point approximation for extreme values of $x$ very close to the endpoints, see
ref.~\cite{doi:10.1142/S0219530514500389} for details.


For the other integrals in our basis, it follows from equations~\eqref{OL_11w} 
and~\eqref{W11w} that $\OL(1,1,w,g^2)$ and $\WL(1,1,w,g^2)$ are also exponentially suppressed for all $x<1$ ($w>0$), whereas we recall that $\tilde\Omega$ vanishes identically on this line.

In more general kinematics, we can analyze the integral representation in \eqref{result double penta}, which we reproduce here for convenience:
\be \label{recopied double penta}
\Omega(u,v,w,g^2)
= \int_{-\infty}^\infty \frac{d\nu}{2i} \, z^{i\nu/2} \,
 \frac{F_{+\nu}^{j(\nu)}(x)F_{+\nu}^{j(\nu)}(y)-F_{-\nu}^{j(\nu)}(x)F_{-\nu}^{j(\nu)}(y)}{\sinh(\pi \nu)}\,.
\ee
By plotting the integrand for various values of $x,y,z$ we
find that it is generally dominated by the region near the origin,
where $|\nu| \sim 1 \ll g$.
In this regime we can use the limit in \eqn{eq:Fstrong} to deduce the following exponential suppression of the integrand:
\be
F_{+\nu}^{j(\nu)}(x)F_{+\nu}^{j(\nu)}(y) \sim e^{-g(\phi(x)+\phi(y))},
\ee
where we have focused on the exponent.
Thus, assuming that the region $|\nu|\sim 1$ indeed dominates as suggested by the numerics,
$\Omega$ is itself suppressed by at least the same factor:
\be
 \Big| \Omega(u,v,w,g^2\to\infty) \Big| \lsim e^{-g(\phi(x)+\phi(y))}.
\ee
Remarkably, the true behavior of $\Omega$ for generic $x,y$
appears to be even more strongly suppressed. This can be seen from the fact that
the asymptotic expansion (\ref{eq:Fstrong}) to all orders in $1/g$ turns out to contain only even powers of $\nu$,
so there is a cancellation between the two terms
in the numerator of \eqref{recopied double penta}, causing the dominant behavior to come from subleading
exponential corrections to \eqref{eq:Fstrong}.
While these could in principle be analyzed using formulas in ref.~\cite{doi:10.1142/S0219530514500389},
we will simply conclude this subsection with the observation that $\Omega$ is exponentially suppressed at strong coupling.


\section{The \texorpdfstring{$\Omega$}{Omega} Space}
\label{sec:spacesect}

In this section we analyze the $\Omega$ integrals in general kinematics, working perturbatively in the coupling.
We define the $\Omega$ space of functions to be that containing all
iterated derivatives (more precisely, all iterated $\{n-1,1\}$ coproducts) of the $\WL$, $\Omega$, $\tilde\Omega$ and $\OL$
integrals to arbitrary loop orders.  We are interested in studying
the $\Omega$ space primarily as a model for the full space of
Steinmann hexagon functions~\cite{Caron-Huot:2016owq,Caron-Huot:six_loops}.
If we can characterize the functions that appear in the $\Omega$ integrals
to all orders, we will have encompassed a substantial slice of the full space
of Steinmann hexagon functions, and this may give hints as to their overall
structure.  We will achieve this by first showing that a certain discontinuity of the $\Omega$ integrals is simple,
then using this insight to build the full $\Omega$ space.

\subsection{Coproduct formalism and hexagon function space}

Like MHV and NMHV amplitudes in planar ${\cal N}= 4$ sYM, the integrals defined in section~\ref{sec:integral_definitions} are expected to evaluate to multiple polylogarithms.
This implies that they are endowed with a Hopf algebra.
In particular, there is a coproduct operation
which breaks functions apart into simpler ones.
This yields various concrete representations as iterated integrals.

For a more complete review of how Hopf algebras and the coproduct show up in amplitudes, see ref.~\cite{Duhr:2014woa}. The key property for us will be that derivatives only act on objects appearing in the second entry of the
coproduct $\Delta$:\footnote{%
We use the term ``coproduct'' somewhat loosely; in many cases ``coaction'' would be more appropriate because the spaces to the left and right of ``$\otimes$'' are actually different.  Note also that here $\Delta$ denotes the coproduct, and not the kinematical quantity defined below \eqn{definexy}.}
\be
\Delta \frac{\del}{\del z} F = \left(\text{id} \otimes \frac{\del}{\del z} \right) \Delta F.
\ee
In particular, we can define a set of functions $F^s$ by the coproduct action  
\be
\Delta_{n-1,1} F = \sum_{s \in \cal{S}} F^s \otimes \ln s
\ee 
for any multiple polylogarithm $F$ of weight $n$ that involves symbol letters in the set ${\cal S}$. The functions $F^s$ are then iterated integrals of weight $n-1$.
The derivative of $F$ with respect to an underlying kinematic variable $z$ (say) is
\be
\del_z F = \sum_{s \in \cal{S}} F^s \del_z \ln s.  \label{del coproduct}
\ee 
By repeating this operation $n$ times and integrating along various paths with appropriate boundary conditions,
one obtains concrete integral representations of $F$.  The coproduct
unifies these representations in a canonical way.

In six-particle kinematics,
the traditional symbol alphabet is given
by~\cite{Goncharov:2010jf,Dixon:2011pw}
\be
{\cal S}_\text{hex} = \left\{u, v, w, 1-u, 1-v, 1-w, y_u, y_v, y_w \right\} .
\label{hex_letters}
\ee
The list of iterated integrals with this alphabet at any given weight is finite, and is called the hexagon function space.
However, constructing this space is nontrivial and currently unsolved
beyond weight 12, even after imposing the Steinmann conditions~\cite{Caron-Huot:2016owq} and further restricting to the level of the symbol.

For our discussion it will be convenient to parametrize the same space with a new alphabet
\be
{\cal S}^{\prime}_\text{hex} =
\left\{a, b, c, m_u, m_v, m_w, y_u, y_v, y_w \right\} ,
\label{hex_letters_abc}
\ee
where $a=\frac{u}{v w}$, $m_u=\frac{1-u}{u}$, and the others are defined by
cyclic permutations of $u,v,w$, as in \eqn{abcdef}.
The letters $a,b,c$ are physically significant
due to the Steinmann relations, which state that amplitudes (or individual Feynman integrals) can't have simultaneous discontinuities in overlapping channels.
Each contains a single three-particle Mandelstam invariant: $a\propto (x_{25}^2)^2$, $b\propto (x_{36}^2)^2$, $c\propto (x_{14}^2)^2$
(see the definitions of the cross ratios in \eqn{uvw_def}).
In the new alphabet the Steinmann relations state simply that
$a$ can never appear next to $b$ in the first two entries of the symbol
(or $b$ next to $c$, or $a$ next to $c$)~\cite{Caron-Huot:six_loops}. We discuss this condition further in appendix \ref{ext_steinmann}.

From the kinematic relations~\eqref{uvwyfromxyz}
and \eqref{abcdef} in appendix~\ref{hexvarappendix}, we see that five
of the nine hexagon letters can be taken to be
simple combinations of the variables $x,y,z$ which simplify the ladders:
\be
c   = (1-x)(1-y), \quad
m_u = \sqrt{\frac{xy}{z}} \,, \quad
m_v = \sqrt{xyz}, \quad
y_u y_v = \frac{y}{x} \,, \quad
y_w = \frac{x(1-y)}{y(1-x)} \,.
\label{fiveniceletters}
\ee
These five letters are equivalent under multiplication to
\be
{\cal S}_{\rm disc} = \{x,1-x,y,1-y,z\},
\label{fivenicelettersALT}
\ee
which are the only letters appearing in the matrix $M$ introduced in
\eqn{matrix section3}. As we will now show, after taking a discontinuity in $c$, the ladder integrals collapse to a space with just these five letters, which will enable their complete description.


\subsection{The box ladders and their discontinuities}
\label{boxladdersubsect}

It is helpful to first describe the analogous, simpler, space for the box ladder integrals.
The pure functions entering these integrals are given explicitly in terms of
classical polylogarithms in \eqn{boxladderL}.
An alternate representation, which exposes their coproduct structure a bit
better, is in terms of single-valued harmonic polylogarithms
(SVHPLs)~\cite{BrownSVHPLs}.

Recall that the ordinary HPLs~\cite{Remiddi:1999ew}
with uncompressed arguments $\vec{w}$, $w_i\in \{0,1\}$,
obey the differential relations,
\be
\label{eq:HPLdef_1}
\frac{\partial}{\partial z}H_{0\vec{w}}(z) = \frac{H_{\vec{w}}(z)}{z}  \,,
\qquad
\frac{\partial}{\partial z}H_{1\vec{w}}(z) = \frac{H_{\vec{w}}(z)}{1-z}\,,
\ee
along with the ``initial conditions''
\begin{equation}
H_{1}(z) = -\ln(1-z),\qquad 
H_{\fwboxL{27pt}{{\underbrace{0,\ldots,0}_{n}}}}(z) = \frac{1}{n!}\ln^n z \,.
\end{equation}
The SVHPLs, $\cL_{\vec{w}}$, $w_i \in \{0,1\}$,
are functions of $z$ and $\zb$ that are linear
combinations of products $\sim H_{\vec{w}}(z)H_{\vec{w}^\prime}(\zb)$ and are
real analytic in the complex plane minus the punctures at $0,1,\infty$.
They satisfy a similar set of differential relations,
\begin{equation}\label{eq:Lzdiffeq}
\frac{\partial}{\partial z}\cL_{0\vec{w}}(z,\zb)
= \frac{\cL_{\vec{w}}(z,\zb)}{z} \,,
\qquad
\frac{\partial}{\partial z}\cL_{1\vec{w}}(z,\zb)
= \frac{\cL_{\vec{w}}(z,\zb)}{1-z}\,,
\end{equation}
with
\begin{equation} \label{eq:LzRegularization}
\cL_{1}(z,\zb) = - \ln|1-z|^2 \,, \qquad
\cL_{\fwboxL{27pt}{{\underbrace{0,\ldots,0}_{n}}}}(z,\zb) = \frac{1}{n!}\ln^n |z|^2 \,.
\end{equation}
Their symbol alphabet is
\be
{\cal S}_f = \{ z, 1-z, \zb, 1-\zb \},
\label{boxladderalphabet}
\ee
along with the single-valuedness requirement that the first entry
is either $z\zb$ or $(1-z)(1-\zb)$.

In terms of the $\cL_{\vec{w}}$ functions, the box ladders
become~\cite{Drummond:2013nda,Basso:2017jwq},
\be
f^{(L)}(z, \bar z) =
(-1)^L \Bigl[ \cL_{\fwboxL{27pt}{{\underbrace{0,\ldots,0,1,0,0,\ldots,0}_{2L}}}}\hspace{1.72cm}
            - \cL_{\fwboxL{27pt}{{\underbrace{0,\ldots,0,0,1,0,\ldots,0}_{2L}}}}\hspace{1.72cm} \Bigr] \,.
\label{boxladderSVHPLform}
\ee
These are depth 1 SVHPLs~\cite{Drummond:2012bg}, where the single ``1'' in the weight vector
appears in one of the two middlemost slots.

Now we ask, what is the $f$ space of functions that contains all coproducts
of the $f^{(L)}(z,\bar z)$ as $L\to\infty$?  This is a minimal space in which
we can construct all $f^{(L)}$'s as iterated integrals.
Taking the derivative $z\del_z$
simply clips off the first ``0''.  Taking the anti-holomorphic derivative
$\zb\del_{\zb}$ can be a bit complicated for a generic SVHPL, but
at depth 1 it just clips off the last ``0''.  Iterating this procedure,
the ``1'' can slide forward and backward to any location in the string,
until it reaches either end, where it is clipped off by taking a $1-z$ or
$1-\zb$ coproduct instead of a $z$ or $\zb$ one.
In summary, by taking coproducts using the letters in
${\cal S}_f$, we generate all depth 1 weight $n$ SVHPLs,
of which there are $n$, as well as the single depth 0 function at
this weight, $\cL_{0,\ldots,0}$.  The dimension of this
space at weight $n$ is $n+1$, as illustrated in table~\ref{tab:boxl}.

\renewcommand{\arraystretch}{1.25}
\begin{table}[!t]
\centering
\begin{tabular}[t]{l c c c c c c c c c}
\hline
weight $n$                   & 0 & 1 & 2 & 3 &  4 & 5 & 6 & 7 & 8  \\\hline\hline
$f_n$ functions             & 1 & 2 & 3 & 4 &  5 & 6 & 7 & 8 & 9  \\\hline
${\rm Disc}\,f_n$ functions & 0& 1 & 2 & 3 & 4 &  5 & 6 & 7 & 8  \\\hline
\end{tabular}
\caption{Number of weight $n$ functions in the $f$ space, and in the space of functions
after taking a discontinuity in $1-z$, holding $1-\zb$ fixed.  Only a single function gets lost in the process.}
\label{tab:boxl}
\end{table}

The last line in table~\ref{tab:boxl} shows what happens to
the dimension if we
take the discontinuity in $1-z$, holding $1-\zb$ fixed,
for all functions in the $f$ space.
This discontinuity is associated with the cut in the channel carrying
momentum along the ladder~\cite{Basso:2017jwq}.
It will be analogous to the $c$-discontinuity of the pentaladder integrals.
The discontinuity of $f^{(L)}$ (defined as the difference of its value for $z>1$ taken above or below the branch cut) is given by
\be
\frac{1}{2\pi i} {\rm Disc}\, f^{(L)}(z,\zb) \ =\
\frac{(-1)^L}{L!(L-1)!} \, \ln(z/\zb) \, ( \ln z \ln\zb )^{L-1} \,.
\label{discf}
\ee
It is depth 0, since the discontinuity removed the ``1''.
That is, the symbol entries belong to
\be
{\cal S}_{{\rm Disc}\,f} = \{ z, \zb \}.
\label{discboxladderalphabet}
\ee
The discontinuity ${\rm Disc}\, f^{(L)}$ is itself not a single-valued function.
Because of this, when we take derivatives to fill out the full 
${\rm Disc}\,f$ space, using the fact that discontinuities commute
with derivatives, we get all $n$ monomials at weight $n-1$:
$\ln^k z \ln^{n-1-k} \zb$, $k=0,1,\ldots n-1$.
Thus the dimension of the space $f$ is reduced by only one in passing to
${\rm Disc}\,f$, even though the symbol alphabet is halved in size.

This means there is very little loss of information in going to ${\rm Disc}\,f$:
any function in $f$ can be recovered from its discontinuity at the price of a single boundary condition.
Indeed the only combinations of SVHPLs with vanishing $z=1$ discontinuity are the simple logarithms $\ln^{k}(z\zb)$.
The ladder integral $f(z,\zb)$ can be characterized as the unique combination of SVHPLs with the discontinuity (\ref{discf})
and which vanishes at $z,\zb\to 0$.

Before we discuss the pentaladders, it is instructive to understand this simplification from the perspective of the integral representation in \eqn{result box ladder}.
We fix $\zb$ to a value between 0 and 1 while we analytically continue to $z>1$.
The argument of the sine is then complex, $\pi-\phi\equiv \pi-\tfrac{1}{2i}\ln(z/\zb)$,
with a real part that approaches $\pm \pi$ depending on whether we approach the cut from above or from below.
The discontinuity thus gives a simpler integral with the $\sin(\pi j)$ denominator canceled:
\be
 \frac{1}{2\pi i} {\rm Disc}\, f(z,\zb,g^2) = -\int_{-\infty}^\infty \frac{d\nu}{2\pi} (z\zb)^{i\nu/2} \left[ \left(\frac{z}{\zb}\right)^{j(\nu)/2}+\left(\frac{z}{\zb}\right)^{-j(\nu)/2} \right]\,. \label{box disc integral}
\ee
The integrand is now an entire function of $\nu$. This is so despite the branch points of $j(\nu)=i\sqrt{\nu^2+4g^2}$ at $\nu=\pm 2ig$, because it is even in $j$.
It is convenient to use this property to shift the contour slightly below the two branch points. 
One can then integrate over the two terms separately:
closing the contour in the upper-half plane for the first term, where it decays, and below for the second term.
The discontinuity thus reduces to a small contour integral over the cut in the first term: letting $s=i\nu/2$,
\be
 \frac{1}{2\pi i} {\rm Disc}\, f(z,\zb,g^2) = -\oint_{[-g,g]} \frac{ds}{i \pi} z^{s+\sqrt{s^2-g^2}}\zb^{s-\sqrt{s^2-g^2}}\,.
\label{bdi2}
\ee
It is now easy to see why the discontinuity involves only powers of $\ln z$ and $\ln \zb$ in perturbation theory: both exponents are uniformly small over the integration contour, and so can be series-expanded.
Expanding the integrand at small coupling and taking the coefficient of $(-g^2)^L$,
we indeed find a pole in $1/s$ whose residue reproduces
\eqn{discf}.\footnote{
The integral can be computed exactly as a Bessel function,
$g\ln(z/\bar{z})/\sqrt{\ln z\ln\bar{z}}\times I_1(2g\sqrt{\ln z\ln\bar{z}})$,
reproducing the resummation of \eqn{discf}.}

Notice that an alternative evaluation of \eqn{box disc integral} would have been to move the contour to slightly above the two branch points, closing it in the lower half-plane, and picking up the contribution from the second term in \eqn{box disc integral} instead.  In this version, we would have the same formula with $j\to -j$ and an opposite overall sign from the reversed contour orientation.  Thus \eqn{bdi2} must be odd under $j=2\sqrt{s^2-g^2} \to -j$.  Since this symmetry exchanges $z$ and $\bar{z}$, it is consistent with \eqn{discf} being odd under $z\lr\bar{z}$.

\subsection{The pentaladder integrals and their \texorpdfstring{$c$}{c}-discontinuity}

By investigating the first few loop orders, we observed that the $c$-discontinuity of the pentaladders is similarly
very simple.  Here we will derive such a simplification directly from the integral representation of section \ref{sec:resummed_result}, where the discontinuity will collapse the $\nu$ integral onto a small circle as in the preceding example.

The $c$-discontinuity represents a cut along the channel carrying
momentum along the ladder, from switching the sign of $x_{14}^2$ from spacelike to timelike.  In terms of the cross ratios $u,v,w$ in \eqn{uvw_def}, it can be implemented by
\be
u\to e^{-i\pi} \, u, \qquad v\to e^{-i\pi} \, v, \qquad w\to w.
\label{uvwforcdisc}
\ee
Normalizing it so that $\ln c$ has discontinuity 1,
the discontinuity is properly defined as
\be
 \cDisc \Omega(a,b,c) \equiv \frac{1}{4\pi i} 
\Bigl[ \Omega(a,b,e^{2\pi i}c)-\Omega(a,b,e^{-2\pi i}c) \Bigr].
\ee
To translate this to the $x,y,z$ variables we use \eqns{zasuvw}{omxomy}:
\be
 x y = \frac{(1-u)(1-v)}{uv}, \qquad (1-x)(1-y) = c, \qquad z= \frac{u(1-v)}{v(1-u)}\,.
\ee
After the continuation we have $u,v<0$, which corresponds to $x>1$ and $y>1$,
with $x$ just below the $x>1$ branch cut in the $e^{2\pi i}c$ term,
and similarly for $y$. Because both $u$ and $v$ rotate in the same way 
in \eqn{uvwforcdisc}, we also need $xy$ to acquire the phase $2\pi i$.
This phase can be put in either $x$ or $y$, with $z$ not rotating;
the result will be physically equivalent due to the single-valuedness
constraint enforced in section \ref{sec:doublepentasubsection}.
We conclude that a valid continuation path for the $e^{2\pi i}c$ term is to take $x>1$ and $y>1$
below the cut, but with a prior rotation of $y$ around the origin.

The discontinuity in the finite-coupling formula~\eqref{result double penta} for $\Omega$ is trickier to compute than that for the box ladder $f$ because $x>1$ ends up outside the radius of convergence of the hypergeometric series.
We deal with this complication by using the standard hypergeometric transformation law under $x\to 1/x$ to express the result
in terms of new functions with argument $1/x$.
Applied to \eqn{eq:defF}, this transformation yields hypergeometric functions
with spin and dimension effectively interchanged:
\be
 \tilde{F}_j^{\nu}(x) \equiv F_{ij}^{i\nu}(1/x)
 = \frac{\Gamma(1-\frac{j+i\nu}{2})\Gamma(1-\frac{j-i\nu}{2})}{\Gamma(1-j)}\,x^{j/2}\ {}_2F_1(-\tfrac{j+i\nu}{2},\tfrac{-j+i\nu}{2},1-j,\tfrac{1}{x}).
\label{Fnujswapped}
\ee
(Note that the transformation $\nu\to ij$, $j\to i\nu$ is equivalent
to $\nu^2 \to \nu^2 + 4g^2$, $g^2 \to -g^2$ in \eqn{Omdiffeqxy}, which
is also equivalent to letting $x\to 1/x$ in the differential operator.
See ref.~\cite{Kravchuk:2018htv} for further examples of such transformations.)

The analytic continuation of $F$ to $x>1-i0$ is then written as
\be
 F_\nu^j(x) \to \frac{e^{i\pi(j-i\nu)/2} \sin(\pi\frac{j+i\nu}{2})}{\sin(\pi j)} \tilde{F}_j^\nu(x)
+\frac{e^{-i\pi(j+i\nu)/2} \sin(\pi\frac{j-i\nu}{2})}{\sin(\pi j)} \tilde{F}_{-j}^\nu(x)\,,
\ee
while for $y$ we get an additional factor of $e^{-\pi\nu}$, accounting for the rotation around the origin.
For the $e^{-2\pi i}c$ term the phases just get reversed, allowing us to compute the discontinuity:
\be
\cDisc \,F_\nu^j(x)F_\nu^j(y) = 
\frac{\sin^2(\pi\frac{j+i\nu}{2})}{2\pi\sin(\pi j)}
 \tilde{F}_j^\nu(x)\tilde{F}_j^\nu(y)
- \frac{\sin^2(\pi\frac{j-i\nu}{2})}{2\pi\sin(\pi j)}
 \tilde{F}_{-j}^\nu(x)\tilde{F}_{-j}^\nu(y)\,.
\ee
When we subtract the $\nu\to -\nu$ term in the integral~\eqref{result double penta}, after noting that $\tilde{F}_j^\nu|_{\nu\to-\nu} = \tilde{F}_j^\nu$,
the trigonometric factors simplify dramatically and we end up with only:
\be
 \cDisc\, \Omega(x,y,z,g^2) = \int_{-\infty}^\infty \frac{d\nu}{4\pi} z^{i\nu/2}
\left(
\tilde{F}^\nu_{j(\nu)}(x)\tilde{F}^\nu_{j(\nu)}(y) + \tilde{F}^\nu_{-j(\nu)}(x)\tilde{F}^\nu_{-j(\nu)}(y)\right).
\ee
This result should be compared with \eqn{box disc integral}: the power laws for the box ladders have simply been replaced by hypergeometric functions.

A further simplification, as in the box ladder case, is that the integral can be rewritten as a contour integral.
Because the integrand is symmetrical in $j=i\sqrt{\nu^2+4g^2} \lr -j$, it does not have branch points at $\nu=\pm 2ig$, but to discuss its terms separately we need to choose a branch. We pick the one with $j\sim i\nu$ at large $|\nu|$.
By the analysis leading to \eqn{eq:AllLoopOmegaSum}, the function $\tilde{F}^\nu_{j(\nu)}(x)$
then only has poles in the lower-half-plane and is analytic in the upper half-plane (for $x>1$).
The function $\tilde{F}^\nu_{-j(\nu)}(x)$ has the opposite properties.
Shifting the contour to an imaginary part just below $-2ig$, and closing the contour as in the box ladder
case, we conclude that the $c$-discontinuity is saturated by the integral over the short cut from $-2ig$ to $2ig$ of the first term:
\be
\boxed{
 \cDisc\, \Omega(x,y,z,g^2) = \oint_{[-2ig,2ig]} \frac{d\nu}{4\pi} z^{i\nu/2}
\tilde{F}^{\nu}_{j(\nu)}(x)\tilde{F}^{\nu}_{j(\nu)}(y).
    }
\label{disc contour}
\ee
This formula is the main result of this subsection.
An immediate consequence is that the dependence on $z$ in perturbation theory occurs solely through powers of $\ln z$.

More generally, at weak coupling, the integrand of \eqn{disc contour} can be series expanded in $\nu$ and $j$, which are uniformly small over the contour.
Just as for the box ladder discontinuity~\eqref{bdi2}, only odd powers of $j$ contribute to the integral.  It is helpful to explicitly pick out the odd part and divide it by $j$. At this point we also recall that the $\Omega$ pentaladders were identified in \eqn{Videf} as one component of a four-vector.  Repeating the calculation for the other integrals, we write
\be
 \cDisc\, \{ \WL(g^2),\Omega(g^2),\Omt_{\rm e}(g^2),\OL(g^2)\} = \oint_{[-g,g]} \frac{ds}{i \pi} \, \sqrt{s^2-g^2}\, \cDisc\,\VL_i(s,g^2)
\label{disc from V}
\ee
where we have set $s=i\nu/2$ for future convenience.
The result \eqn{disc contour} then implies that
\be
\cDisc\,\VL_2(s,g^2)\equiv
 \frac{z^{s}}{4\sqrt{s^2-g^2}}
\left(\tilde{F}^{-2is}_{j}(x)\tilde{F}^{-2is}_{j}(y) - (j\to -j)\right)_{j=2\sqrt{s^2-g^2}}\,.
\ee
The other entries share the generic form
\be
\cDisc\,\VL_i(s,g^2) = \frac{z^{s}}{4\sqrt{s^2-g^2}} \left( X^{-2is}_{i,j} - X^{-2is}_{i,-j}\right)_{j=2\sqrt{s^2-g^2}}\,,
\label{def disc V}
\ee
and are given explicitly as
\begin{align}
 X^\nu_{1,j} &= g^2\left(\tilde{F}^{\nu\prime}_{j}(x)\tilde{F}^{\nu}_{j}(y) +
  \tilde{F}^{\nu}_{j}(x)\tilde{F}^{-\nu\prime}_{j}(y)\right),
&
 X^\nu_{2,j} &= \tilde{F}^{\nu}_{j}(x)\tilde{F}^{\nu}_{j}(y), \label{X_j_1}
 \\
 X^\nu_{3,j} &= \frac{g^2}{2}\left(\tilde{F}^{\nu\prime}_{j}(x)\tilde{F}^{-\nu\prime}_{j}(y)+\tilde{F}^{-\nu\prime}_{j}(x)\tilde{F}^{\nu\prime}_{j}(y)\right),
& X^\nu_{4,j} &= \tilde{F}^{\nu\prime}_{j}(x)\tilde{F}^{\nu}_{j}(y)
 -\tilde{F}^{\nu}_{j}(x)\tilde{F}^{\nu\prime}_{j}(y)\,, \label{X_j_2}
\end{align}
where $\tilde{F}^{\nu\prime}_{j}$ is the hypergeometric function entering
the $x\to 1/x$ transform of \eqn{eq:defFprime}:
\be
 \tilde{F}^{\nu\prime}_{j} \equiv \frac{\Gamma(1-\tfrac{j+i\nu}{2})\Gamma(\tfrac{-j+i\nu}{2})}{\Gamma(1-j)}
x^{j/2}\ {}_2F_1(1-\tfrac{j+i\nu}{2},\tfrac{-j+i\nu}{2},1-j,\tfrac{1}{x}).
\label{Fpnujswapped}
\ee
We now describe this result explicitly at weak coupling.

\subsection{Perturbative expansions and coproducts of \texorpdfstring{$c$}{c}-discontinuities}

The discontinuity integrand in \eqn{disc from V} can be doubly Taylor-expanded in $s$ and $g^2$:
\be
\VLt_i(s,g^2) \equiv \cDisc\,\VL_i(s,g^2) 
 = \sum_{L=1}^\infty\sum_{k=0}^\infty  (-g^2)^{L-1} s^k  \VLt^{(L)}_{i,k}\,.
 \label{disc_expand}
\ee
In general, we expect each term to involve powers of $\ln z$, as mentioned,
as well as polylogarithms of $x$ and $y$ originating from the expansion of the hypergeometric functions.
For example, taking the $g^2\to 0$, $s\to 0$ limit of \eqn{def disc V} using the methods of section \ref{sec:sumsandpolys}, we get
\be
\VLt^{(1)}_{1,0} =1, \qquad
\VLt^{(1)}_{2,0} = \tfrac12 \ln(x y), \qquad
\VLt^{(1)}_{3,0} = -\tfrac12\ln c, \label{disc1a}
\ee
and
\be
\VLt^{(1)}_{4,0} = \Li_2(1-x)-\Li_2(1-y)
+\tfrac12 \ln \tfrac{x}{y}\ln c \,,\label{disc1b}
\ee
where as usual $c = (1-x)(1-y)$. More generally, we find that the functions have uniform transcendental weights
\be
 \mbox{weight} \, \VLt_{i,k}^{(L)} = 2L+k + \{-2,-1,-1,0\}\,.
\label{VWeight}
\ee
A good way to see these weights and to compute the higher-order terms is to
use the differential equation that the $\VLt_{i,k}^{(L)}$ satisfy.
Before taking the discontinuity, by using the properties of the
hypergeometric functions we found a set of four coupled first-order equations,
eqs.~\eqref{dVisMV} and \eqref{matrix section3}.
Not surprisingly, since discontinuities and derivatives commute,
$\cDisc\,\VL_i(s,g^2)$ satisfies precisely the same coupled equations.
Using \eqn{del coproduct} we can rewrite the system as a coaction
\be
 \Delta_{\bullet,1} \left(\VLt_i(s,g^2) \right) = \left( \VLt_j(s,g^2) \right) \otimes M_{ij}(s,g^2),
 \label{Delta of VLt}
\ee
where we recall the definition
\be
M_{ij}(s,g^2)= \left(\begin{array}{c@{\hspace{4mm}}c@{\hspace{4mm}}c@{\hspace{4mm}}c}
s \ln z& -g^2 \ln c +2 s^2\ln(xy)& g^2\ln(xy) & 0 \\
\frac{1}{2}\ln(xy) & s\ln z & 0 & \frac{g^2}{2} \ln \frac{x}{y} \\
-\frac{1}{2}\ln c & 0 & s\ln z & \frac{g^2}{2}\ln \frac{x-1}{y-1} -s^2\ln\frac{x}{y} \\
0 & -\ln \frac{x-1}{y-1} & -\ln\frac{x}{y} & s\ln z \end{array}\right). \label{matrix}
\ee
Although the entries of this matrix are all weight one transcendental functions, we see that the relative weights of the $\VLt_{i,k}^{(L)}$ functions are encoded in the powers of $g$ and $s$. Namely, if we assign these expansion parameters both transcendental weight minus one, the diagonal terms in the matrix have weight zero, the upper-triangular entries have weight minus one, and the lower-triangular terms have weight one. For fixed $L$ and $k$, coacting on $\VLt_{j,k}^{(L)}$ with $M_{ij}(s,g^2)$ will thus increase the weight of the iterated integrals in each entry by one, but the resulting entries should be interpreted as multiplying different powers of $s$ and $g$ in the expansion~\eqref{disc_expand}.

This coaction can be used to construct the functions $\VLt^{(L)}_{i,k}$ iteratively,
using a single boundary condition, which we will describe shortly.
One begins with the transcendental weight 0 vector
and coacts on it using the matrix $M$ to get the complete weight-one component of $\VLt_i(s,g^2)$:
\be
 \VLt_{i}(s,g^2) \Big|_{\rm weight\, 0} = \begin{pmatrix} 1\\0\\0\\0 \end{pmatrix}
 \quad\Rightarrow\quad
\Delta_{0,1} \VLt_i(s,g^2) = \begin{pmatrix} s (1 \otimes \ln z) \\ \frac{1}{2}(1 \otimes \ln(x y)) \\ -\frac{1}{2} ( 1 \otimes \ln c ) \\ 0 \end{pmatrix}. \label{coaction_01}
\ee
Reading off the second and third element, we reproduce $\VLt^{(1)}_{2,0} = \tfrac12 \ln(x y)$ and $\VLt^{(1)}_{3,0} = -\tfrac12\ln c$. The first element has a factor of $s$, so it corresponds to a term with $k=1$, in particular $\VLt^{(1)}_{1,1} = \ln z$. 

Following this procedure further, one can construct the $c$-discontinuity of our basis functions to any weight by iteratively coacting with matrix $M$. At each step one has to supplement the information from the coproduct
with one boundary condition, because (being effectively a derivative operator) $\Delta_{\bullet,1}$ kills all constants.
A convenient limit can be given at $(x,y,z)=(1,1,1)$ (corresponding to $(u,v,w)=({-}\infty,{-}\infty,1)$).
There we find from \eqn{def disc V} that $\VLt_{2}$ and $\VLt_4$ vanish, while
\ba
\lim_{x,y,z\to 1}\VLt_{1}(s,g^2) &=&\frac{2\pi g^2\sin(\pi j)}{j (\cos(\pi j)-\cos(2\pi s))}, \\
\lim_{x,y,z\to 1}\frac{\VLt_{3}(s,g^2)}{\VLt_{1}(s,g^2)} &=&
-\tfrac12\left(\psi(s+\tfrac{j}{2})+\psi(s-\tfrac{j}{2})+\psi(-s+\tfrac{j}{2})+\psi(-s-\tfrac{j}{2})+4\gamma_E\right), \label{limit VLt}
\ea
where $j=2\sqrt{s^2-g^2}$ as above, and $\gamma_E$ is the Euler-Mascheroni constant.  For $\VLt_{3}$ we have dropped the singular logarithm $\ln c$
in this limit, in order to focus on the constant piece.
Up to weight 4, these can be expanded explicitly as
\be
 \lim_{x,y,z\to 1}\left[\VLt_{i}(s,g^2)\right]\ =
 \begin{pmatrix} 1+ 2g^2\zeta_2 +(8g^2s^2-2g^4)\zeta_4+ \ldots\\ 0 \\ (4s^2-2g^2)\zeta_3+ \ldots \\ 0 \end{pmatrix} . \label{wt4_boundary_condition}
\ee
In general, we find that the top row and ratio of the first and third can be expanded into even and odd zeta values respectively:
\ba
\lim_{x,y,z\to 1}\VLt_{1}(s,g^2) &=&1-2\sum_{k=1}^\infty \zeta_{2k} \sum_{L=1}^{k} (-g^2)^{L} (2s)^{2k-2L}\ \frac{(2k-L-1)!}{(2k-2L)!(L-1)!},
\\
\lim_{x,y,z\to 1}\frac{\VLt_{3}(s,g^2)}{\VLt_{1}(s,g^2)} &=&
2\sum_{k=1}^\infty \zeta_{2k+1} \sum_{L=0}^{k} (-g^2)^{L} (2s)^{2k-2L}\ \frac{(2k-L-1)!k}{(2k-2L)!L!}.
\ea
Referring back to eq.~\eqref{disc1a}, we see that the functions there match \eqn{wt4_boundary_condition}, modulo $\ln c$ terms, in the limit $x,y,z\rightarrow1$ without the addition of any constants, as needed (remember that the first entry in~\eqref{coaction_01} should be compared to the $g^0 s^1$ term in~\eqref{wt4_boundary_condition}).

To get $\VLt^{(1)}_{4,0}$ and the weight-two contributions to the other functions, we now coact with $M$ on the weight-one vector in~\eqref{coaction_01}, which we have promoted to a vector of full functions. This gives
\be
\Delta_{1,1} \VLt_i(s,g^2) = 
\begin{pmatrix} s^2 \big(\ln(x y) \otimes \ln(x y) + \ln z \otimes \ln z \big) - \frac{g^2}{2} \big(\ln(x y) \otimes \ln c + \ln c \otimes \ln(x y)\big) \\ \frac{s}{2} \big(\ln(x y) \otimes \ln z + \ln z \otimes \ln(x y)\big) \\ - \frac{s}{2}\big(\ln c \otimes \ln z + \ln z \otimes \ln c \big) \\ \frac{1}{2}\left( \ln c \otimes\ln \frac{x}{y} - \ln(x y)\otimes\ln \frac{x-1}{y-1} \right)\end{pmatrix}, \label{coaction_11}
\ee
the fourth component of which is indeed the coproduct of \eqn{disc1b}. The other three entries can be promoted to products of logs, which can be matched to the boundary condition~\eqref{wt4_boundary_condition} to give
\bea
\VLt^{(1)}_{1,2} &=& \tfrac{1}{2} \ln^2 (x y) + \tfrac{1}{2}  \ln^2 z \,, \quad \VLt^{(1)}_{2,1} = \tfrac{1}{2} \ln z \ln (x y)\,, \quad \VLt^{(1)}_{3,1} = - \tfrac{1}{2} \ln z \ln c \,, \nonumber \\
\VLt^{(2)}_{1,0} &=& \tfrac{1}{2} \ln (x y) \ln c - 2 \zeta_2 \,.
\eea

The space of functions generated by this procedure turns out to be one we have already encountered---the space of SVHPLs introduced in eqns.~\eqref{eq:Lzdiffeq} and~\eqref{eq:LzRegularization}, where $z$ and $\bar z$ are now equal to $1-x$ and $1-y$. For instance, we can rewrite all the functions we have computed above as
\bea
\VLt^{(1)}_{2,0} = - \tfrac12 \mathcal{L}_1\,, \qquad \VLt^{(1)}_{3,0} = -\tfrac12 \mathcal{L}_0\,, \qquad \VLt^{(1)}_{1,2} = \mathcal{L}_{1,1} + \tfrac12 \ln^2 z\,, \qquad
\VLt^{(1)}_{2,1} =  - \tfrac12 \mathcal{L}_{1} \ln z\,, \nonumber \\
\VLt^{(1)}_{3,1} = -\tfrac12 \mathcal{L}_{0} \ln z\,, \qquad \VLt^{(1)}_{4,0} = \tfrac12 \mathcal{L}_{0,1} -\tfrac12 \mathcal{L}_{1,0}\,, \qquad
\VLt^{(2)}_{1,0} = - \tfrac12 \mathcal{L}_{0,1} - \tfrac12 \mathcal{L}_{1,0} - 2\zeta_2 \,,~~ 
\eea
where we have left the SVHPL arguments $\{z, \bar{z}\} = \{1-x,1-y\}$ implicit.

We wish to show that the dependence on $x$ and $y$ for arbitrary weight is captured by SVHPLs with arguments $1-x$ and $1-y$.  Given the letters ${\cal S}_{\rm disc}$ in \eqn{fivenicelettersALT}, the main issue is to show that the functions $\VLt^{(L)}_{i,k}(x,y,z)$ are single-valued at $0,1,\infty$ in the complex plane for $(x,y)$.  It is sufficient to look at two of the three limits, say where $x$ and $y$ both approach 1 or both approach $\infty$. We can probe the first limit with the boundary conditions~\eqref{wt4_boundary_condition}, which tell us that the monodromies around this point are dictated by the first and third columns of the matrix $M$ in \eqn{matrix}.  The first and third columns contain only $\ln z$, $\ln (x y)$, $\ln(x/y)$ and $\ln c = \ln [(x-1)(y-1)]$.  The first three of these functions are smooth or vanish as $(x,y)\to(1,1)$, and the fourth is real analytic (single valued).  In other words, the potentially problematic entry $\ln[(x-1)/(y-1)]$ in the second and fourth columns is killed by the boundary condition~\eqref{wt4_boundary_condition}. (This boundary condition is for $z=1$, but the $z$ dependence factorizes.) The single-valuedness at $x,y\rightarrow \infty$ can be seen by considering eqns.~\eqref{def disc V}, \eqref{X_j_1}, and \eqref{X_j_2}, as well as the expansions~\eqref{Fnujswapped} and \eqref{Fpnujswapped}, whereby the dependence on $x$ and $y$ in $\cDisc\,\VL_i(s,g^2)$ takes the form $(x y)^{\pm j}$ times a regular expansion in powers of $1/x$ and $1/y$ in each term.  Thus, the functions in $\cDisc\,\VL_i(s,g^2)$ are single-valued in $1-x$ and $1-y$, making them SVHPLs.

In an ancillary file, \texttt{omegacdiscwt0-12.m}, we provide the SVHPL
representation of all the $c$-discontinuity functions $\VLt^{(L)}_{i,k}(x,y,z)$
through weight 12.

Finally, to return to the ladder integrals themselves,
we insert the series expansion of $\cDisc \VL_i(s,g^2)$ in \eqn{disc_expand} and the series expansion of the square root in terms of Catalan numbers,
\be
s-\sqrt{s^2-g^2}  = \frac{g^2}{2s}\sum_{n=0}^\infty
 \left( -\frac{1}{4} \right)^n C_n \left(\frac{-g^2}{s^2}\right)^n \,,
\ee
into \eqn{disc from V}. Performing the contour integral in $s$ by residues at the origin, we obtain an expression for the $c$-discontinuity of all pentaladder integrals in terms of the $\VLt_{i,k}^{(L)}$:
\be
 \cDisc\, \{ \WL^{(L)},\Omega^{(L)},\Omt_{\rm e}^{(L)},\OL^{(L)}\}
 = \sum_{n=0}^{L-1} \left(-\frac{1}{4} \right)^n
 C_n \VLt^{(L-n)}_{i,2n}\,.  \label{catalan_disc}
\ee
The chief advantage of the enlarged set of $\VLt_{i,k}^{(L)}$, as opposed to looking only at the
combinations in \eqn{catalan_disc}, is that this set is closed under the coaction.  This allows the $\VLt_{i,k}^{(L)}$ to be computed recursively in an efficient manner, and will be critical to ``undoing'' the discontinuity.

\subsection{The \texorpdfstring{$\Omega$}{Omega}-functions in the coproduct formalism}
\label{omegacopsubsec}

Having now constructed a basis of functions describing the $c$-discontinuities of the $\Omega$ system and their derivatives,
our next task is to ``undo'' the discontinuity to get the full functions. 
As in the box ladder example, the key is that Steinmann hexagon functions without $c$-discontinuity are extremely constrained.

A function with no $c$-discontinuity must have first entries in $\{a,b\}$. Using \eqns{oddintegpairs}{evenintegpairs} one can check that such functions can only have symbol letters in the set $\{a,b,m_w\}$.
Functions of this type were classified in ref.~\cite{Caron-Huot:2016owq}; they are a subset of the functions called $K$ functions there. These $K$ functions can be expressed simply as products of logarithms $\ln^k(a/b)$ and HPLs in $m_w$. Thus, the $c$-discontinuity uniquely fixes the coproducts of our functions of interest, up to a few such $K$ functions!

Since all the non-log, harmonic-polylogarithmic behavior of these $K$ functions
depends on $w$,
they are naturally probed by values of the $i$ loop integral $\Omega^{(i)}$
on the line where $u=v=1$, namely $\Omega^{(i)}(1,1,w)$ as given in \eqn{Om_11w}.
In fact, with a bit of trial and error, we find that a special combination
always occurs,
\be
f_k(u,v,w) = -\sum_{i=0}^{\lfloor k/2\rfloor} \frac{1}{(k-2i)!} \ln^{k-2i}\left(\frac{u}{v}\right) \Omega^{(i)}(1,1,w) \,,
\ee
where $\lfloor x \rfloor$ is the greatest integer less than or equal to $x$,
and $f_0(u,v,w) = 1$.  Note that $\ln(u/v) = \tfrac12 \ln(a/b)$.
For the $\Omega$ space, we need only two additional transcendental $K$ functions
at each weight:
\be
 \kappa_k= f_k+ (1-w)\partial_w f_{k+1},\qquad
 \tilde{\kappa}_k= -f_k+(1-w)\partial_w f_{k+1}\,.
\ee
These two functions can be constructed recursively from their nonvanishing
coproducts:
\be
 \kappa_k^a =  \kappa_{k-1},\qquad  \tilde{\kappa}_k^b = -\tilde{\kappa}_{k-1},
 \qquad
\kappa_k^{m_w} = -\tilde{\kappa}_k^{m_w}
= - \tfrac12 \left( \kappa_{k-1} + \tilde{\kappa}_{k-1} \right)\,,
\ee
together with the boundary condition at $(u,v,w)=(1,1,1)$:
\be
 f_k(1,1,1) = \kappa_k(1,1,1) = -\tilde{\kappa}_k(1,1,1) =
\begin{cases}
0& k \quad\textrm{odd,}\\
-\Omega^{(k/2)}(1,1,1)=(2k-2) \zeta_{k}& k \quad\textrm{even.}
\end{cases}
\ee
For $k=0$, $\kappa_0 = - \tilde{\kappa}_0 = 1$, since
$f_0(1,1,1)=1$ and the derivatives $(1-w)\partial_w f_{k+1}$ vanish uniformly at this point.

The functions
$\VL^{(L)}_{i,k} \equiv \{ \WL_k^{(L)},\Omega_k^{(L)},\Omt_{{\rm e},k}^{(L)},\OL_k^{(L)}\}$
are defined for $k\geq0$ and $L\geq1$ (if $k<0$ or $L<1$, they are set to zero, with the exception of $\Omega_0^{(0)}\equiv 1$).
Given the formula~\eqref{VWeight} for the weight of these functions,
the complete set of functions appearing at weight $n$ is:
\bea
&&\kappa_n,\quad \tilde{\kappa}_n, \nonumber\\
&&\WL_{n+1-2L}^{(L)},\ \ L=1,2,\ldots,\lfloor\tfrac{n+1}{2}\rfloor, \nonumber\\
&&\Omega_{n-2L}^{(L)},\ \ L=1,2,\ldots,\lfloor\tfrac{n}{2}\rfloor, \nonumber\\ 
&&\Omt_{{\rm e},n-2L}^{(L)},\ \ L=1,2,\ldots,\lfloor\tfrac{n}{2}\rfloor, \nonumber\\ 
&&\OL_{n-1-2L}^{(L)},\ \ L=1,2,\ldots,\lfloor\tfrac{n-1}{2}\rfloor.
\label{OmegaSpaceBasis}
\eea
The dimension of the space is
\be
2 + \lfloor\tfrac{n+1}{2}\rfloor + 2 \times \lfloor\tfrac{n}{2}\rfloor
  + \lfloor\tfrac{n-1}{2}\rfloor = 2n+1.
\label{OmegaSpaceDimension}
\ee

The coproducts of the $\VL^{(L)}_{i,k}$ functions contain two types of terms.
The first type involves other functions in $\VL$; they are determined by the matrix~\eqref{matrix}, which in the alphabet~\eqref{hex_letters_abc} reads
\be
 M(s,g^2)= \left(\begin{array}{c@{\hspace{3mm}}c@{\hspace{3mm}}c@{\hspace{3mm}}c}
    s \ln \frac{m_v}{m_u} & -g^2 \ln c +2 s^2\ln(m_um_v) & g^2 \ln(m_um_v) & 0 \\
    \frac{1}{2}\ln(m_um_v) & s \ln \frac{m_v}{m_u}&0& -\frac{g^2}{2}\ln(y_uy_v)\\
    -\frac{1}{2}\ln c&0& s \ln \frac{m_v}{m_u}& -\frac{g^2}{2}\ln(y_uy_vy_w)+s^2\ln(y_uy_v)\\
    0&\ln(y_uy_vy_w)&\ln(y_uy_v)& s \ln \frac{m_v}{m_u} 
 \end{array}\right). \label{matrix abc}
\ee
The second type of terms involves $\kappa$ and $\tilde{\kappa}$, and we have determined them by solving
integrability conditions.
(By ``integrability'' here we refer to the commutativity of partial derivatives, not to be confused with quantum integrability.)

Given the nonvanishing coproducts of these functions, one can define the $\Omega$ system recursively.
The most complicated of these involve $m_u$ and $m_v$ for the even functions,
and $y_u$ and $y_v$ for the odd functions:
\begin{eqnarray} \label{Omega space}
 \WL^{(L)m_u}_k &=&  -\WL^{(L)}_{k-1} + 2\Omega^{(L)}_{k-2} -\Omt^{(L-1)}_{{\rm e},k}
+ c^{(L-2)}_k \tilde{\kappa}_{2L+k-2},
\\
 \Omega^{(L)m_u}_k &=&  -\Omega^{(L)}_{k-1} + \frac{1}{2}\WL^{(L)}_k + c^{(L-1)}_k \tilde{\kappa}_{2L+k-1},
  \\
 \Omt^{(L)m_u}_{{\rm e},k} &=& -\Omt^{(L)}_{{\rm e},k-1}+ c^{(L)}_{k-1} \tilde{\kappa}_{2L+k-1},
  \\
  \OL^{(L)y_u}_k &=& \Omega^{(L)}_k+\Omt^{(L)}_{{\rm e},k}  + c^{(L)}_k\tilde{\kappa}_{2L+k},
\end{eqnarray}
and
\begin{eqnarray}
 \WL^{(L)m_v}_k &=&  \WL^{(L)}_{k-1} + 2\Omega^{(L)}_{k-2} -\Omt^{(L-1)}_{{\rm e},k}
-c^{(L-2)}_k \kappa_{2L+k-2}
\\
  \Omega^{(L)m_v}_k &=& \Omega^{(L)}_{k-1} +\frac{1}{2} \WL^{(L)}_k +c^{(L-1)}_k \kappa_{2L+k-1}
  \\
   \Omt^{(L)m_v}_{{\rm e},k} &=& \Omt^{(L)}_{{\rm e},k-1} +c^{(L)}_{k-1}\kappa_{2L+k-1}
  \\
  \OL^{(L)y_v}_k &=& \Omega^{(L)}_k+\Omt^{(L)}_{{\rm e},k}  - c^{(L)}_k \kappa_{2L+k}\,.
\end{eqnarray}
The first three terms on the first line of these, for example, come from the $m_u$ and $m_v$
terms in the first row of \eqn{matrix abc}.
The remaining nonvanishing coproducts are then:
\begin{align} \label{Omega space remaining}
 \WL^{(L)c}_k &= \Omega^{(L-1)}_k,&
  \Omega^{(L)y_u}_k &= \Omega^{(L)y_v}_k = \tfrac12\OL^{(L-1)}_k,&& \\
 \Omt^{(L)_c}_{{\rm e},k} &= - \frac{1}{2}\WL^{(L)}_k, &
  \Omt^{(L)y_u}_{{\rm e},k} &=\ \Omt^{(L)y_v}_{{\rm e},k} =\tfrac12\OL^{(L-1)}_k+\OL^{(L)}_{k-2},\\
  \Omt^{(L)m_w}_{{\rm e},k} &= \tfrac12c^{(L-1)}_k \big(\kappa_{2L+k-1}+\tilde{\kappa}_{2L+k-1}\big), &
 \Omt^{(L)y_w}_{{\rm e},k} &= \tfrac12\OL^{(L-1)}_k,\\
  \OL^{(L)m_u}_k &=  -\OL^{(L)m_v}_k = -\OL^{(L)}_{k-1}, &
 \OL^{(L)y_w}_k &= \Omega^{(L)}_{k} -c^{(L)}_k \big(\kappa_{2L+k}-\tilde{\kappa}_{2L+k}\big).
\end{align}
The binomial coefficients $c^{(L)}_k$,
which intertwine the $\VL$ and $\kappa$ systems, are:
\be
 c^{(L)}_k = 2^{k-1} \, \binom{k+L-1}{k}.
\label{cGeneral}
\ee
There are a few exceptional cases at low weights:
\be
c^{(-1)}_1=-1, \quad c^{(0)}_0=\frac12 \, ;
\quad \hbox{otherwise $c^{(L)}_k = 0$ for $k<0$ or $L<1$}.
\label{cExceptional}
\ee
This concludes the complete recursive definition of the $\Omega$ space of functions to all weights.

Remarkably, when evaluated on the line $(1,1,w)$, {\it every} function in the
$\Omega$ space approaches an integer multiple of either $\Omega^{(m)}(1,1,w)$
(for even weight $2m$) or $(1-w) d\Omega^{(m)}(1,1,w)/dw$
(for odd weight $2m-1$).
The integer multiples are given by the binomial coefficients $c^{(L)}_k$:
\bea
\kappa_k &\to& -1, \quad \tilde{\kappa}_k \to (-1)^k \nonumber\\
\{ \WL^{(L)}_k, \Omega^{(L)}_k, \Omt^{(L)}_{{\rm e},k}, \OL^{(L)}_k \} &\to&
 \{ 0, 0, 0, 0 \}, \qquad k\ \hbox{odd,} \nonumber\\
\{ \WL^{(L)}_k, \Omega^{(L)}_k, \Omt^{(L)}_{{\rm e},k}, \OL^{(L)}_k \} &\to&
 \{ - c^{(L-2)}_{k+1}, c^{(L-1)}_{k+1}, c^{(L)}_{k}, c^{(L)}_{k+1} \},
 \qquad k\ \hbox{even.} 
\label{limits11w}
\eea
In other words, we can fix the boundary conditions for the coproduct
description along the entire line $(1,1,w)$, not just at the point $(1,1,1)$ given in eqs.~(\ref{Omt11wvanish}),
(\ref{Om_111}) and (\ref{W11w}).

Finally, let us be explicit as to how the actual ladder integrals sit inside this basis, as Catalan-weighted sums along the same lines as \eqn{catalan_disc}:
\be
\{ \WL^{(L)},\ \Omega^{(L)},\ \Omt^{(L)}_{\rm e},\ \OL^{(L)} \}
= \sum_{n=0}^{L-1} \left( -\frac{1}{4} \right)^n \, C_n \,
\{ \WL^{(L-n)}_{2n},\ \Omega^{(L-n)}_{2n},\ \Omt^{(L-n)}_{{\rm e},2n},\ \OL^{(L-n)}_{2n} \},~~
\label{IntegralsInBasis}
\ee
and
\be
\Omt^{(L)}_{\rm o}
= - \sum_{n=1}^{L-1} \left( -\frac{1}{4} \right)^n \, C_n \, \OL^{(L-n)}_{2n-1}.
\label{OmtoInBasis}
\ee
Note that $\Omt_{\rm e}^{(1)}$ is not pure, so we should use \eqn{Omt1}, not
\eqn{IntegralsInBasis} for that case.  Also,
the first two instances of $\WL^{(L)}$ and the first instance of $\Omega^{(L)}$
are exceptional, needing additional $\kappa$ contributions:
\begin{align}
\WL^{(1)} &= \WL^{(1)}_0 + v\, \kappa_1  + u\, \tilde{\kappa}_1 , &\quad
\WL^{(2)} &= \WL^{(2)}_0 - \tfrac14 \WL^{(1)}_2
         + \tfrac12 ( \kappa_3  + \tilde{\kappa}_3 ),
         \\
 \Omega^{(1)}&= \Omega^{(1)}_0+\tfrac12(\tilde{\kappa}_2-\kappa_2). &&
\label{Wexceptions}
\end{align}
Apart from these exceptions, \eqns{IntegralsInBasis}{OmtoInBasis}
locate the five integrals per loop perfectly inside the $\Omega$ space for all $L\geq 1$.

The space of $\Psi$ functions for the pentabox ladders has an analogous
description, which is not surprising since the $\Psi$ integral is obtained from the $\Omega$ integral by letting $w\to0$, $\Psi^{(L)}(u,v) = \Omega^{(L)}(u,v,0)$.
However, not all of the integrals are nonsingular in this limit.
Instead of the $(2n+1)$-dimensional space~\eqref{OmegaSpaceBasis} at weight $n$,
the following subspace survives,
\bea
&&\hat\kappa_n \equiv \tfrac12 ( \kappa_n - \tilde{\kappa}_n ), \nonumber\\
&&\hat\WL_{n+1-2L}^{(L)} \equiv
\WL_{n+1-2L}^{(L)} - \OL_{n+1-2L}^{(L-1)}
+ c_{n+1-2L}^{(L-1)} ( \kappa_n + \tilde{\kappa}_n ),
\ \ L=1,2,\ldots,\lfloor\tfrac{n+1}{2}\rfloor, \nonumber\\
&&\hat\Omega_{n-2L}^{(L)} \equiv \Omega_{n-2L}^{(L)},
\ \ L=1,2,\ldots,\lfloor\tfrac{n}{2}\rfloor, 
\label{PsiSpaceBasis}
\eea
with a total dimension of
\be
1 + \lfloor\tfrac{n+1}{2}\rfloor + \lfloor\tfrac{n}{2}\rfloor
= n+1.
\label{PsiSpaceDimension}
\ee
Note that, while this dimension matches the size of the box ladder space, these spaces cannot be isomorphic because they involve a different number of symbol letters.

The $4\times 4$ matrix $M$ collapses to a $2\times 2$ matrix $\hat{M}$ acting
on $(\hat{W}, \hat\Omega)$:
\be
 \hat{M}(s,g^2)= \left(\begin{array}{c@{\hspace{3mm}}c}
    s \ln z & -g^2 \ln(1-x) + s^2\ln x \\
    \frac{1}{2}\ln x & s \ln z 
 \end{array}\right). \label{hatmatrix}
\ee
The reduction to a two-dimensional matrix occurs because there is only a single
hypergeometric function of $x$ in the finite-coupling formula~\eqref{result penta} for $\Psi$, in contrast to the product of functions of $x$ and $y$ in the corresponding formula for $\Omega$.

We remark that the perturbative $\Omega$ space is much smaller at each weight than what would be obtained solely by imposing proper branch cuts (at weight one) and the constraints of the Steinmann relations at weight two.  As we describe in appendix \ref{appendix:coproduct_relations}, there are additional constraints on pairs of adjacent letters in the $\Omega$ space, that are reminiscent of the Steinmann relations.  In appendix~\ref{ext_steinmann}, we mention that there are similar ``extended Steinmann relations'' that apply to the more general space of hexagon functions~\cite{Caron-Huot:six_loops}, and are related to the cluster adjacency principle~\cite{Drummond:2017ssj}.


\subsection{Other \texorpdfstring{$\Omega$}{Omega} space properties and
embedding into hexagon function space}
\label{countsect}

Although we have given a complete construction of the $\Omega$ space
in the previous subsection, we can also ask how many functions can be
obtained just as coproducts of a single function.  This enumeration was
useful for our initial understanding of the $\Omega$ space, before
the above construction was discovered.

In particular, we can examine all the coproducts of the $L$ loop
odd ladder integral $\OL^{(L)}$.  Of the five ladder integrals at $L$ loops,
it has the highest weight, $2L+1$.  We iteratively construct all of the
$\{n,1,1,\ldots,1\}$ coproducts of $\OL^{(L)}$ at weight $n$.
These coproducts are highly degenerate, so we only keep the linearly
independent span of them at each weight. Then we differentiate each of those
functions to go to the next lower weight, and again keep only
the linearly independent ones, and so on. The results for the dimensions
of these spaces, and for just the parity-odd subspaces,
are tabulated in table~\ref{tab:OLLcoprods} for each $L\leq6$.

\renewcommand{\arraystretch}{1.25}
\begin{table}[!t]
\centering
\begin{tabular}[t]{l c c c c c c c c c c c c c c}
\hline
weight $n$    & 0 & 1 & 2 & 3 & 4 & 5 &  6  &  7  &  8  &  9 
&  10 &  11 &  12 & 13 \\\hline\hline
$L=2$         & 1 & 3 & 5 & 5 & 3 & 1 & $-$ & $-$ & $-$ & $-$
& $-$ & $-$ & $-$ & $-$ \\\hline
$L=2$, P odd  & 0 & 0 & 0 & 1 & 1 & 1 & $-$ & $-$ & $-$ & $-$
& $-$ & $-$ & $-$ & $-$ \\\hline\hline
$L=3$       & 1 & 3 & 5 & 7 & 7 & 5 &  3  &  1  & $-$ & $-$
& $-$ & $-$ & $-$ & $-$ \\\hline
$L=3$, P odd  & 0 & 0 & 0 & 1 & 1 & 2 &  1  &  1  & $-$ & $-$
& $-$ & $-$ & $-$ & $-$ \\\hline\hline
$L=4$         & 1 & 3 & 5 & 7 & 9 & 9 &  7  &  5  &  3  &  1
& $-$ & $-$ & $-$ & $-$ \\\hline
$L=4$, P odd  & 0 & 0 & 0 & 1 & 1 & 2 &  2  &  2  &  1  &  1
& $-$ & $-$ & $-$ & $-$ \\\hline\hline
$L=5$       & 1 & 3 & 5 & 7 & 9 & 11 &  11  &  9  &  7  &  5
&  3  &  1  & $-$ & $-$ \\\hline
$L=5$, P odd  & 0 & 0 & 0 & 1 & 1 & 2 &  2  &  3  &  2  &  2
&  1  &  1  & $-$ & $-$ \\\hline\hline
$L=6$       & 1 & 3 & 5 & 7 & 9 & 11 & 13  & 13  & 11  &  9
&  7  &  5  &  3  &  1 \\\hline
$L=6$, P odd  & 0 & 0 & 0 & 1 & 1 & 2 &  2  &  3  &  3  &  3
&  2  &  2  &  1  &  1  \\\hline

\end{tabular}
\caption{The dimensions of the spaces of $\{n,1,\ldots,1\}$ coproducts
of the odd ladder integral $\OL^{(L)}$, and also that of the
parity odd subspace.  They are both palindromic sequences.}
\label{tab:OLLcoprods}
\end{table}

Table~\ref{tab:OLLcoprods} shows a few interesting properties.
First of all, the dimensions are ``palindromic'':  The number of
independent functions increases by two with each successive differentiation,
tracing the odd natural numbers, until it peaks and then declines again by two
at each step, tracing out the same set of numbers.\footnote{%
The space of coproducts of the $L$-loop box ladder integral,
which lives in the space enumerated in table~\ref{tab:boxl},
has the same palindromic property, except using all natural numbers
instead of just the odd ones.}  The same palindromic property
holds for just the subspace that is odd under parity P,
although the peak position is shifted up in weight.

Secondly, once the number of functions has reached its peak for
a given $L$, the dimensions for weights below that peak equal
the dimension of the full $\Omega$ space at that weight.
We say that the space of coproducts becomes ``saturated'' below a given
weight $n_s^{\rm e,o}(L)$, which depends on whether the parity is
even (e) or odd (o).  Higher loop orders do not give additional functions
for weight $n\leq n_s^{\rm e,o}(L)$, and the dimension $n_\Omega(n)$
of the full $\Omega$ space can be read off. 
From table~\ref{tab:OLLcoprods}, we see that the dimensions saturate for
even (or all) and odd functions at
\be
n_s^{\rm e}(L) = L, \qquad n_s^{\rm o}(L) = L+2.
\label{Nsaturated}
\ee
The total number of $\Omega$ functions at weight $n$ is seen to be
$n_\Omega(n) = 2n+1$, matching the number in \eqn{OmegaSpaceDimension}.

\renewcommand{\arraystretch}{1.25}
\begin{table}[!t]
\centering
\begin{tabular}[t]{l c c c c c c c c c c c c c c}
\hline\hline
weight $n$
& 0 & 1 & 2 & 3 & 4 &  5 &  6 &  7 &  8 &  9 & 10 & 11 & 12 & 13 \\
\hline\hline
$\Omega$ space dimension
& 1 & 3 & 5 & 7 & 9 & 11 & 13 & 15 & 17 & 19 & 21 & 23 & 25 & 27 \\
\hline\hline
P even 
& 1 & 3 & 5 & 6 & 8 &  9 & 11 & 12 & 14 & 15 & 17 & 18 & 20 & 21 \\\hline
P even, $K$
& 1 & 3 & 5 & 6 & 6 &  6 &  6 &  6 &  6 &  6 &  6 &  6 &  6 &  6 \\\hline\hline
P odd
& 0 & 0 & 0 & 1 & 1 &  2 &  2 &  3 &  3 &  4 &  4 &  5 &  5 &  6 \\\hline\hline
\end{tabular}
\caption{Dimension of the full weight-$n$ $\Omega$ space, and decomposed
into even and odd subspaces under parity P.  In the P-even sector
we also list the number of $K$ functions, which have no $y_u,y_v,y_w$
letters in their symbol.}
\label{tab:Omcoprods}
\end{table}

In table~\ref{tab:Omcoprods} we list the dimensions of the
even and odd subspaces for weight $n\leq13$.
Parity-odd functions necessarily contain the parity-odd letters
$y_i \equiv \{y_u,y_v,y_w\}$ in their symbols.
The P-even subspace has a further subspace of ``$K$''
functions~\cite{Caron-Huot:2016owq} whose symbols contain {\it no}
parity-odd letters. These functions are simply
HPLs with arguments $1-1/u$, $1-1/v$, and $1-1/w$ in some cases
combined with logarithms.  We have already identified two of them,
$\kappa_n$ and $\tilde{\kappa}_n$, but there
are four more ``secret'' $K$ functions in the $\Omega$ space,
for a total of six at each weight $n$ (for $n>3$):
\be
\{ \kappa_n,\ \tilde{\kappa}_n,\
\WL^{(1)}_{n-1},\ \WL^{(2)}_{n-3},\
\Omega^{(1)}_{n-2},\ \Omega^{(2)}_{n-4} - \tfrac12 \Omt^{(1)}_{{\rm e},n-2} \}.
\label{allKfunctionsinOmega}
\ee
In terms of HPLs, the four secret $K$ functions at weight $n$ are
\bea
&&H_n\Bigl(1-\frac{1}{u}\Bigr) \,,  \qquad H_n\Bigl(1-\frac{1}{v}\Bigr) \,,
\nonumber\\
&&\ln\frac{v}{w} \, H_{n-1}\Bigl(1-\frac{1}{u}\Bigr)
                   + \sum_{i=1}^{n-2} H_{i,n-i}\Bigl(1-\frac{1}{u}\Bigr) \,,
\nonumber\\
&&\ln\frac{u}{w} \, H_{n-1}\Bigl(1-\frac{1}{v}\Bigr)
                   + \sum_{i=1}^{n-2} H_{i,n-i}\Bigl(1-\frac{1}{v}\Bigr) \,.
\label{secretK}
\eea

We have referred to the $\Omega$ space as a prototype or model for the
full space of Steinmann hexagon functions ${\cal H}$.
How many of the functions in ${\cal H}$ does it capture or miss,
as we go up in weight?  The $\Omega$ space has a particular orientation,
while ${\cal H}$ is closed under all permutations of $(u,v,w)$.
We define $\Omega_{\rm cyc}$ to also include cyclic permutations
of the $\Omega$ space functions under $(u,v,w)\to(v,w,u)$
and $(u,v,w)\to(w,u,v)$.  For the most part, these permuted functions
are independent.  However, the top line of \eqn{secretK} has two $K$
functions, which after including cyclic permutations, become only three
$K$ functions in all, $H_n(1-1/u_i)$, $i=1,2,3$.
So we lose three $K$ functions at each weight,
and the number of $K$ functions in $\Omega_{\rm cyc}$
is $3\times6-3 = 15$ at weight 6 and above.
There is a similar degeneracy under cyclic permutation for the few
parity-odd functions at weights 3 and 4, and for the non-$K$
parity-even functions at weights 4 and 5. (In the latter case,
certain linear combinations of cyclic permutations of non-$K$ functions
are actually $K$ functions.)  Beyond weight 5 there are no
non-$K$ degeneracies, and so the dimension of the weight $n$ part of
$\Omega_{\rm cyc}$ is $3(2n+1)-3=6n$ for $n\geq6$.  This dimension grows only
linearly with $n$, whereas the dimension of the weight $n$ part of ${\cal N}$
grows much faster, roughly like $1.7^n$.

\renewcommand{\arraystretch}{1.25}
\begin{table}[!t]
\centering
\begin{tabular}[t]{l c c c c c c c c c}
\hline\hline
weight $n$
& 0 & 1 & 2 & 3 & 4 &  5 &  6 &  7 &  8 \\\hline\hline
${\cal H}$, P even, $K$
& 1 & 3 & 6 & 12 & 22 & 39 & 67 & 114 & 190 \\\hline
$\Omega_{\rm cyc}$, P even, $K$
& 1 & 3 & 6 & 12 & 18 & 16 &  15 & 15 & 15 \\\hline\hline
${\cal H}$, P even, non-$K$
& 0 & 0 & 0 & 0 & 3 & 9 & 25 & 56 & 123 \\\hline
$\Omega_{\rm cyc}$, P even, non-$K$
& 0 & 0 & 0 & 0 & 3 & 8 & 15 & 18 &  24 \\\hline\hline
${\cal H}$, P odd
& 0 & 0 & 0 &  1 &  2 &  6 & 13 &  30 & 59 \\\hline
$\Omega_{\rm cyc}$, P odd
& 0 & 0 & 0 &  1 &  2 & 6 &  6 & 9 & 9 \\\hline\hline
\end{tabular}
\caption{Dimension of the full weight-$n$ hexagon function space ${\cal H}$,
decomposed into even and odd subspaces under parity P,
and compared with the corresponding dimensions for $\Omega_{\rm cyc}$.
In the P-even sector we divide the space into non-$K$ and $K$ functions.}
\label{tab:OmcycvsH}
\end{table}

In table~\ref{tab:OmcycvsH} we compare the dimensions of the full hexagon
function space ${\cal H}$, which has been trimmed to remove all inessential
functions~\cite{Caron-Huot:six_loops}, with the dimensions of $\Omega_{\rm cyc}$
through weight 8.  We have split the functions into P even and P odd,
and we have further split the P-even functions into $K$ functions
and non-$K$ functions.  The space $\Omega_{\rm cyc}$ spans the full hexagon
function space through weight 3.  At weight 4 it only misses 4 $K$ functions,
one of which is the constant $\zeta_4$.  At weight 5 it misses numerous $K$
functions, and a single P-even, non-$K$ function, but it still captures
all the P-odd functions at weight 5.
The single missing P-even, non-$K$ function evaluates
to $5\zeta_5-2\zeta_2\zeta_3$ at $(u,v,w)=(1,1,1)$, while the
weight 5 part of $\Omega_{\rm cyc}$ vanishes at $(1,1,1)$.
Presumably this weight 5 function is a $\WL$-like seed for non-ladder
DCI integral topologies beginning at three loops, in which three pentagons
with appropriate numerators are joined together at a common vertex.


\subsection{A nonperturbative coaction}
\label{nonpertcoactionsect}

That the dimensions of the $\Omega$ space
saturate has an interesting implication: it allows us to define the coaction
nonperturbatively.  This can be illustrated by returning to the discontinuity functions $\cDisc \VL_i(s,g^2)$,
defined nonperturbatively in \eqn{def disc V}.
Using the differential equation these functions satisfy, we see they can also be defined by a path-ordered exponential, where the argument of the exponential is a $4\times 4$ matrix:
\be
 \mathcal{U}_{ij}(s,g^2; x,y,z) = \left[\mathcal{P} \exp \left(\int^{(x,y,z)}_{(1,1,1)} dM^T(s,g^2; x,y,z)\right)\right]_{ij} \,. \label{path order}
\ee
Dotting $\mathcal{U}_{ij}$ on the left with the initial condition in \eqn{wt4_boundary_condition} reproduces the vector $\cDisc \VL_i$,
but this matrix contains additional
transcendental functions. Roughly speaking, we expect this space of functions to be necessary and sufficient to
describe all the possible analytic continuations of the $\VLt_i$.
A nonperturbative coaction can then be defined simply as a matrix product:
\be
 \Delta \mathcal{U}_{ij}(s,g^2;x,y,z)
 = \sum_k \big[ \mathcal{U}_{ik}(s,g^2; x,y,z)\big] \otimes \big[ \mathcal{U}_{kj}(s,g^2; x,y,z)\big]. \label{non pert coprod}
\ee
(See refs.~\cite{Anastasiou:2013mca,Abreu:2017enx} for related constructions also involving hypergeometric functions.)
The $\Delta_{\bullet,1}$ coproduct component from \eqn{non pert coprod} is easily seen to reproduce \eqn{Delta of VLt},
and perturbatively the $\Delta_{1,\ldots,1}$ components reproduce the symbol of these functions.
However, this equation defines $\Delta$ nonperturbatively, and in particular for all
$\Delta_{\bullet,k}$ and $\Delta_{k,\bullet}$ for $k\geq 1$.  (We haven't discussed boundary conditions, and
in principle the definition in \eqn{path order} might need to be ``twisted'' by $\zeta$-valued constants
so as to match the ordinary coproduct of polylogarithms. We leave this exploration to future work.)

Note that, by construction, the coaction~\eqref{non pert coprod} satisfies a coaction principle~\cite{Schnetz:2013hqa,Brown:2015fyf,Panzer:2016snt}. 
That is, the first entry of the coaction is always contained within the original space of functions. A similar (perturbative) coaction principle has been 
observed in the full space of Steinmann hexagon functions, where it constrains the transcendental constants that can appear in the first entry in addition to restricting the symbols of these objects~\cite{Caron-Huot:2016owq,Brown:2015fyf}. The consequences of this coaction principle thus extend beyond what is currently understood in terms of physical principles (since only the symbol-level constraints are understood in terms of allowed branch cuts of Feynman integrals), as will be described in more detail in a forthcoming work~\cite{Caron-Huot:six_loops}. The first entry of the coaction~\eqref{non pert coprod} manifestly realizes this same property. In particular, the first entry of the coaction maps to the same space of functions for any initial conditions one dots into $\mathcal{U}_{ij}$ (after which the coaction principle more closely resembles those discussed in~\cite{Brown:2015fyf}, since the second entry of the coaction will in general map to a larger space).  

The nonperturbative coaction for the $\Omega$ system can be defined in an analogous fashion, writing in matrix form the general solution to the differential equations following from eqs.~\eqref{Omega space}--\eqref{cExceptional}. The appearance of the $\kappa$ and $\tilde\kappa$ functions in this space implies that $\mathcal{U}$ will now appear as a $4\times 4$ sub-block of a bigger matrix.

It is remarkable that the set of all coproducts of the $\Omega$ integrals
(loosely speaking, the space of all their possible derivatives and analytic continuations) can be encoded nonperturbatively in a single matrix.


\section{Conclusions}
\label{ConclusionsSection}

We have investigated a class of integrals, $\Omega^{(L)}$ and $\tilde\Omega^{(L)}$, that have representatives at each loop order. In doing so, we have found something remarkable: that their all-orders behavior can be expressed in terms of beautifully simple integral formulas,
given in \eqns{result double penta}{result double penta 2}.
Using these expressions, we can extract any desired loop coefficient, and have control over the full behavior of the functions via infinite sums. 

We have also investigated the coproducts of these functions to all orders, allowing us to characterize the complete space of polylogarithms that envelops the $\Omega^{(L)}$ and $\tilde\Omega^{(L)}$ integrals and their derivatives. As a consequence, we can now efficiently construct a subspace of the Steinmann hexagon functions to arbitrarily high weight. This space is equipped with a coaction both perturbatively and at finite coupling, and obeys a coaction principle.

The inevitable next question is, can we characterize the full Steinmann hexagon space ${\cal H}$ in a similar way?  For example, can we find a systematic definition of the hexagon function coproducts, analogous to eqs.~\eqref{Omega space}--\eqref{Omega space remaining}, which solves the integrability conditions to all orders?  Or perhaps are there other subspaces of ${\cal H}$, larger than $\Omega$, that we can describe to all orders, that capture 
a wider set of functions that are not in $\Omega_{\rm cyc}$?
Does the amplitude itself have a form like the $\Omega^{(L)}$ and $\tilde\Omega^{(L)}$ integrals, and could it be written as a finite-coupling expression involving (several) Mellin integrals?  We suspect that this might be possible, although presumably it will have to include the full flux-tube dispersion relations~\cite{Basso:2013vsa,Basso:2015uxa}.

One reason for suspecting this is that the radius of convergence of the perturbative expansion of the $\Omega$ integrals appears to be much larger than that for amplitudes, as discussed in section~\ref{Line11wsubsection}.  The radius of convergence for amplitudes is relatively close to that for the cusp anomalous dimension, which also controls the behavior of the flux-tube expansion at finite excitation number.  Thus it seems likely that the large-order behavior of six-point amplitudes is controlled by other families of integrals that grow more quickly than the ladders.

In the past, several of the authors have observed differential equations linking the amplitude at different loop orders~\cite{Dixon:2014iba,Caron-Huot:2016owq}. While some of these relations do not hold at higher orders~\cite{Caron-Huot:six_loops} they still suggest that a larger piece of the Steinmann hexagon space has an iterative or recursive structure which awaits exploitation.


\vskip0.5cm
\noindent {\large\bf Acknowledgments}
\vskip0.3cm

\noindent We are grateful to Francis Brown, Bob Cahn, Einan Gardi,
Enrico Herrmann and Jaroslav Trnka for many illuminating discussions.
This research was supported in part by the National Science Foundation under
Grant No.\ NSF PHY17-48958, by the US Department of Energy under contract
DE--AC02--76SF00515, by the Munich Institute for Astro- and Particle
Physics (MIAPP) of the DFG cluster of excellence
``Origin and Structure of the Universe'', by the Perimeter Institute for Theoretical Physics, and by the Danish National Research Foundation (DNRF91), a grant from the Villum Fonden, and a Starting Grant \mbox{(No.\ 757978)} from the European Research Council. SCH's research is supported by the National Science and Engineering Council of Canada.
LD thanks the Perimeter Institute, LPTENS, the Institut de Physique Th\'eorique 
Philippe Meyer, the Higgs Centre at U.~Edinburgh, the Simons
Foundation and the Hausdorff Institute for Mathematics for hospitality.
AM is grateful to the Higgs Centre at U.~Edinburgh for hospitality, and
LD, MvH, AM, and GP thank the Kavli Institute for Theoretical Physics for hospitality.  Research at Perimeter Institute is supported by the Government of Canada through Industry Canada and by the Province of Ontario through the Ministry of Economic Development and Innovation.

\appendix


\section{Hexagon Variables}
\label{hexvarappendix}

The three cross ratios $u,v,w$ used to describe hexagon functions are,
\be\label{uvw_def_app}
u = \frac{x_{13}^2\,x_{46}^2}{x_{14}^2\,x_{36}^2}\,, 
\qquad v = \frac{x_{24}^2\,x_{51}^2}{x_{25}^2\,x_{41}^2}\,, \qquad
w = \frac{x_{35}^2\,x_{62}^2}{x_{36}^2\,x_{52}^2}\,.
\ee
The hexagon-function symbol alphabet is given by
\be
{\cal S}_\text{hex} =
\left\{u, v, w, 1-u, 1-v, 1-w, y_u, y_v, y_w \right\} \label{hex_letters_app}
\ee
where~\cite{Dixon:2011pw}
\be
y_u = \frac{u-z_+}{u-z_-}\,, \qquad y_v = \frac{v-z_+}{v-z_-}\,,
\qquad y_w = \frac{w - z_+}{w - z_-}\, ,
\label{yfromu_app}
\ee
and
\be
z_\pm = \frac{1}{2}\Bigl[-1+u+v+w \pm \sqrt{\Delta}\Bigr],
\qquad
\Delta = (1-u-v-w)^2 - 4 u v w .
\ee
The cross ratios $u,v,w$ are rational in terms of
$y_u,y_v,y_w$,
\bea
u &=& \frac{y_u (1 - y_v) (1 - y_w)}{(1 - y_u y_v) (1 - y_u y_w)}\,, 
\quad v = \frac{y_v (1 - y_w) (1 - y_u)}{(1 - y_v y_w) (1 - y_v y_u)}\,, 
\quad w = \frac{y_w (1 - y_u) (1 - y_v)}{(1 - y_w y_u) (1 - y_w y_v)}\,,
\nonumber\\
1-u &=& \frac{(1-y_u) (1-y_u y_v y_w)}{(1 - y_u y_v) (1 - y_u y_w)} \,,
\ \ {\rm etc.},
\quad
\sqrt{\Delta} = \frac{(1-y_u) (1-y_v) (1-y_w) (1-y_u y_v y_w)}
                       {(1-y_u y_v) (1-y_v y_w) (1-y_w y_u)} \,.
\label{u_from_y}
\eea
The corresponding momentum-twistor representations are
\bea
u &=& \frac{\l 6123\r \l 3456\r}{\l 6134\r \l 2356\r} \,,
\qquad 
1-u = \frac{\l 6135\r \l 2346\r}{\l 6134\r \l 2356\r} \,,
\nonumber\\
y_u &=& \frac{\l 1345\r \l 2456\r \l 1236\r}
             {\l 1235\r \l 3456\r \l 1246\r} \,,
\quad
\sqrt{\Delta} = \frac{ \l 1234\r \l 1256\r \l 3456\r
                     - \l 2345\r \l 1236\r \l 1456\r }
                     {\l 2356\r \l 1346\r \l 1245\r} \,.
\label{momtwistorreps}
\eea
The representations for $v$, $1-v$, $y_v$, and so on, can be obtained
by cycling $Z_i \to Z_{i+1}$, remembering that under this transformation
$u\to v\to w\to u$, while $y_u \to 1/y_v \to y_w \to 1/y_u$.

As discussed in section~\ref{sec:resummed_result},
for many purposes, a better set of variables for the $\Omega$ integrals
is $\{x,y,z\}$, where

\begin{align}
x &= 1+\frac{1-u-v-w+\sqrt{\Delta}}{2uv}
   = \frac{1 - y_u y_v y_w}{y_u y_v(1-y_w)} \,, \label{defx}\\
y &= 1+\frac{1-u-v-w-\sqrt{\Delta}}{2uv}
   = \frac{1 - y_u y_v y_w}{(1-y_w)} \,, \label{defy}\\
z &= \frac{u(1-v)}{v(1-u)}
   = \frac{y_u(1-y_v)^2}{y_v(1-y_u)^2}\,. \label{defz}
\end{align}
The inverse relations are,
\bea
u &=& \frac{1}{1+\sqrt{xy/z}} \,, \quad
v = \frac{1}{1+\sqrt{xyz}} \,, \quad
w = \frac{(1-x)(1-y)}{(1+\sqrt{xy/z})(1+\sqrt{xyz})} \,, \nonumber\\
y_u &=& \frac{1+\sqrt{y/(xz)}}{1+\sqrt{x/(yz)}} \,, \quad
y_v = \frac{1+\sqrt{yz/x}}{1+\sqrt{xz/y}} \,, \quad
y_w = \frac{x(1-y)}{y(1-x)} \,.  \label{uvwyfromxyz}
\eea
Notice that $x$ and $y$ depend only on $y_uy_v$ and $y_w$. That is,
the only dependence on $y_u/y_v$ is through $z$.

Parity sends the dual coordinates $x_i\to x_{i+3}$ (mod 6) and the momentum twistors $Z_i \to Z_{i+3}$ (mod 6). Parity does not affect the cross ratios $u,v,w$ but it exchanges $\sqrt\Delta \lr -\sqrt\Delta$, so that $z_+ \lr z_-$
and $y_u,y_v,y_w$ are inverted: $y_i \lr 1/y_i$.
Eqs.~\eqref{defx}--\eqref{uvwyfromxyz} show that parity exchanges
$x$ and $y$, leaving $z$ invariant.

The $u$ derivative of a function $F$, holding $v,w$ fixed, is given
in terms of first coproducts by
\be
\frac{\del F}{\del u} =
\frac{F^u}{u} - \frac{F^{1-u}}{1-u}
+ \frac{1 - u - v - w}{u \sqrt{\Delta}} F^{y_u}
+ \frac{1 - u - v + w}{(1-u) \sqrt{\Delta}} F^{y_v}
+ \frac{1 - u + v - w}{(1-u) \sqrt{\Delta}} F^{y_w} \,, \label{dFu_app}
\ee
and derivatives with respect to $v$ and $w$ are obtained by cyclic
permutations of this equation.  Derivatives with respect to $x,y,z$
are related to these derivatives by,
\bea
x\del_x + y\del_y &=&
- u(1-u) \del_u - v(1-v) \del_v + (1-u-v)(1-w) \del_w \,,
\label{xdxplusydy}\\
x\del_x - y\del_y &=& - \sqrt{\Delta} \del_w \,,
\label{xdxminusydy}\\
z \del_z &=& \frac{1}{2} \Bigl[
u(1-u) \del_u - v(1-v) \del_v - w(u-v) \del_w \Bigr] \,.
\label{zdz}
\eea

In discussing the $\Omega$ space of functions in 
section~\ref{sec:spacesect},
it is very useful to change the hexagon alphabet from
${\cal S}_\text{hex}$ in \eqn{hex_letters_app} to
\be
{\cal S}^{\prime}_\text{hex} =
\left\{a, b, c, m_u, m_v, m_w, y_u, y_v, y_w \right\} \,,
\label{hex_letters_abc_app}
\ee
where
\be
a = \frac{u}{vw} \,, \quad
b = \frac{v}{wu} \,, \quad
c = \frac{w}{uv} \,, \quad
m_u = \frac{1-u}{u} \,, \quad
m_v = \frac{1-v}{v} \,, \quad
m_w = \frac{1-w}{w} \,.
\label{abcdef}
\ee
Given coproducts labelled using ${\cal S}^{\prime}_\text{hex}$,
we can convert them to those using ${\cal S}_\text{hex}$, by
\be
F^u = F^a - F^b - F^c - F^{m_u} \,,
\qquad F^{1-u} = F^{m_u} \,,
\label{coprodsfromatou}
\ee
plus the relations obtained by cyclic permutations.
To go in the opposite direction, we use,
\be
F^a = - \frac{1}{2} ( F^v + F^{1-v} + F^w + F^{1-w} ) \,,
\qquad F^{m_u} = F^{1-u} \,,
\label{coprodsfromutoa}
\ee
plus the cyclic relations.

For example, the $z$ derivative, using~\eqn{zdz}
and expressed in terms of coproducts using
the alphabet ${\cal S}^{\prime}_\text{hex}$, is
\be
z \frac{\del F}{\del z} =
(1-v) F^a - (1-u) F^b - \frac{1}{2} ( F^{m_u} - F^{m_v} )
+ \frac{u-v}{2(1-w)} F^{m_w}
+ \frac{\sqrt{\Delta}}{2(1-w)} ( F^{y_u} - F^{y_v} ) \,.
\label{zdz_abc_app}
\ee
The $x$ and $y$ derivatives are more complicated,
but are a bit more simply expressed
in terms of the $y_i$ variables and using the
following combinations with opposite parity:
\bea
(x\del_x - y\del_y) F &=&
\frac{(1-y_w)(1-y_uy_vy_w)}{y_w(1-y_uy_v)} (F^a+F^b-F^c)
+ \frac{(1-y_uy_w)(1-y_vy_w)}{y_w(1-y_uy_v)}  F^{m_w}
\nonumber\\&&\null
- F^{y_u} - F^{y_v}
+ \frac{y_u-y_v}{1-y_uy_v} (F^{y_u} - F^{y_v})
- \frac{(1-y_w)(1+y_uy_vy_w)}{y_w(1-y_uy_v)} F^{y_w}  \,,
\label{xdxminusydy_abc}\\
(x\del_x + y\del_y) F &=&
- \frac{(1+y_vy_w)(1-y_w)(1-y_uy_vy_w)}{y_w(1-y_uy_v)(1-y_vy_w)} F^{a}
- \frac{(1+y_uy_w)(1-y_w)(1-y_uy_vy_w)}{y_w(1-y_uy_v)(1-y_uy_w)} F^{b}
\nonumber\\&&\null
+ \frac{(1+y_w)(1-y_uy_vy_w)}{y_w(1-y_uy_v)} F^{c}
\nonumber\\&&\null
+ F^{m_u} + F^{m_v}
- \frac{1-y_uy_vy_w^2}{y_w(1-y_uy_v)} F^{m_w}
+ \frac{(1-y_w)(1-y_uy_vy_w)}{y_w(1-y_uy_v)} F^{y_w} \,.
\label{xdxplusydy_abc}
\eea


\section{Extended Steinmann Relations in the Full Hexagon Function Space}
\label{ext_steinmann}

In this appendix we discuss properties of adjacent symbol entries
in the full hexagon function space ${\cal H}$, as a prelude to
a similar discussion for the $\Omega$ space and its $c$-discontinuity
in the following appendix.

One advantage of the alphabet ${\cal S}^{\prime}_\text{hex}$ is that
the Steinmann relations are made transparent, insofar as each letter $a,b,c$ contains a unique three-particle invariant~\cite{Caron-Huot:2016owq}:
\be
a = \frac{x_{13}^2x_{46}^2}{x_{24}^2 x_{35}^2 x_{51}^2 x_{62}^2} (x_{25}^2)^2 \,,
\qquad
b = \frac{x_{24}^2x_{51}^2}{x_{35}^2 x_{46}^2 x_{62}^2 x_{13}^2} (x_{36}^2)^2 \,,
\qquad
c = \frac{x_{35}^2x_{62}^2}{x_{46}^2 x_{51}^2 x_{13}^2 x_{24}^2} (x_{41}^2)^2 \,.
\label{abc3particle}
\ee
A similar simplification occurs in the heptagon letters used in~\cite{Drummond:2014ffa,Dixon:2016nkn}, where each three-particle invariant only appears in a single letter $a_{1j}$. Thus the Steinmann constraints~\cite{Caron-Huot:2016owq},
\be
{\rm Disc}_{x_{25}^2} ({\rm Disc}_{x_{36}^2} A_6) = 0,
\label{Steinmann6}
\ee
and permutations thereof, are solved (at symbol level) simply by requiring
that the first two entries of the symbol do {\it not} contain any
of the six combinations,
\begin{align}
&a\otimes b \otimes \ldots, \quad
b\otimes c \otimes \ldots, \quad
c\otimes a \otimes \ldots, \nonumber\\
&b\otimes a \otimes \ldots, \quad
c\otimes b \otimes \ldots, \quad
a\otimes c \otimes \ldots.
\label{Steinmannabc}
\end{align}
However, by examining the double coproducts of functions obtained
by taking multiple coproducts of high loop six-point amplitudes,
we have found that the same constraint also holds deeper into the symbol.
That is, the combinations
\begin{align}
&\ldots \otimes a\otimes b \otimes \ldots, \quad
\ldots \otimes b\otimes c \otimes \ldots, \quad
\ldots \otimes c\otimes a \otimes \ldots, \nonumber\\
&\ldots \otimes b\otimes a \otimes \ldots, \quad
\ldots \otimes c\otimes b \otimes \ldots, \quad
\ldots \otimes a\otimes c \otimes \ldots
\label{ExtSteinmannabc}
\end{align}
never appear~\cite{Caron-Huot:six_loops}. We refer to this condition as the extended Steinmann constraints. 

There are also 26 independent constraints on double coproducts
from function-level integrability.  Expressed in
the alphabet ${\cal S}_\text{hex}$, they are contained in the following,
\begin{align}
F^{[u_i,u_j]} &= - F^{[y_i,y_j]} \,, \label{integrabilitycA}\\
F^{[1-u_i,1-u_j]} &= F^{[y_i,y_j]} + F^{[y_j,y_k]} + F^{[y_k,y_i]} \,, \\
F^{[u_i,1-u_j]} &= - F^{[y_k,y_i]} \,, \\
F^{[u_i,y_i]} &= 0 \,, \\
F^{[u_i,y_j]} &= F^{[u_j,y_i]} \,, \\
F^{[1-u_i,y_i]} &= F^{[1-u_j,y_j]} - F^{[u_j,y_k]} + F^{[u_k,y_i]} \,, \\
F^{[1-u_i,y_j]} &= - F^{[u_k,y_j]} \,,
\label{integrabilityc}
\end{align}
for all $i\neq j \neq k \in \{1,2,3\}$, where $F^{[x,y]} \equiv F^{x,y}-F^{y,x}$.

When we combine the constraint~\eqref{ExtSteinmannabc} with the integrability
constraints, we find 52 independent pairs of double coproducts.
However, when we construct the space of (extended) Steinmann hexagon
functions iteratively in the weight, imposing correct branch cuts along with the additional
constraint~\eqref{ExtSteinmannabc}, we find only 40 independent
pairs~\cite{Caron-Huot:six_loops}. Of these pairs, 24 are parity even
and 16 parity odd. Interestingly, these pairs also match those provided by the ``cluster adjacency'' principle described in ref.~\cite{Drummond:2017ssj}, once integrability is imposed.

In order to show linear combinations of symbol entry pairs more clearly, we denote a pair of allowed final entries using the following notation (not to be confused with a similar notation used in superscripts in eqs.~\eqref{integrabilitycA}--\eqref{integrabilityc}):
\be
\ldots \otimes x\otimes y \rightarrow [x,y]\,,
\ee
so that a sum of $[x,y]$ denotes symbols of the form
\be
e_1\otimes\ldots\otimes e_j\otimes x\otimes y + e_1\otimes\ldots\otimes e_j\otimes z\otimes w \rightarrow [x,y]+[z,w]\,.
\ee
We also use the multiplicative property of the symbol and make the following abbreviations,
\be
[xy,z] \equiv [x,z] + [y,z], \qquad [x/y,z] \equiv [x,z] - [y,z].
\ee
To denote cyclic classes, we write $a_i\in\{a,b,c\}$, $m_i\in\{m_u,m_v,m_w\}$, and $y_i\in\{y_u,y_v,y_w\}$. Again, $i\neq j\neq k$. In this notation, the 16 odd pairs are
\bea
[a_i,y_i]+[y_i,a_i],\nonumber\\ 
{[}a_i,y_j y_k]+[y_j y_k,a_i],\nonumber\\ 
{[}m_j/m_k,y_i]+[y_i,m_j/m_k],\nonumber\\ 
{[}m_i,y_u y_v y_w]+[y_u y_v y_w,m_i],\nonumber\\ 
{[}a_i m_i,y_j y_k]-[m_j, y_j]-[m_k,y_k]-[y_j y_k,a_i m_i]+[y_j,m_j]+[y_k,m_k],\nonumber\\ 
{[}m_u,y_v y_w]+[m_v,y_u y_w]+[m_w,y_u y_v]-[y_v y_w,m_u]-[y_u y_w,m_v]-[y_u y_v,m_w],
\label{oddintegpairs}
\eea
while the 24 even pairs are
\bea
[a_i,a_i],\nonumber\\
{[}m_i,m_i],\nonumber\\
{[}a_i,m_j]+[m_j,a_i],\quad [a_i a_j,m_k],\nonumber\\
{[}m_j,m_k]+[m_k,m_j]-[y_i,y_i],\nonumber\\
{[}a_i,m_j m_k]+[y_i,y_u y_v y_w],\nonumber\\
{[}y_u,y_u^2 y_v y_w]+[y_u^2 y_v y_w,y_u],\quad [y_v,y_u y_v^2 y_w]+[y_u y_v^2 y_w,y_v],\nonumber\\
{[}a,m_v]+[m_u,m_v]-[m_w,b]+[m_w,m_u]-[m_w,m_v]+[y_v,y_w].
\label{evenintegpairs}
\eea
%


\section{Coproduct Relations in the \texorpdfstring{$\Omega$}{Omega} Space} \label{appendix:coproduct_relations}

\subsection{Coproduct relations for general \texorpdfstring{$\Omega$}{Omega} functions}

The spaces of single and double coproducts are much smaller in the
$\Omega$ subspace than in the full space of Steinmann hexagon functions described in appendix~\ref{ext_steinmann}.
At the single coproduct level, parity even functions in the $\Omega$
space are observed to 
have only 8, not 9 final entries: $y_u$ and $y_v$ do not appear
separately, but only the combination $y_uy_v$.  That is,
$E^{y_u} - E^{y_v} = 0$ if $E\in\Omega$ and $E$ is parity even.
Notice from \eqn{zdz_abc_app} that the last, $\sqrt{\Delta}$-containing
term vanishes for the $z$ derivatives of all even $\Omega$ functions.

Parity odd functions are found to be even more restricted;
they have only the 4 final entries $\{ m_u/m_v, y_u, y_v, y_w \}$.
That is, $O^a = O^b = O^c = O^{m_w} = O^{m_u} + O^{m_v} = 0$
if $O\in\Omega$ and $O$ is parity odd.

Parity-odd functions $O$ in the $\Omega$ space have 8 allowed final entry pairs, 2 pairs of letters that are even under parity and 6 that are odd. They are
\bea
[m_u/m_v,m_u/m_v]+[y_u y_v,y_u y_v],\nonumber\\
2 [y_u y_v,y_u y_v]+[y_u y_v,y_w]+[y_w,y_u y_v],
\label{OddFnsOddNFE}
\eea
and
\bea
[m_u/m_v,y_w]+[y_w,m_u/m_v],\nonumber\\
{[}m_u/m_v,y_u y_v]+[y_u y_v,m_u/m_v],\nonumber\\
2 [m_u m_v,y_w]+2 [m_w,y_u y_v]+[m_u m_v,y_u y_v]+[y_u/y_v,m_u/m_v],\nonumber\\
{[}a,y_v]+[a,y_w]+[m_u,y_w]-[m_w,y_w]-[y_v,m_u]+[y_v,m_v],\nonumber\\
{[}b,y_u]+[b,y_w]+[m_v,y_w]-[m_w,y_w]+[y_u,m_u]-[y_u,m_v],\nonumber\\
{[}c,y_u y_v]-[m_u m_v,y_u y_v]-[m_u m_v,y_w].
\label{OddFnsEvenNFE}
\eea

Parity-even functions $E$ in the $\Omega$ space have
20 allowed final entry pairs, 17 even and 3 odd. They are
\bea
[m_u,m_u],\quad [m_v,m_v],\nonumber\\
{[}c,c]-[a,a],\quad [c,c]-[b,b],\nonumber\\
{[}m_w,a]-[b,m_w],\quad [a,m_w]-[m_w,b],\nonumber\\
{[}c,m_v]+[m_v,c],\quad [c,m_u]+[m_u,c],\nonumber\\
{[}a,m_v]-[m_v,c],\quad [b,m_u]-[m_u,c],\nonumber\\
{[}m_w,a]+[b,m_w]+[a,m_w]+[m_w,b]-4 [c,c],\nonumber\\
2 [m_u,c]+2 [m_v,c]+[y_u y_v,y_w]-[y_w,y_u y_v],\nonumber\\
2 [c,c]-[y_u y_v,y_w]-[y_w,y_u y_v]-2 [y_w,y_w],\nonumber\\
2 [m_u,m_v]+2 [m_v,m_u]+[y_u y_v,y_w]+[y_w,y_u y_v],\nonumber\\
-[a/b,m_w]+[m_u/m_v,c]-[y_u/y_v,y_u y_v]-[y_u/y_v,y_w],\nonumber\\
-[a/b,m_w]-[m_u/m_v,c]-[m_u m_v,m_u/m_v]+2[m_w,m_u/m_v]-[y_u/y_v,y_w],\nonumber\\
2 [y_u y_v, y_u y_v]+[y_u y_v, y_w]+[y_w,y_u y_v],
\label{EvenFnsEvenNFE}
\eea
and
\bea
[m_u/m_v,y_w]+[y_w,m_u/m_v],\nonumber\\
{[}m_u/m_v,y_u y_v]+[y_u y_v,m_u/m_v],\nonumber\\
{[}y_u y_v,c]-[y_u y_v,m_u m_v]-[y_w,m_u m_v].
\label{EvenFnsOddNFE}
\eea
%


\subsection{Coproduct relations for the \texorpdfstring{$c$}{c} Discontinuity}
\label{cDiscAppendix}

We define the space $\Omega_c$ to be the discontinuity in the $c$ variable
of all the functions in the $\Omega$ space.  In $\Omega_c$,
the set of allowed single and double coproducts
shrinks even further.  Also, one of the three derivatives simplifies
considerably.  At the same time, we lose almost no information,
because only the functions $\kappa$ and $\tilde{\kappa}$ are set to zero,
while the remaining functions are still linearly independent.

In particular, there are only five letters in the alphabet for $\Omega_c$:
\be
\{ c, m_u, m_v, y_u y_v, y_w \}.
\label{discOletters}
\ee
Odd functions in $\Omega_c$ are further restricted to have only three
final entries, $\{ m_u/m_v, y_u y_v, y_w \}$.

Notice from \eqn{zdz_abc_app} that the $z$ derivative of the $c$-discontinuity
$F_c$ of a function $F$ simplifies greatly, to
\be
z \frac{\del F_c}{\del z} = - F_c^{m_u} \,.
\label{zdz_abc_discc}
\ee
In the finite-coupling solution~\eqref{result double penta},
the $z$ derivative is also simple, in that it does not touch
the hypergeometric functions.

On the $c$-discontinuity, the 17 even pairs of final entries for parity-even
functions $E$ in \eqn{EvenFnsEvenNFE} reduce to 8 final entry pairs
\bea
[m_u,m_u],\quad [m_v,m_v],\nonumber\\
{[}c,m_v]+[m_v,c],\quad [c,m_u]+[m_u,c],\nonumber\\
2 [m_u,c]+2 [m_v,c]+[y_u y_v,y_w]-[y_w,y_u y_v],\nonumber\\
2 [c,c]-[y_u y_v,y_w]-[y_w,y_u y_v]-2 [y_w,y_w],\nonumber\\
2 [m_u,m_v]+2 [m_v,m_u]+[y_u y_v,y_w]+[y_w,y_u y_v],\nonumber\\
2 [y_u y_v, y_u y_v]+[y_u y_v, y_w]+[y_w,y_u y_v],
\label{EvenFnsEvenNFEdisc}
\eea
while the 3 odd final entry pairs for parity-even
functions in \eqn{EvenFnsOddNFE} remain the same.

On the $c$-discontinuity, the 2 parity-even pairs of final entries for parity-odd
functions $O$, in \eqn{OddFnsOddNFE},
remain the same, while the 6 parity-odd final entry pairs in \eqn{OddFnsEvenNFE}
reduce to 3 final entry pairs,
\bea
[m_u/m_v,y_w]+[y_w,m_u/m_v],\nonumber\\
{[}m_u/m_v,y_u y_v]+[y_u y_v,m_u/m_v],\nonumber\\
{[}c,y_u y_v]-[m_u m_v,y_u y_v]-[m_u m_v,y_w].
\label{OddFnsEvenNFEdisc}
\eea
%


\subsection{Coproduct relations for \texorpdfstring{$\Omega$, $\tilde{\Omega}$, and $\OL$}{Omega, OmegaTilde, and O}}
\label{ExtraCoproductRelationsSubAppendix}

In this subsection we provide coproduct relations between the integrals $\Omega^{(L)}$, $\tilde\Omega^{(L)}$ and $\OL^{(L)}$.  While it is possible to read off all such relations from results in section \ref{omegacopsubsec}, we can also derive many of them directly from the differential equations they satisfy. These relations were useful when constructing $\Omega^{(L)}$ at higher loops in an earlier stage of this work, and they serve to illustrate the structure of the $\Omega$ functions in the coproduct formalism.  The relations are all valid at sufficiently high loop order, using starting at either $L=2$ or 3.

In appendix \ref{hexvarappendix}, the $x,y,z$ derivatives of any functions $F$
are expressed in terms of
coproducts for the alphabet ${\cal S}^{\prime}_\text{hex}$.
Consider, for example, the operator $z\partial_z$. 
Its action can be written as
\be
z \frac{\del F}{\del z} =
(1-v) F^a - (1-u) F^b - \frac{1}{2} ( F^{m_u} - F^{m_v} )
+ \frac{u-v}{2(1-w)} F^{m_w}
+ \frac{\sqrt{\Delta}}{2(1-w)} ( F^{y_u} - F^{y_v} ) \,.
\label{zdz_abc}
\ee
When applying a second-order differential operator,
such as those appearing in the differential equations for the
weight-$2L$ transcendental function $\Omega^{(L)}$,
the weight can be reduced either by one or two.
The former case occurs when the second derivative hits the rational
factor in \eqn{zdz_abc} instead of the transcendental function.
In the case of \eqn{Omdiffxfin}, using the analogous expression for $x\partial_x$, we find that these weight-$(2L-1)$ terms combine to
\be
\frac{1-x}{x} \Bigl( (x\del_x)^2 - (z\del_z)^2 \Bigr) \Omega^{(L)} \Big|_{2L-1} = 
\frac{1}{1-x} \left( \Omega^a + \Omega^b -\Omega^c + \Omega^{m_w}+ \Omega^{y_w}\right).
\ee
Since the action of the operator should give $\Omega^{(L-1)}$, which has
uniform weight $2L-2$, the right-hand side should vanish.
This condition, together with the parity conjugate relation~\eqref{Omtdiffyfin},
implies that
\be
 \Omega^a+\Omega^b-\Omega^c +\Omega^{m_w} =0, \qquad \Omega^{y_w} =0.
\ee
Furthermore, combining \eqn{Odef} with \eqns{OmtefromOL}{OmtofromOL}
gives second-order equations
relating $\Omega^{(L)}$ to pure functions of weight $2L-2$,
$\Omte^{(L-1)}$ and $\Omto^{(L-1)}$.
By canceling the wrong-weight terms we find two more relations:
\be
 \Omega^{m_w}=0, \qquad \Omega^{y_u}=\Omega^{y_v}\,.
\ee
Substituting the second of these equations into the derivative~\eqref{Odef} that defines $\OL$, we learn that $\OL$ is a pure function and that\footnote{
We have suppressed the $(L)$ superscript for coproducts of $L$ loop functions, but include a reminder in this equation that the odd ladder integral evaluated at one lower loop order, $\OL^{(L-1)}$.}

\be
\Omega^{y_u} = \Omega^{y_v} = \frac{1}{2} \OL^{(L-1)}\,.
\label{OmOLm1}
\ee

Now we look at derivatives of $\OL$.  We start with the relation $\Omto = -z\del_z \OL$ and set $F=\OL$ in \eqn{zdz_abc}.  Because $\Omto$ is a pure transcendental function with no rational prefactor, all the terms containing non-constant rational prefactors must vanish.
It is easy to see that no linear combinations of $F^a$, $F^b$, $F^{m_w}$ and
$F^{y_u}-F^{y_v}$ can produce a constant prefactor.
Hence we obtain,
\be
\OL^{a} = \OL^{b} = \OL^{m_w} = 0,  \qquad \OL^{y_u} = \OL^{y_v} \,,
\qquad \OL^{m_u} - \OL^{m_v} = 2 \Omto.
\label{OLc_set1}
\ee

Next we insert \eqn{OLc_set1} into \eqn{xdxplusydy_abc} for
$(x\del_x+y\del_y)\OL$, which appears in \eqn{OmfromOL} for $\Omega$:
\bea
\Omega &=& \frac{1+y_w}{1-y_w} \OL^{c}
+ 2 \frac{y_w(1-y_uy_v)}{(1-y_w)(1-y_uy_vy_w)} ( \OL^{m_u} - \Omto )
+ \OL^{y_w} \,.
\label{xdxpydy_OL}
\eea
Purity of $\Omega$ in \eqn{xdxpydy_OL} leads to the additional equations,
\be
\OL^{c} = 0,  \qquad \OL^{m_u} = - \OL^{m_v} = \Omto,  \qquad 
\OL^{y_w} = \Omega.
\label{OLc_set2}
\ee
Substituting \eqns{OLc_set1}{OLc_set2} into \eqn{OmtefromOL} for $\Omte$,
after evaluating it with the help of \eqns{xdxplusydy}{xdxminusydy},
yields
\be
\Omte = \frac{xy}{x-y} \Bigl[ (1-x)\del_x + (1-y)\del_y \Bigr] \OL
 = \OL^{y_u} - \Omega,
\label{OmtefromOLsub}
\ee
so that
\be
\OL^{y_u} = \OL^{y_v} = \Omega + \Omte.
\label{OLyuoryv}
\ee

In summary, since we know all three derivatives of the odd ladder $\OL$,
we can determine all nine of its first coproducts,
\bea
\OL^{a} &=& \OL^{b} = \OL^{c} = \OL^{m_w} = 0,  \qquad 
\OL^{m_u} = - \OL^{m_v} = \Omto, \nonumber\\
\OL^{y_u} &=& \OL^{y_v} = \Omega + \Omte, \qquad
\OL^{y_w} = \Omega.
\label{OLcsummary}
\eea

Returning to the derivatives of $\Omega$, we find empirically that $z\del_z \Omega$ is a pure function.  This fact implies, via \eqn{zdz_abc}, that
\be
\Omega^a = \Omega^b = \Omega^c = 0.
\label{Omegaabcvanish}
\ee
These additional relations imply that
\bea
x\del_x\Omega &=& \frac{1}{2} ( \Omega^{m_u} + \Omega^{m_v} ) - \Omega^{y_u} \,,
\qquad
y\del_y\Omega  = \frac{1}{2} ( \Omega^{m_u} + \Omega^{m_v} ) + \Omega^{y_u} \,,
\label{Omderivcoprodxy}\\
z\del_z\Omega &=& \frac{1}{2} ( - \Omega^{m_u} + \Omega^{m_v} ) \,.
\label{Omderivcoprodz}
\eea

We have also found some first-order coproduct relations for $\Omt$:
\be
0 = \Omt^a = \Omt^b = \Omt^{m_w} = \Omt^{m_u} + \Omt^{m_v}
= \Omt^{y_u} - \Omt^{y_v} = - \Omt^{m_u} + \Omt^{y_u} - \Omt^{y_w} \,.
\label{Omtcrelations}
\ee
These relations are equivalent to
\bea
x\del_x \Omt &=& - \Omt^{m_u} - \frac{x}{1-x} ( \Omt^c - \Omt^{y_w} )\,,
\label{xdxOmt}\\
y\del_y \Omt &=& \Omt^{m_u} - \frac{y}{1-y} ( \Omt^c + \Omt^{y_w} )\,,
\label{ydyOmt}\\
z\del_z \Omt &=& - \Omt^{m_u} \,.
\label{zdzOmt}
\eea
There are also two relations that are specific to $\Omte$ and $\Omto$,
\be
\Omte^{y_w} = \frac{1}{2} \OL^{(L-1)} \,, \qquad
\Omto^c = 0.
\label{Omtcspecial}
\ee

By taking derivatives of the all-orders representation~\eqref{result double penta 2} of $\Omt$, it is possible to show that the quantity $\Omt^{m_u}$ appearing
in the $x$ and $y$ derivatives of $\Omt$ in \eqns{xdxOmt}{xdxOmt} is indeed
the same as the one appearing in the $z$ derivative~\eqref{zdzOmt}.
One can also show to all orders that
\be
(x\del_x + y\del_y)\Omt = \frac{y}{1-y} (x\del_x + z\del_z)\Omega
                        + \frac{x}{1-x} (y\del_y - z\del_z)\Omega \,.
\label{OmtOmdiffrel}
\ee
Inserting the coproduct representations of these derivatives, given above,
we find relations between the first coproducts of $\Omt$ and $\Omega$:
\be
- \Omt^c + \Omt^{y_w} = \Omega^{m_u} + \Omega^{y_u} \,, \qquad
  \Omt^c + \Omt^{y_w} = - \Omega^{m_v} + \Omega^{y_u} \,.
\label{OmtOmcoprel}
\ee

By combining \eqns{OmOLm1}{OLcsummary}, we can write all
three integrals, $\Omega$, $\Omte$ and $\Omto$, as double coproducts
of the $\Omega$ integral at one higher loop order:
\be
\Omega^{(L-1)} = 2 \, \Omega^{y_w,y_u} \,, \qquad
\Omte^{(L-1)} = 2 \, ( \Omega^{y_u,y_u} - \Omega^{y_w,y_u} ) \,, \qquad
\Omto^{(L-1)} = 2 \, \Omega^{m_u,y_u} \,.
\label{fromOm}
\ee
We have found empirically that these integrals can also be
written as double coproducts of $\Omte$ and $\Omto$:
\be
\Omega^{(L-1)} = - 2 \Omte^{c,c} \,, \qquad
\Omte^{(L-1)} = 2 ( \Omte^{m_u,c} + \Omto^{m_u,y_w} ) \,, \qquad
\Omto^{(L-1)} = 2 \Omte^{y_w,m_u} = 2 \Omto^{y_w,y_u} \,,
\label{fromOmt}
\ee
for $L>2$.
In both sets of equations, we suppress the $(L)$ superscript
on the right-hand side for clarity.


\subsection{Improving the MHV-NMHV operator}

Through five loops, the six-point MHV amplitude ${\cal E}$ and
NMHV amplitude $E$ obey a curious relation that connects
these amplitudes at different loop orders~\cite{Caron-Huot:2016owq}.
If we perform a cyclic permutation $u\to v\to w\to u$ on that relation,
in order to give it the same $u \lr v$ symmetry as the $\Omega$ integral,
it becomes
\be
X^{\rm old}[{\cal E}(u,v,w)] = g^2 (2E(v,w,u) - {\cal E}(u,v,w)),
\label{MHVNMHVrelation}
\ee
where
\be
X^{\rm old}[F] \equiv - F^{w,w} - F^{1-w,w} - 3 F^{y_w,y_w}
+ F^{y_u,y_u} + F^{y_v,y_v} + 2  ( F^{y_u,y_w} + F^{y_v,y_w} )
- F^{y_u,y_v} - F^{y_v,y_u}
\label{MOld}
\ee
is written in terms of the old alphabet ${\cal S}_\text{hex}$.
(For an earlier version of this relation, see ref.~\cite{Dixon:2014iba}.)
In fact, this relation fails for amplitudes at six
loops~\cite{Caron-Huot:six_loops}.
However, we will see that a version of it survives to arbitrary loop order
in the pentaladder integrals.

In general, ${\cal E}$ obeys many relations on its double coproducts, which allow the operator $X$ to be rewritten without changing its action on ${\cal E}$. However, its action on the ladder integrals will generically change. It turns out that a better form for $X$, written in terms of the new alphabet ${\cal S}^{\prime}_\text{hex}$, is
\bea
X[F] &=& - 3 F^{y_w,y_w}
+ F^{y_u,y_u} + F^{y_v,y_v} + 2 ( F^{y_w,y_u} + F^{y_w,y_v} )
- F^{y_u,y_v} - F^{y_v,y_u} \nonumber\\
&&\hskip0cm\null
+ F^{a,a} + F^{b,b} + F^{c,c} + F^{a,m_w} + F^{b,m_w} - F^{m_w,a} - F^{m_w,b} \nonumber\\
&&\hskip0cm\null
- 2 ( F^{c,m_u} + F^{c,m_v} - F^{m_u,c} - F^{m_v,c} ) \,.
\label{Mhat}
\eea
This form is better because $X$ now has a very simple action
on the ladder integrals.  We find that
\bea
X[\WL(u,v,w)] &=& X[\WL(v,w,u)] = X[\WL(w,u,v)] = 0\,,
\label{MW}\\
X[\Omega(u,v,w)] &=& X[\Omega(v,w,u)] = X[\Omega(w,u,v)] = 0\,,
\label{MO}\\
X[\tilde\Omega(u,v,w)] &=& -2g^2\,\tilde\Omega(u,v,w), \qquad
X[\tilde\Omega(v,w,u)] = X[\tilde\Omega(w,u,v)] = 0\,,
\label{MtO}\\
X[\OL(u,v,w)] &=& -2g^2\,\OL(u,v,w), \qquad
X[\OL(v,w,u)] = X[\OL(w,u,v)] = 0\,.
\label{MOddLadder}
\eea
There are anomalous terms in the even parts of $X[\tilde\Omega(v,w,u)]$, $X[\tilde\Omega(w,u,v)]$, $X[\Omega(v,w,u)]$ and $X[\Omega(w,u,v)]$ at $L=2$, and in $X[\WL(u,v,w)]$, $X[\WL(v,w,u)]$, and $X[\WL(w,u,v)]$ at both $L=2$ and $L=3$.  But above three loops, there are no anomalies in the action on these integrals to any order. This can be contrasted with the operator's action on ${\cal E}$, which remains anomalous at six loops.

The operator $X$ also has an interesting action on the full $\Omega$ space. In particular, note that for each ladder integral considered above (and ignoring low-weight anomalies), $X[F(v,w,u)] = X[F(w,u,v)] = 0$.  While this is not quite true for the full space, we do find that for a general function $F(u,v,w) \in \Omega$,
\be
X[F(v,w,u)]\,,\, X[F(w,u,v)]\,\, \in \{ \kappa,\tilde{\kappa} \}\,.
\ee
That is, the action of the operator $X$ on cyclic rotations of functions in the $\Omega$ space can be expressed entirely in terms of $\kappa$ and $\tilde{\kappa}$ functions of the appropriate weight, which vanish on the $c$-discontinuity.  In effect, the operator $X$ annihilates the $c$-discontinuity of the cyclic and anti-cyclic rotations of the $\Omega$ functions. This is a surprising property, and one that suggests further investigation.

\bibliographystyle{JHEP}

\bibliography{omega}

\end{fmffile}
\end{document}